\documentclass[]{aa}  

\usepackage{graphicx}
\usepackage{txfonts}
\usepackage{float}
\usepackage{subfigure}
\usepackage{lscape}
\usepackage{color,hyperref}
\hypersetup{colorlinks,breaklinks,
            linkcolor=blue,urlcolor=blue,
            anchorcolor=blue,citecolor=blue}

\usepackage{caption}
\usepackage{longtable}
\DeclareCaptionFormat{cont}{#1 (cont.)#2#3\par}

\begin{document} 

\title{The chemical structure of the very young starless core L1521E}


   \author{Z. Nagy\inst{1}
          \and
          S. Spezzano\inst{1}
          \and
          P. Caselli\inst{1}
          \and
          A. Vasyunin\inst{2,3}
          \and
          M. Tafalla\inst{4}
          \and
          L. Bizzocchi\inst{1}
          \and
          D. Prudenzano\inst{1}
          \and
          E. Redaelli\inst{1}
          }

\institute{Centre for Astrochemical Studies, Max-Planck-Institute for Extraterrestrial Physics, Giessenbachstrasse 1, 85748 Garching,
Germany \\
\email{znagy@mpe.mpg.de}
\and
Ural Federal University, Ekaterinburg 620002, Russia
\and
Visiting Leading Researcher, Engineering Research Institute, 'Ventspils International Radio Astronomy Centre' of Ventspils University of Applied Sciences, In\v{z}enieru 101, VentspilsLV-3601, Latvia
\and
Observatorio Astron\'omico Nacional (IGN), Calle Alfonso XII, 3 Madrid, Spain
}

\date{}

 
\abstract
{L1521E is a dense starless core in Taurus that was found to have relatively low molecular depletion by earlier studies, thus suggesting a recent formation.}
{We aim to characterize the chemical structure of L1521E and compare it to the more evolved L1544 pre-stellar core.}
{We have obtained $\sim$2.5$\times$2.5 arcminute maps toward L1521E using the IRAM-30m telescope in transitions of various species, including C$^{17}$O, CH$_3$OH, \textit{c}-C$_3$H$_2$, CN, SO, H$_2$CS, and CH$_3$CCH. We derived abundances for the observed species and compared them to those obtained toward L1544. We estimated CO depletion factors using the C$^{17}$O IRAM-30m map, an $N$(H$_2$) map derived from \textit{Herschel}/SPIRE data and a 1.2 mm dust continuum emission map obtained with the IRAM-30m telescope.}
{Similarly to L1544, \textit{c}-C$_3$H$_2$ and CH$_3$OH peak at different positions. Most species peak toward the \textit{c}-C$_3$H$_2$ peak: C$_2$S, C$_3$S, HCS$^+$, HC$_3$N, H$_2$CS, CH$_3$CCH, C$^{34}$S. C$^{17}$O and SO peak close to both the \textit{c}-C$_3$H$_2$ and the CH$_3$OH peaks. CN and N$_2$H$^+$ peak close to the \textit{Herschel} dust peak.
We found evidence of CO depletion toward L1521E. The lower limit of the CO depletion factor derived toward the \textit{Herschel} dust peak is 4.3$\pm$1.6, which is about a factor of three lower than toward L1544. We derived abundances for several species toward the dust peaks of L1521E and L1544. 
The abundances of sulfur-bearing molecules such as C$_2$S, HCS$^+$, C$^{34}$S, C$^{33}$S, and SO are higher toward L1521E than toward L1544 by factors of $\sim$2-20, compared to the abundance of A-CH$_3$OH.
The abundance of methanol is very similar toward the two cores.}
{The higher abundances of sulfur-bearing species toward L1521E than toward L1544 suggest that significant sulfur depletion takes place during the dynamical evolution of dense cores, from the starless to pre-stellar stage.
The CO depletion factor measured toward L1521E suggests that CO is more depleted than previously found. Similar CH$_3$OH abundances between L1521E and L1544 hint that methanol is forming at specific physical conditions in the Taurus Molecular Cloud Complex, characterized by densities of a few $\times$10$^4$ cm$^{-3}$ and $N$(H$_2$) $\gtrsim$ 10$^{22}$ cm$^{-2}$, when CO starts to catastrophically freeze-out, while water can still be significantly photodissociated, so that the surfaces of dust grains become rich in solid CO and CH$_3$OH, as already found toward L1544. Methanol can thus provide selective crucial information about the transition region between dense cores and the surrounding parent cloud.
}


\maketitle


\section{Introduction} 

Starless cores with temperatures of about 10~K and densities above 10$^4$ cm$^{-3}$ define the stage just before the onset of low-mass star formation in molecular clouds. When the central H$_2$ density of the cores is above 10$^5$ cm$^{-3}$, starless cores become thermally supercritical and start to collapse. These are the so-called pre-stellar cores.
Probing the physics and chemistry of such regions is essential for understanding the process of star formation. 
The chemistry of starless and pre-stellar cores can be traced by different molecular lines and the dust continuum at millimetre - sub-millimetre wavelengths. In this paper we aim to characterize the chemical structure of a starless core in an early evolutionary phase.

The L1521E starless core is located in the Taurus region, at a distance of 145$^{+12}_{-16}$ pc \citep{yan2019}. It was classified as a very young core by \citet{hirota2002} because of the high abundances of carbon-chain molecules, comparable to those in TMC-1 \citep{aikawa2003}. \citet{tafallasantiago2004} further confirmed that L1521E was a very young core as they found no evidence of C$^{18}$O depletion. They estimate its age to be $\leq$1.5$\times$10$^5$ yr. \citet{fordshirley2011} confirmed the low level of CO depletion by fitting C$^{18}$O line profiles using a non-LTE radiative transfer code.
\citet{hirota2002} derived the peak H$_2$ volume density to be (1.3$-$5.6)$\times$10$^5$ cm$^{-3}$, which is consistent with the value of 2.7$\times$10$^5$ cm$^{-3}$ found by \citet{tafallasantiago2004}.
Based on \textit{Herschel}/SPIRE observations, \citet{makiwa2016} measured the dust temperature to be 9.8$\pm$0.2 K and the core mass to be 1.0$\pm$0.1 $M_\odot$.

In this paper we aim to study the chemical structure and CO depletion of L1521E and compare it to the more evolved and more massive pre-stellar core L1544. L1544 is a well-studied pre-stellar core also located in Taurus (\citealp{spezzano2017} and references therein). Based on a comparison of observations and chemical models of simple deuterated species toward L1544, \citet{kong2015} derived an age of a few 10$^5$ yr. Based on 1.2 mm contimuum data, the $N$(H$_2$) toward the dust peak is (9.4$\pm$1.6)$\times$10$^{22}$ cm$^{-2}$ and the H$_2$ volume density within a radius of (3.2$\pm$0.4)$\times$10$^3$ AU is (1.4$\pm$0.2)$\times$10$^6$ cm$^{-3}$ (\citealp{crapsi2005}, \citealp{chacontanarro2019}).
Comparing the less evolved L1521E to the more evolved and better characterized L1544 will help us to gain understanding on the evolution of dense cores, and hence the process of star formation.
The paper is organized as follows: we describe the observations in Sect. \ref{sect_observations} and in Sect. \ref{sect_dataset}. Results on the spatial distribution and column densities of the molecules, and on CO depletion are presented in Sect. \ref{sect_results}. We discuss the results and compare them to those toward L1544 in Sect. \ref{sect_discussion} and summarize them in Sect. \ref{sect_summary}.

\section{Observations and data reduction}   
\label{sect_observations}

Maps of various molecules were observed using the Eight MIxer Receiver (EMIR) instrument of the IRAM-30m telescope in on-the-fly mode with position switching. The Fourier transform spectrometer (FTS) was used as the backend with a spectral resolution of 50 kHz. The 6.25 square arcminute maps were centered on the RA(J2000)=$04^{\rm{h}}29^{\rm{m}}15.7^{\rm{s}}$ Dec(J2000)=$+26^\circ14'5''$ position, based on \citet{tafallasantiago2004}. 
Part of the observations were carried out in June and August 2016 (PI: S. Spezzano), and additional observations were done in 2018 January and April (PI: Z. Nagy).
The data were reduced using the GILDAS software \citep{pety2005}. The measured intensities were converted to main beam temperature units using a forward efficiency of 0.94 and a main beam efficiency of 0.78 around the frequency of 115 GHz, and a forward efficiency of 0.95 and a beam efficiency of 0.81 around the frequency of 86 GHz. At these frequencies, the angular resolution of the IRAM-30m observations is around 30$''$. When creating the data cubes analyzed in this paper we re-gridded the maps to a pixel size of 4$''$, as done by \citet{spezzano2017}.

We use \textit{Herschel}/SPIRE \citep{griffin2010} data from bands at 250 $\mu$m, 350 $\mu$m, and 500 $\mu$m to derive the H$_2$ column densities. These data were downloaded from the \textit{Herschel} Science Archive and are part of the \textit{Herschel} Gould Belt Survey \citep{andre2010}. We fitted a modified blackbody function with a dust emissivity index of $\beta$=1.5 to each pixel. The 250 and 350 $\mu$m images were smoothed to the resolution of the 500 $\mu$m image ($\sim$40$''$). The images were resampled to the same grid.

In addition to the IRAM-30m and \textit{Herschel}/SPIRE data we also use the 1.2 mm continuum data from the work of \citet{tafallasantiago2004}. These data were measured with the MAMBO-2 instrument of the IRAM-30m telescope in on-the-fly mode. The spatial resolution of the 1.2 mm continuum data is 11$''$.

\section{Dataset}
\label{sect_dataset}

The dataset analyzed in this paper includes maps of molecular emission from 30 species (including isotopologues) detected toward L1521E as shown in Table \ref{table:observations}. Some species, such as isotopologues containing $^{15}$N and deuterated species will be analyzed in a forthcoming paper.
The quantum numbers, frequencies, Einstein $A$ coefficients, and upper level energies are from the Cologne Database for Molecular Spectroscopy (CDMS, \citealp{muller2005})\footnote{\href{http://www.astro.uni-koeln.de/cdms/catalog}{http://www.astro.uni-koeln.de/cdms/catalog}} and Propulsion Laboratory (JPL, \citealp{pickett1998})\footnote{\href{http://spec.jpl.nasa.gov}{http://spec.jpl.nasa.gov}} molecular databases. We included the critical densities for the transitions whenever collisional rates are available for the molecules.
Table \ref{table:observations} also includes information on whether the transitions detected toward L1521E are detected toward L1544. This information is based on the public data release of the `Astrochemical Surveys At IRAM' (ASAI, \citealp{lefloch2018}, \citealp{vastel2018}). A spectrum in the frequency range between 80 and 106 GHz was observed toward the dust peak of L1544, which allows us to directly compare the two sources in Sect. \ref{sect_discussion}. Figure \ref{spatial_distribution_1} shows integrated intensity maps of the molecules detected toward L1521E, which will be further discussed in the next sections.

\begin{flushleft}
\begin{figure*}[ht!]
\includegraphics[width=6.2cm, trim=0cm 6.5cm 1.5cm 8cm,clip=true]{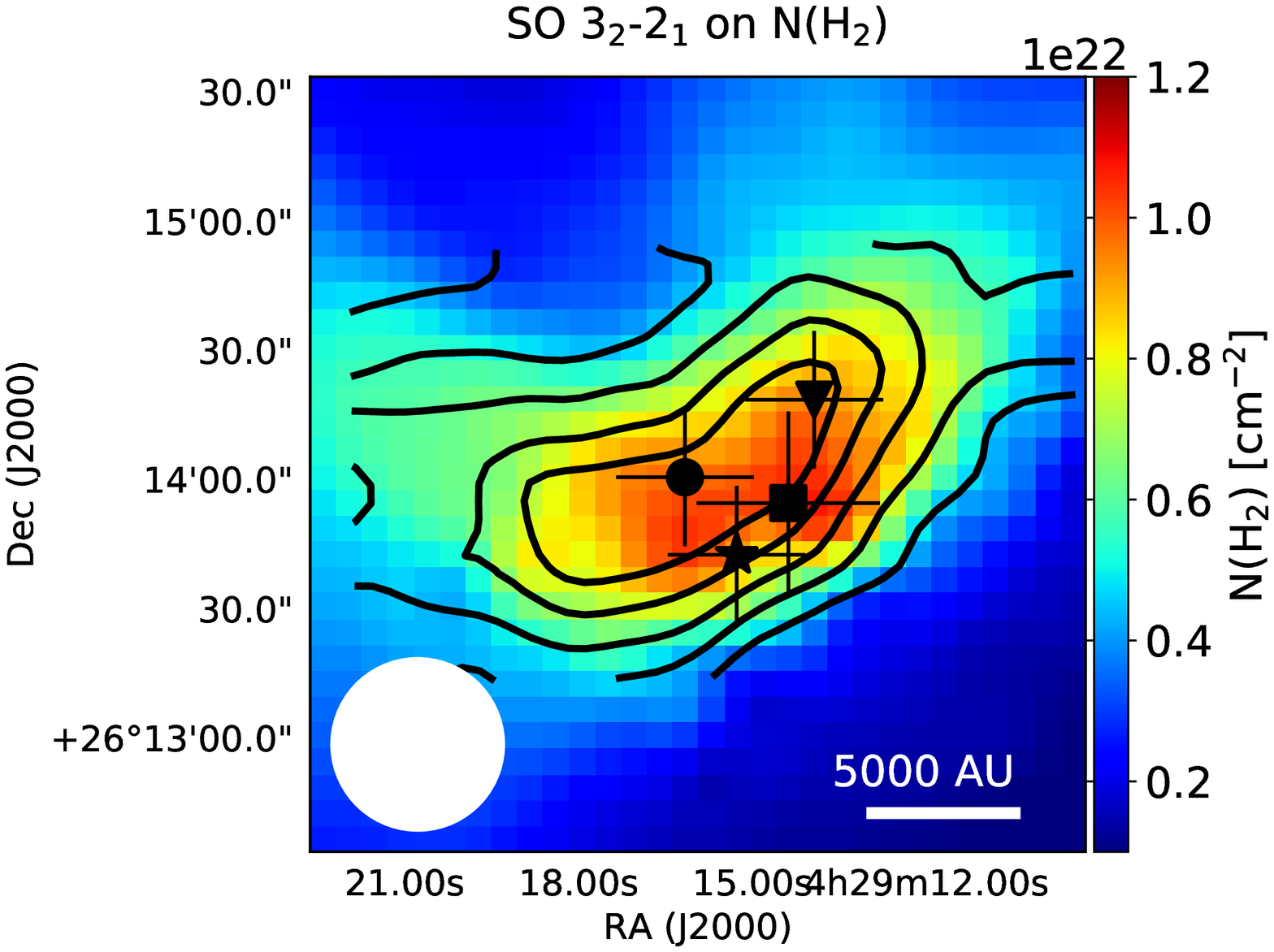} 
\includegraphics[width=6.2cm, trim=0cm 6.5cm 1.5cm 8cm,clip=true]{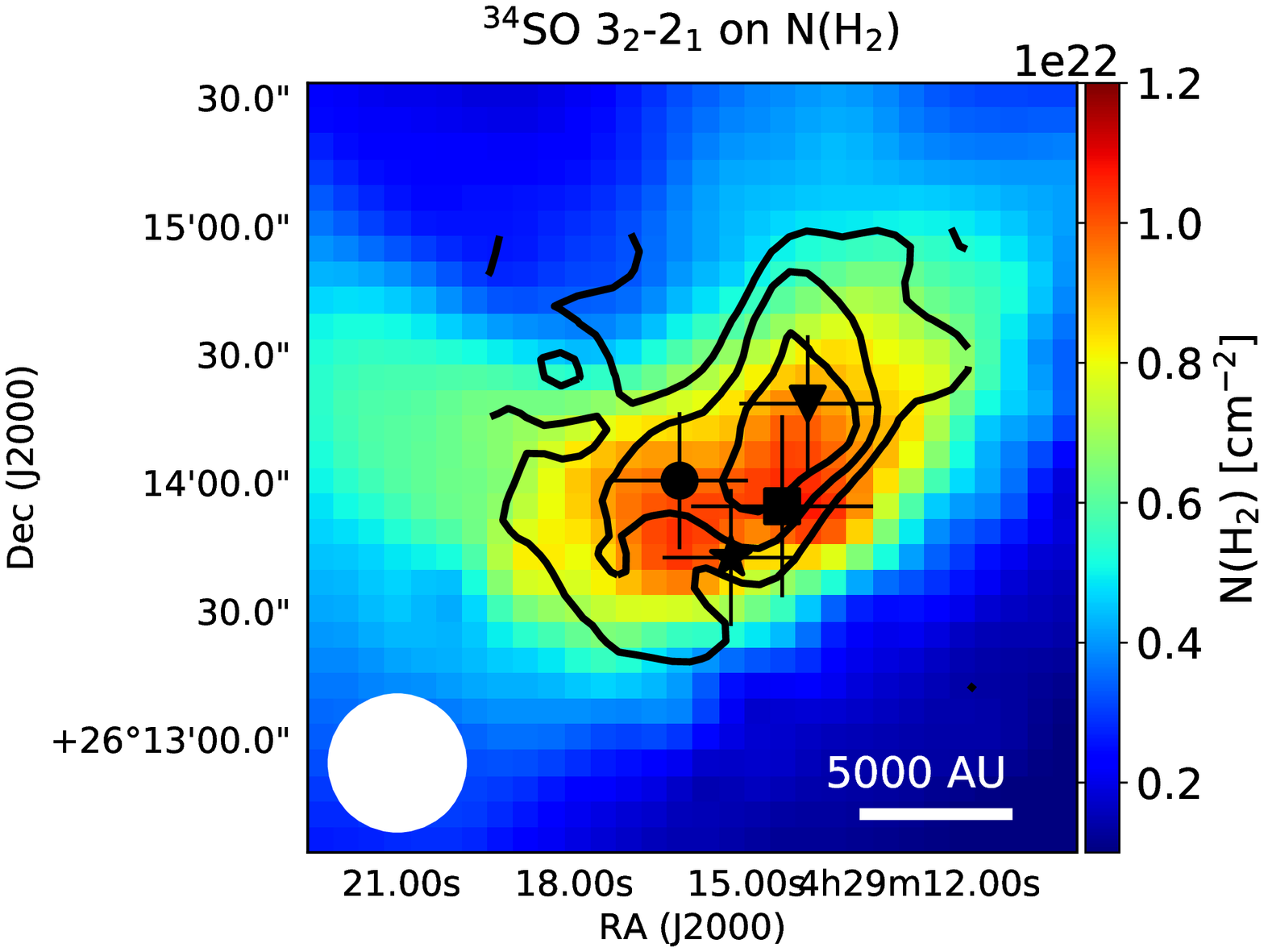} 
\includegraphics[width=6.2cm, trim=0cm 6.5cm 1.5cm 8cm,clip=true]{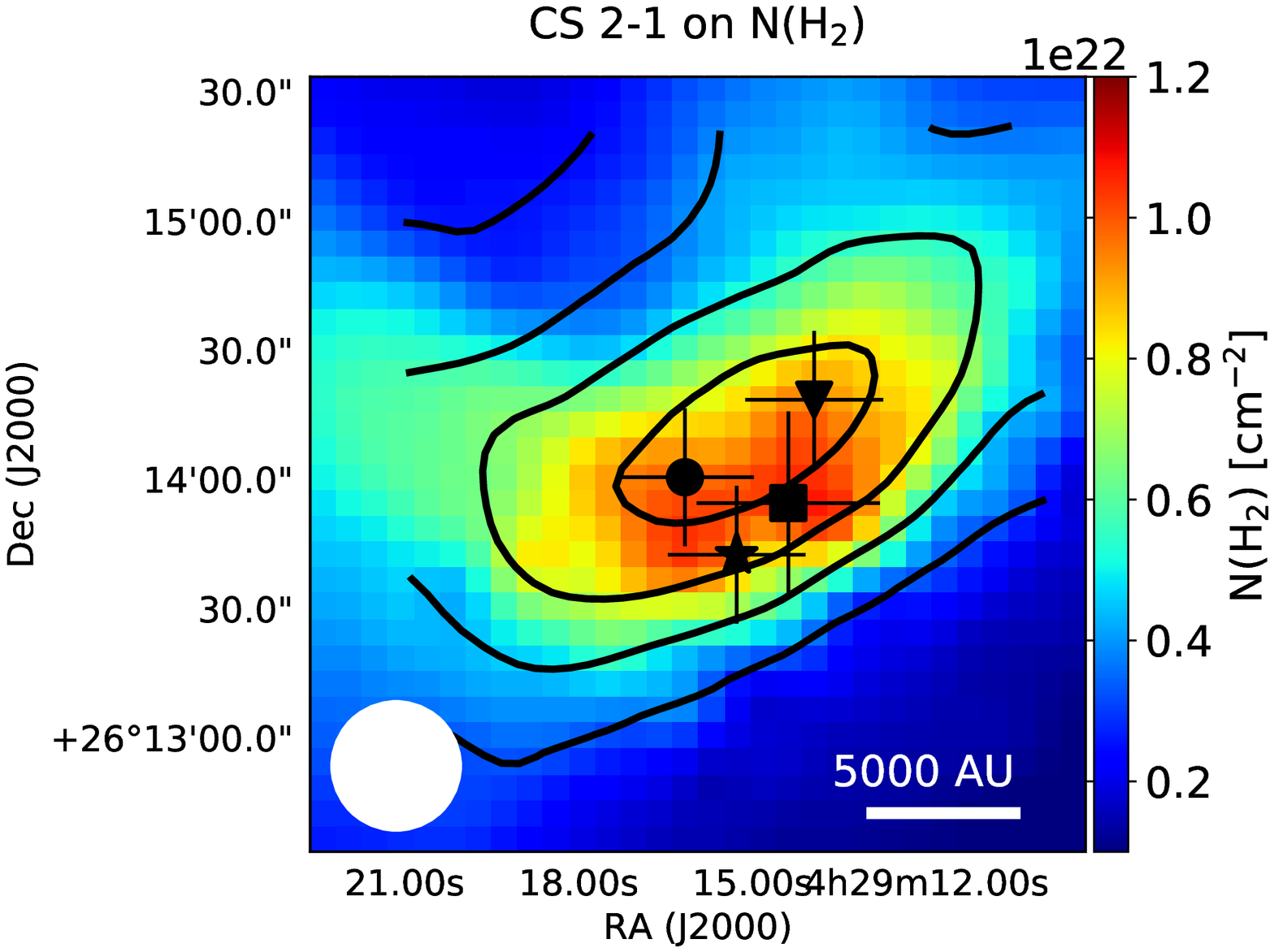} 
\includegraphics[width=6.2cm, trim=0cm 6.5cm 1.5cm 8cm,clip=true]{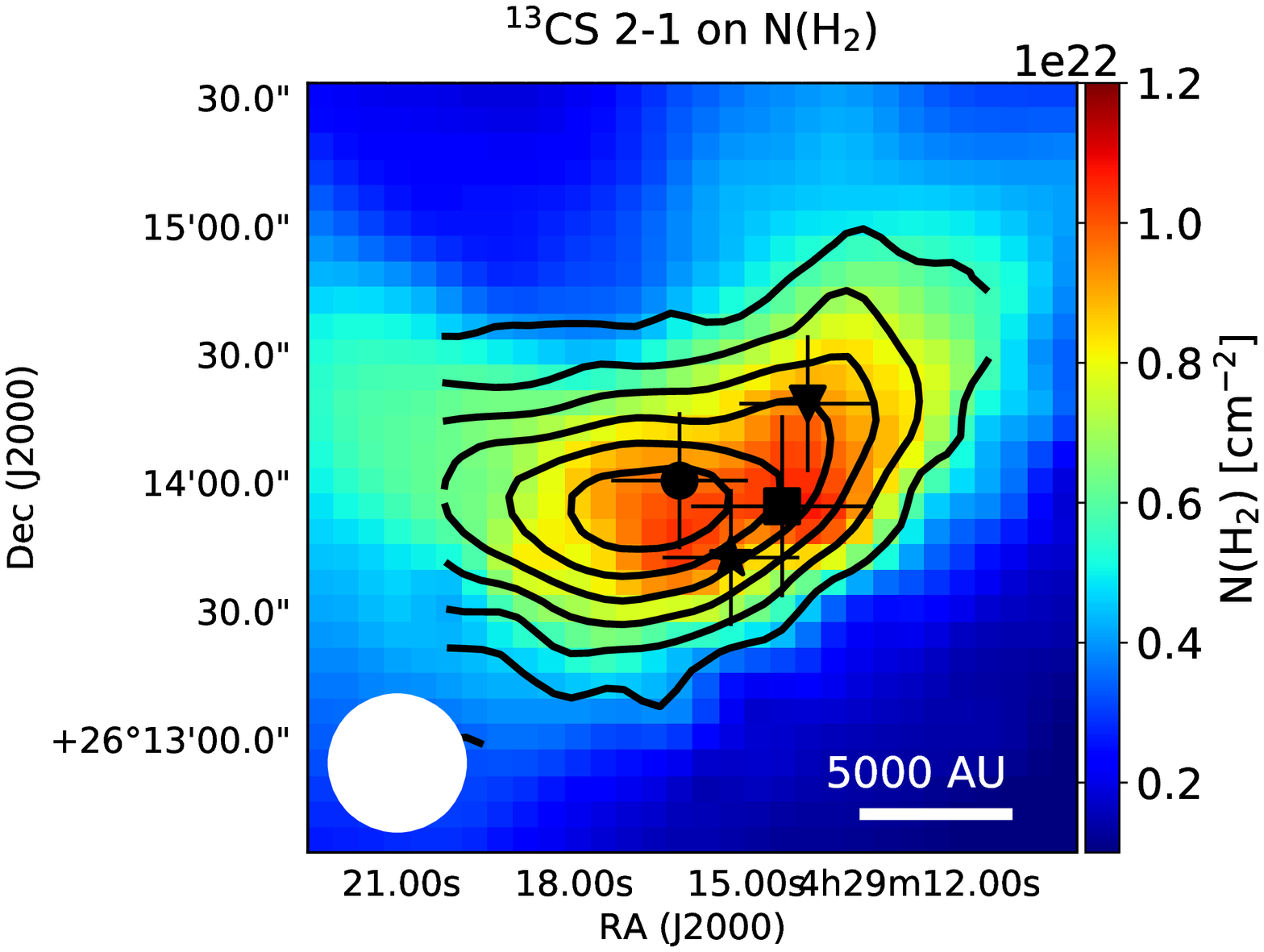} 
\includegraphics[width=6.2cm, trim=0cm 6.5cm 1.5cm 8cm,clip=true]{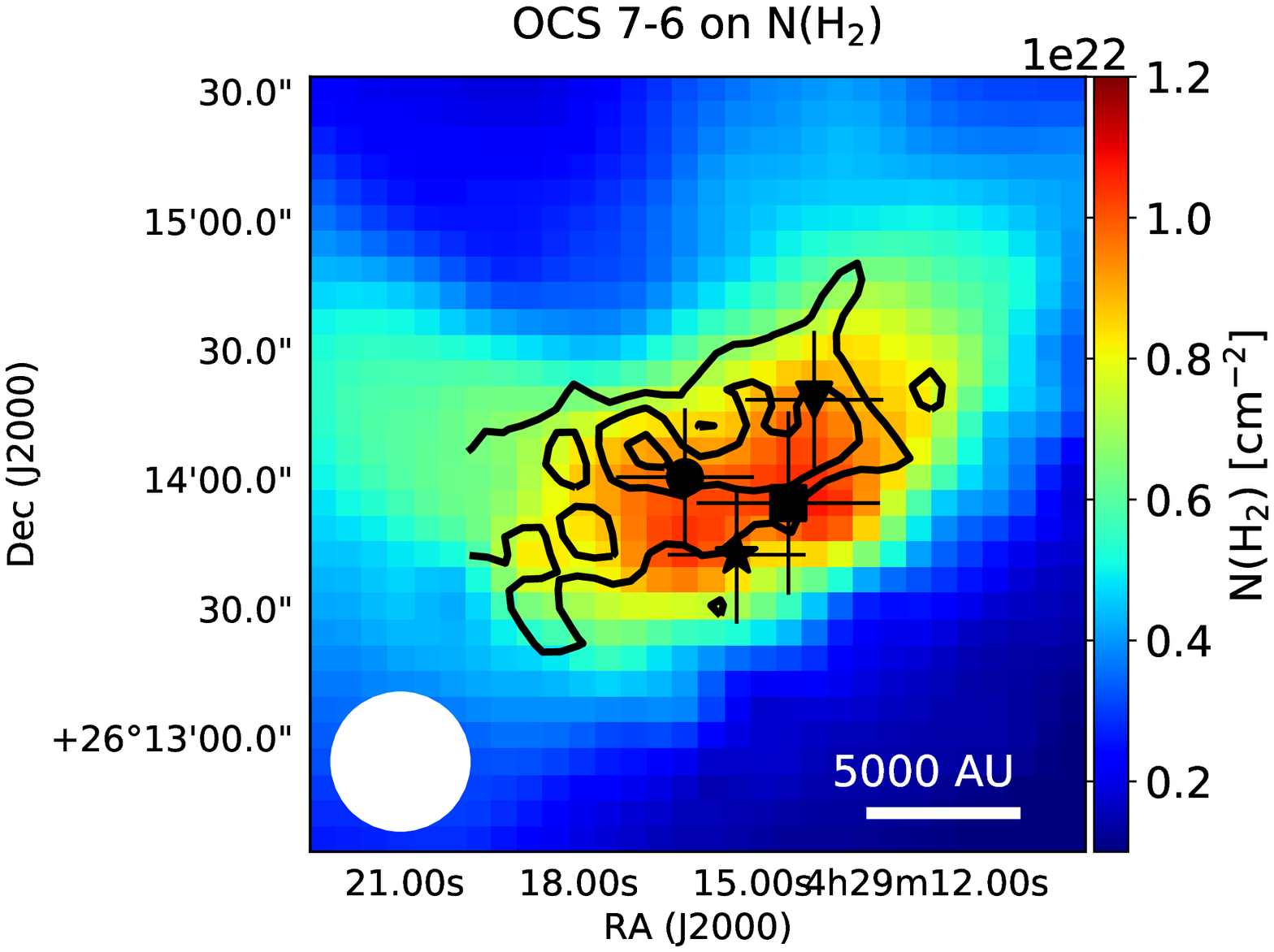}
\includegraphics[width=6.2cm, trim=0cm 6.5cm 1.5cm 8cm,clip=true]{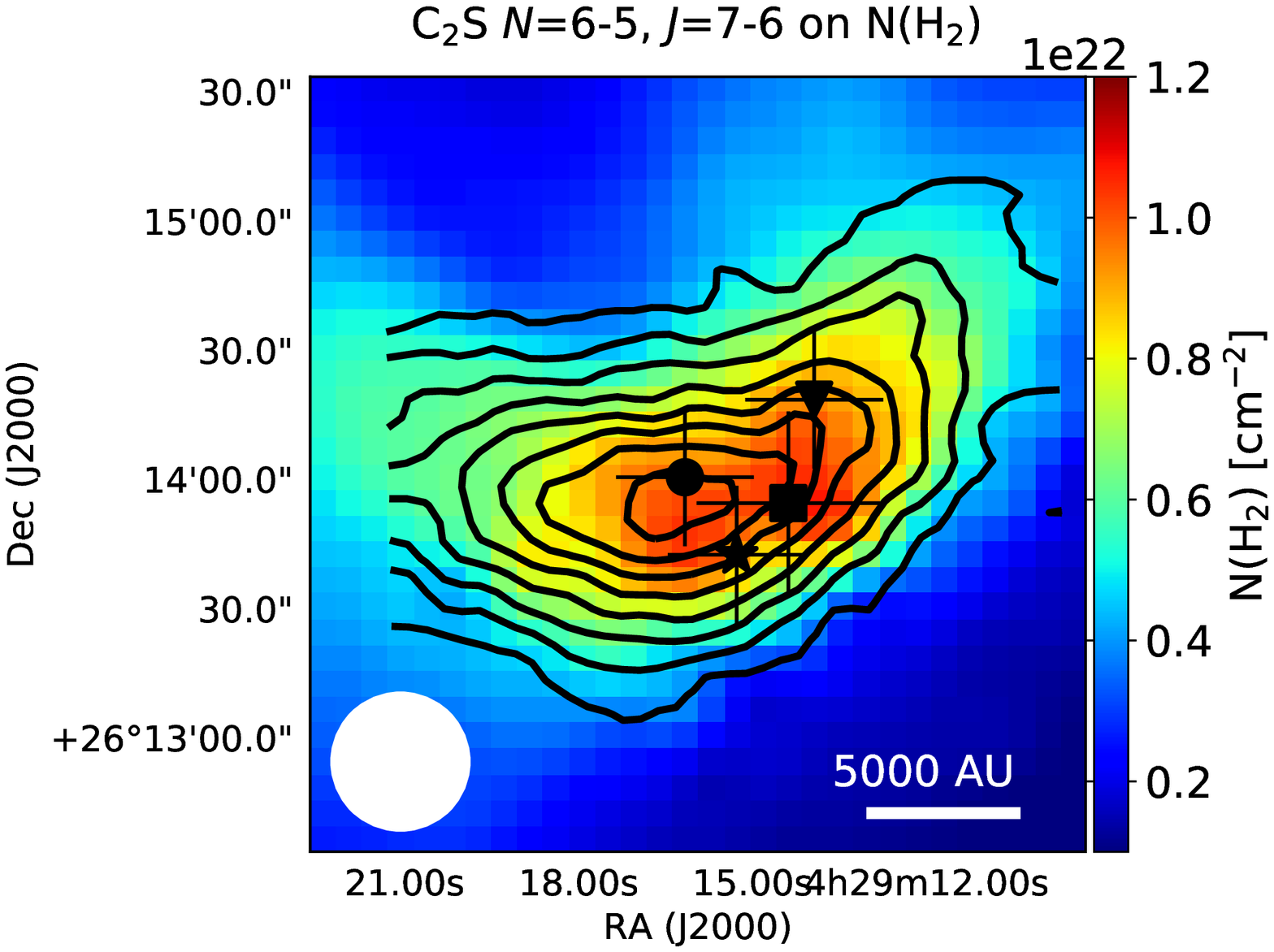}
\includegraphics[width=6.2cm, trim=0cm 6.5cm 1.5cm 8cm,clip=true]{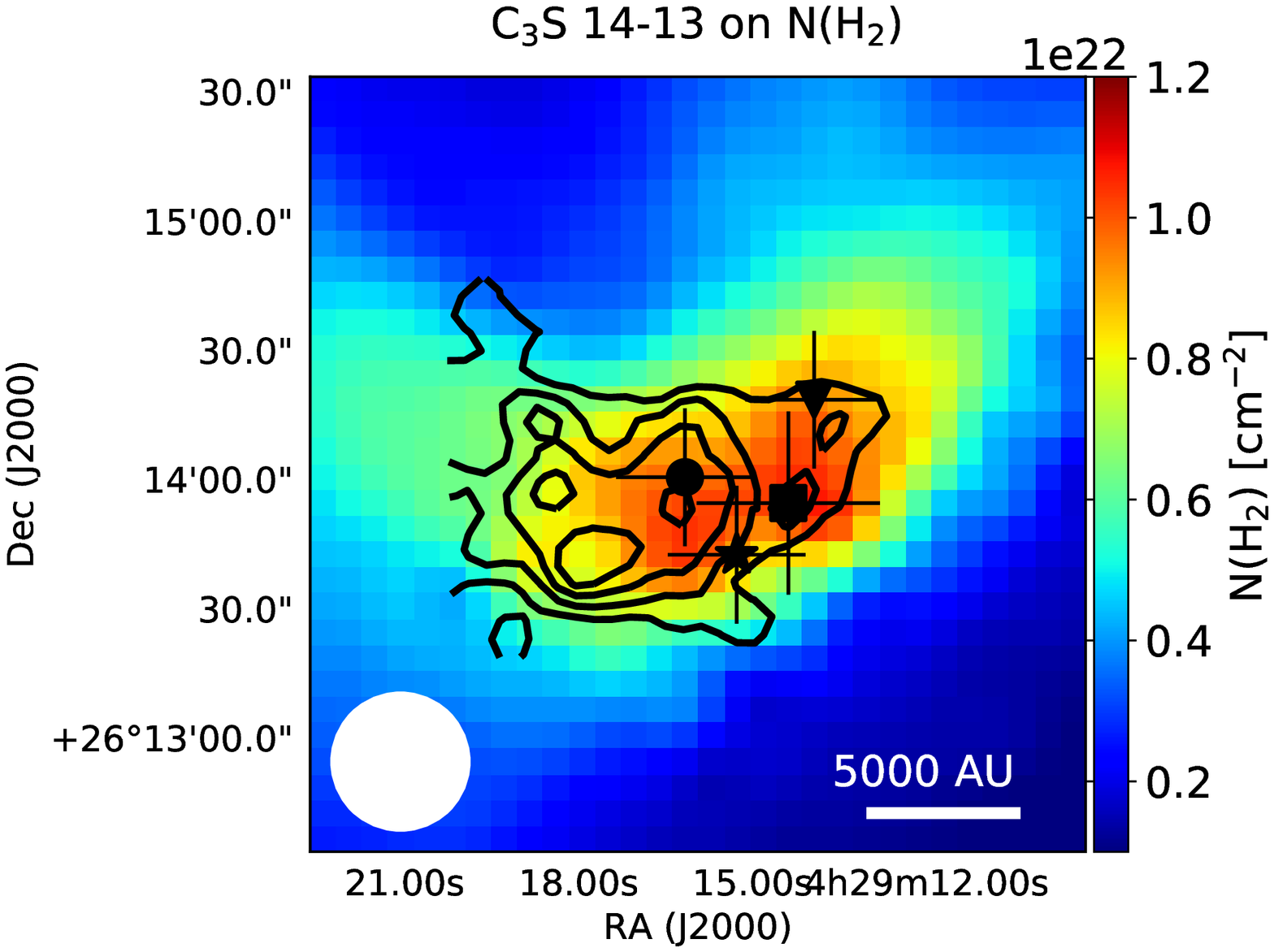}
\includegraphics[width=6.2cm, trim=0cm 6.5cm 1.5cm 8cm,clip=true]{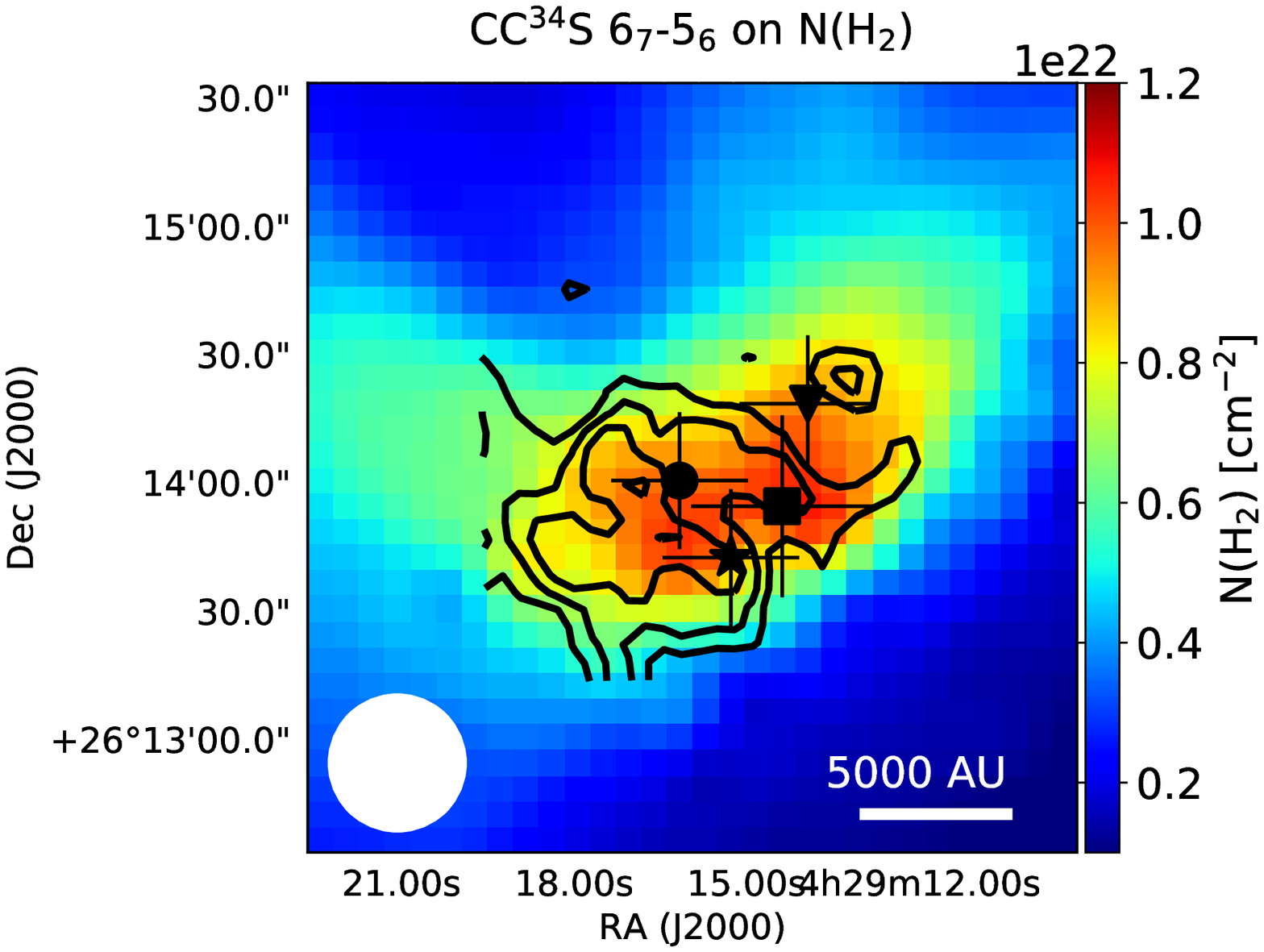} 
\includegraphics[width=6.2cm, trim=0cm 6.5cm 1.5cm 8cm,clip=true]{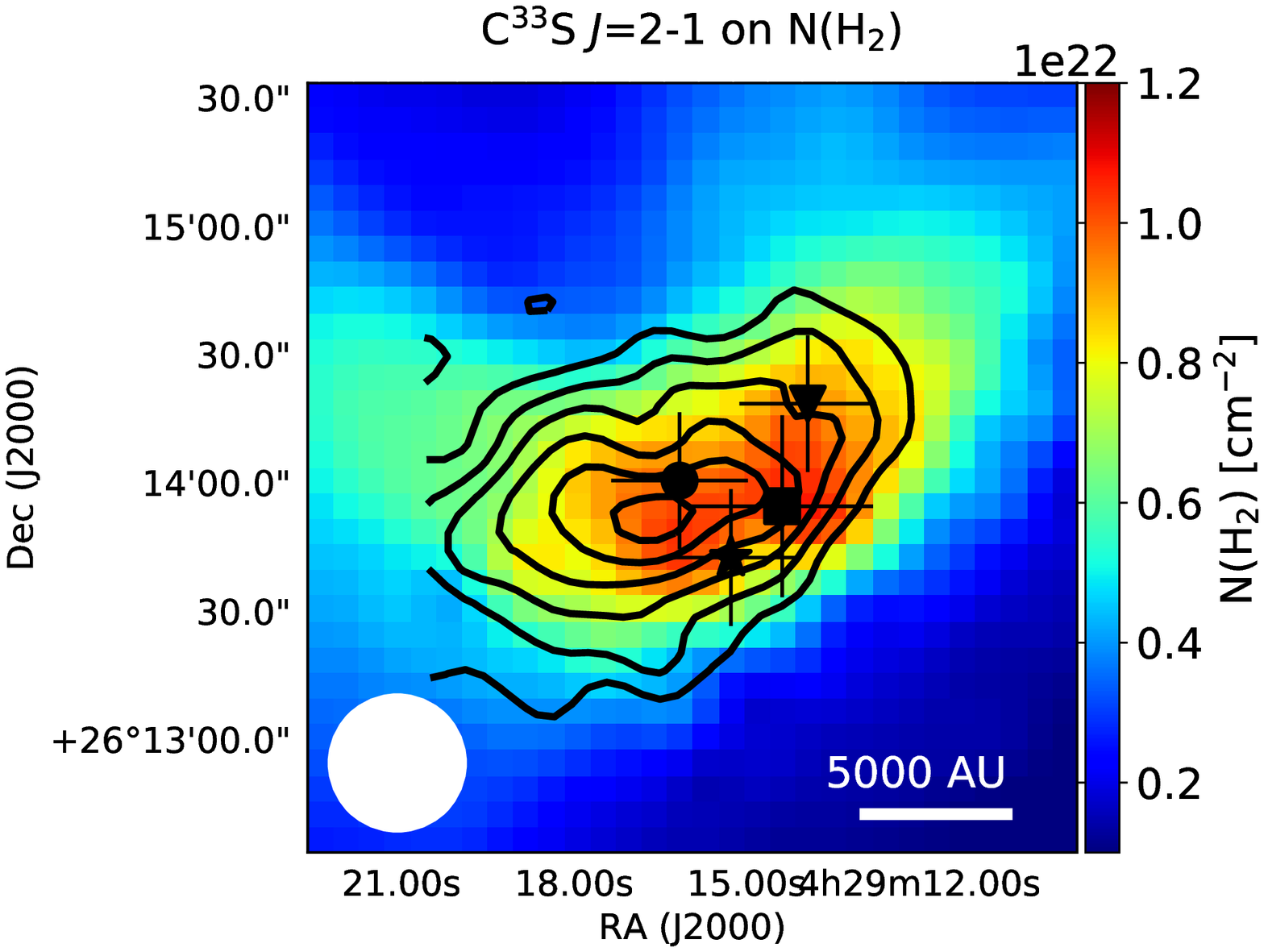}
\includegraphics[width=6.2cm, trim=0cm 6.5cm 1.5cm 8cm,clip=true]{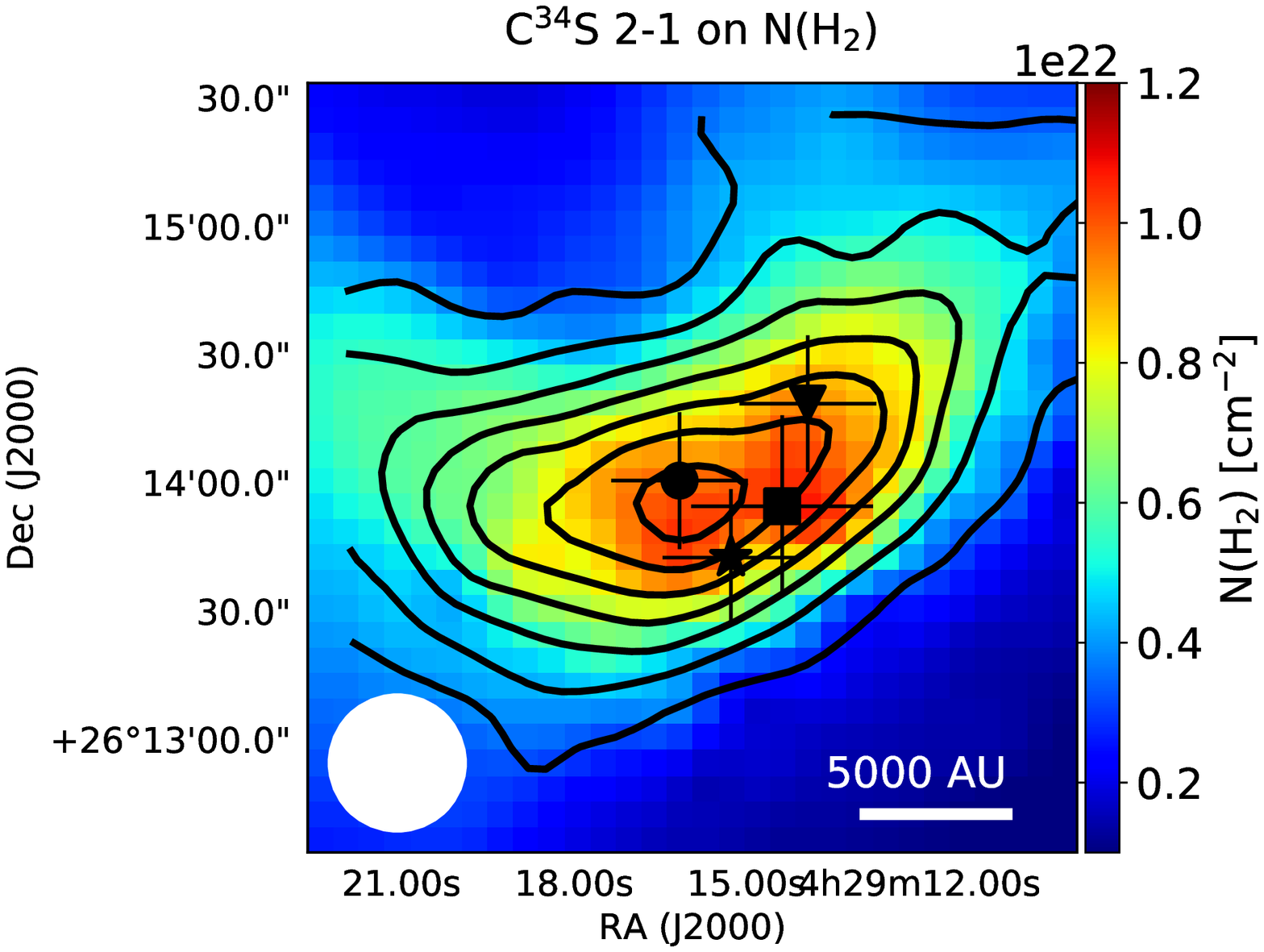} 
\includegraphics[width=6.2cm, trim=0cm 6.5cm 1.5cm 8cm,clip=true]{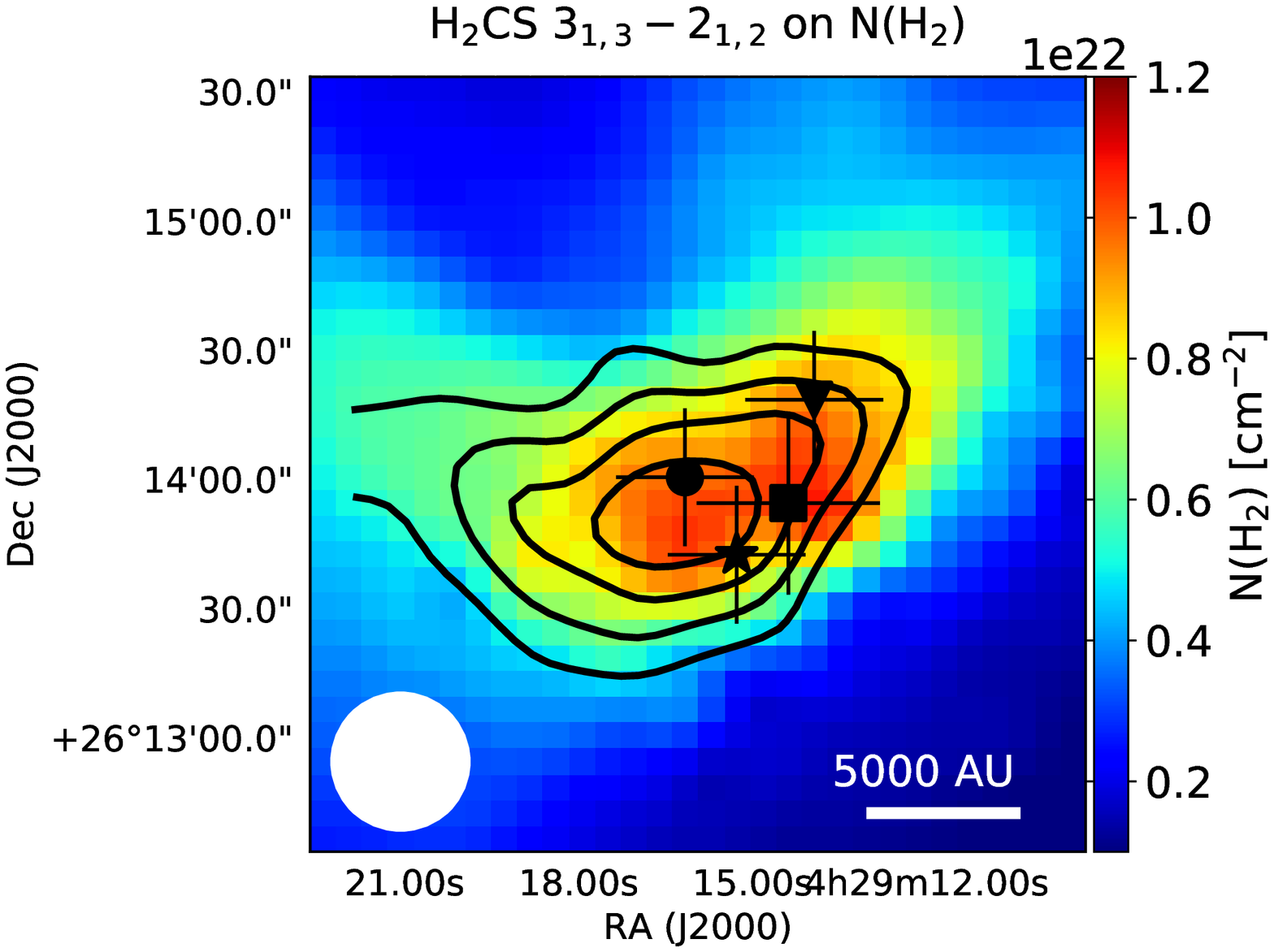}  
\includegraphics[width=6.2cm, trim=0cm 6.5cm 1.5cm 8cm,clip=true]{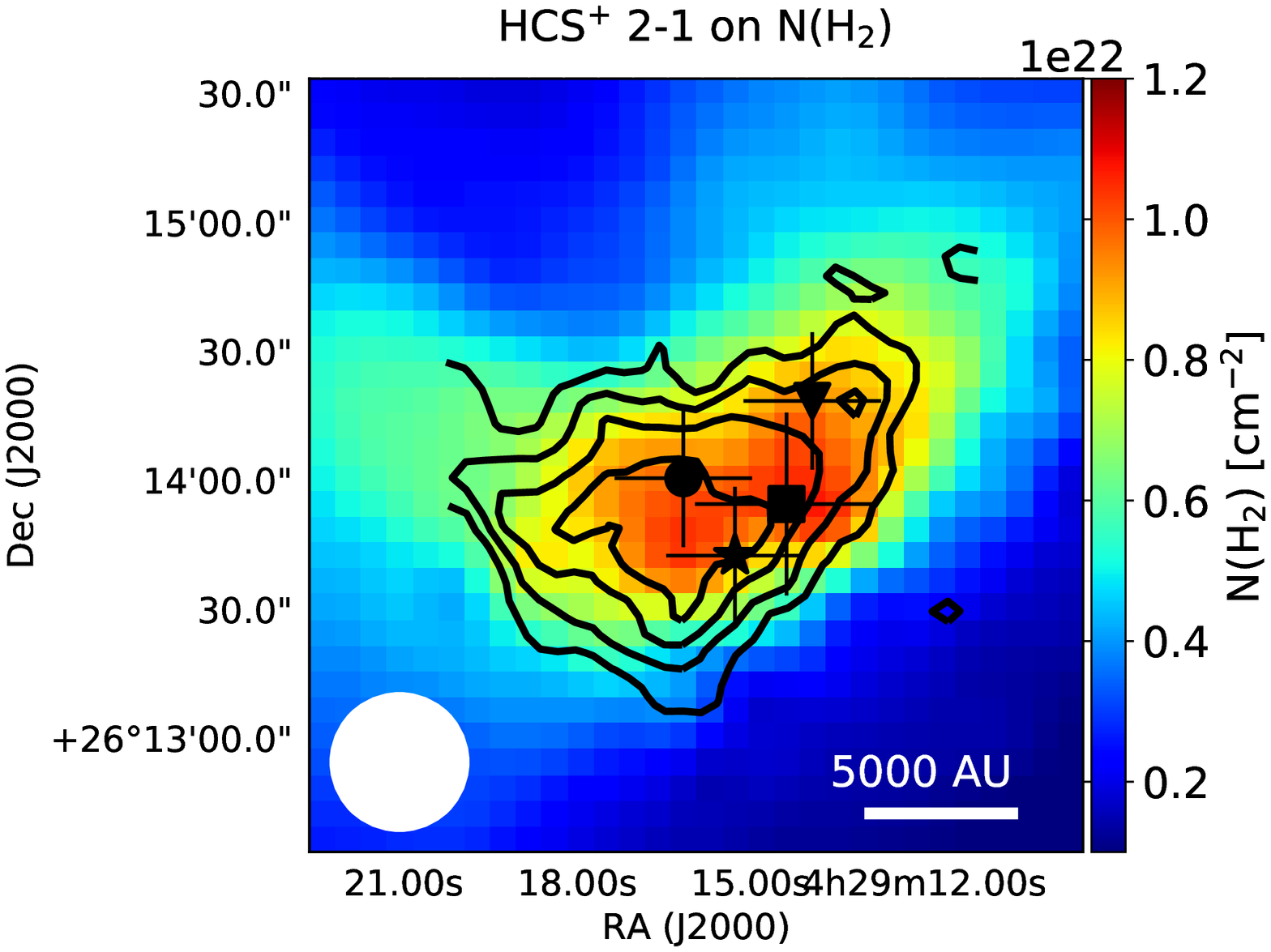} 
\caption{Spatial distributions of molecules detected toward L1521E (black contours) overplotted on the $N$(H$_2$) map (colors) derived from \textit{Herschel}/SPIRE. The black dot, triangle, and asterisk show the \textit{c}-C$_3$H$_2$, CH$_3$OH, and HNCO peaks, respectively. The black square shows the \textit{Herschel} dust peak. 
The beam size overplotted on the SO map shows the beam size of the SPIRE data, for the other maps it is the beam size of the IRAM-30m data.
For SO, CS, $^{13}$CS, C$_2$S, and C$^{34}$S the contour levels start from 3-$\sigma$ rms in steps of 3-$\sigma$ rms with 3-$\sigma$ rms values of 0.13 K km/s, 0.38 K km/s, 0.06 K km/s, 0.12 K km/s, and 0.11 K km/s, respectively. For OCS and C$^{33}$S the contour levels start from 6-$\sigma$ rms in steps of 4-$\sigma$ rms with 3-$\sigma$ rms levels of 0.02 K km/s for both species. For $^{34}$SO, C$^3$S, CC$^{34}$S, H$_2$CS, and HCS$^+$ the contour levels start from 9-$\sigma$ rms in steps of 6-$\sigma$ rms with 3-$\sigma$ rms values of 0.01 K km/s, 0.02 K km/s, 0.01 K km/s, 0.07 K km/s, and 0.03 K km/s, respectively.
} 
\label{spatial_distribution_1}
\end{figure*}
\end{flushleft}

\begin{flushleft}
\begin{figure*}[ht!]
\ContinuedFloat
\captionsetup{list=off,format=cont}
\includegraphics[width=6.2cm, trim=0cm 6.5cm 1.5cm 8cm,clip=true]{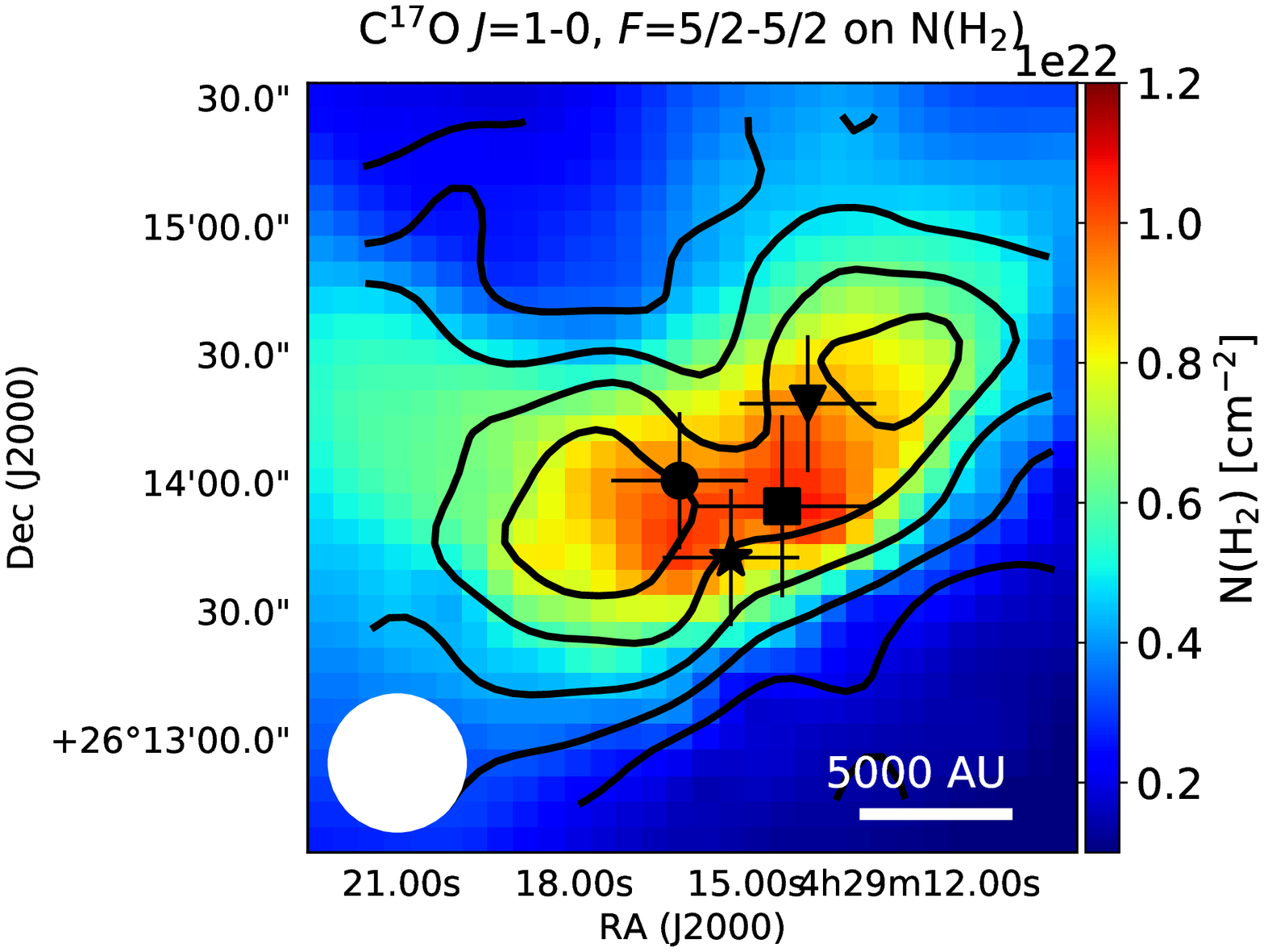}  
\includegraphics[width=6.2cm, trim=0cm 6.5cm 1.5cm 8cm,clip=true]{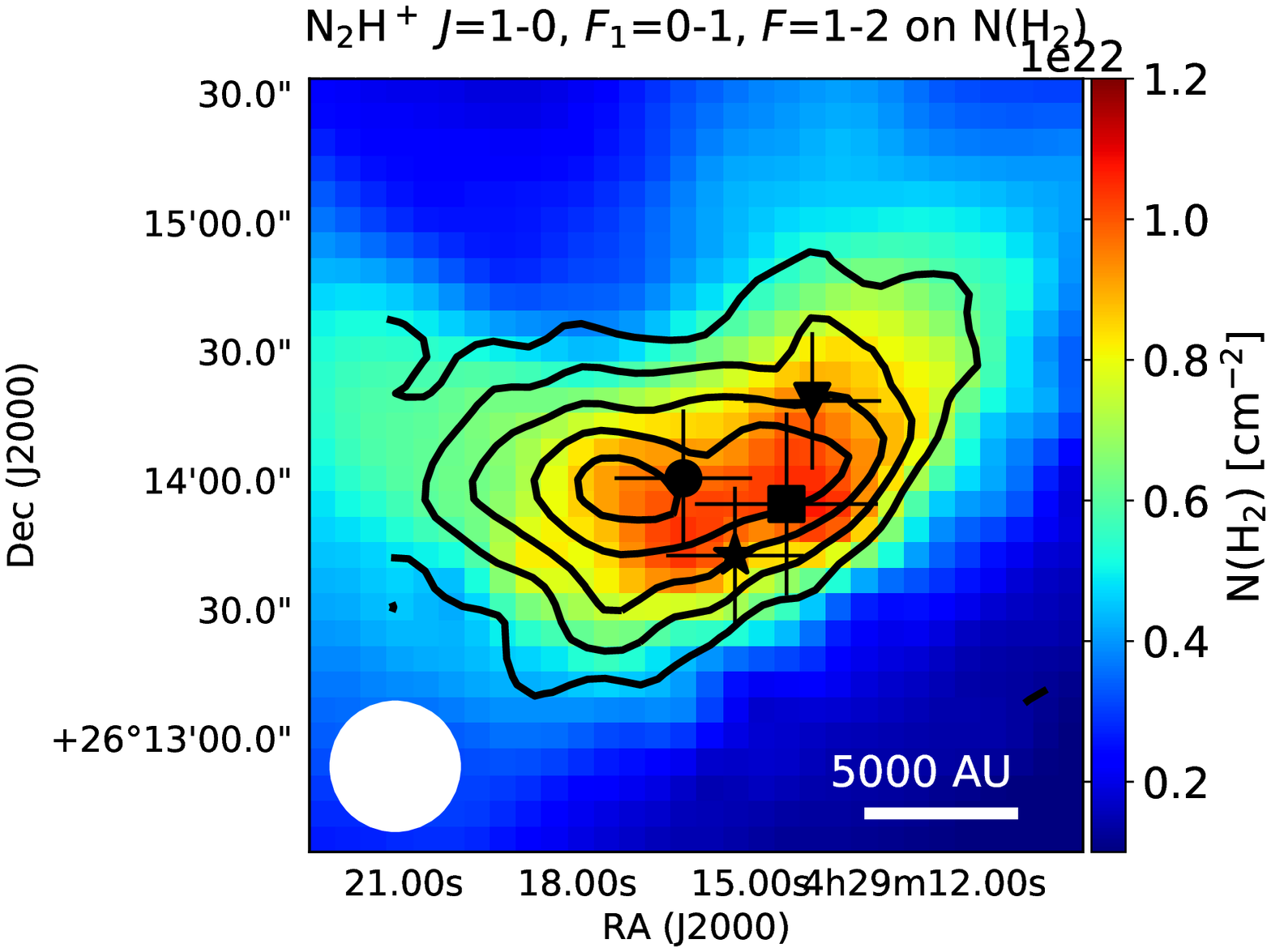}
\includegraphics[width=6.2cm, trim=0cm 6.5cm 1.5cm 8cm,clip=true]{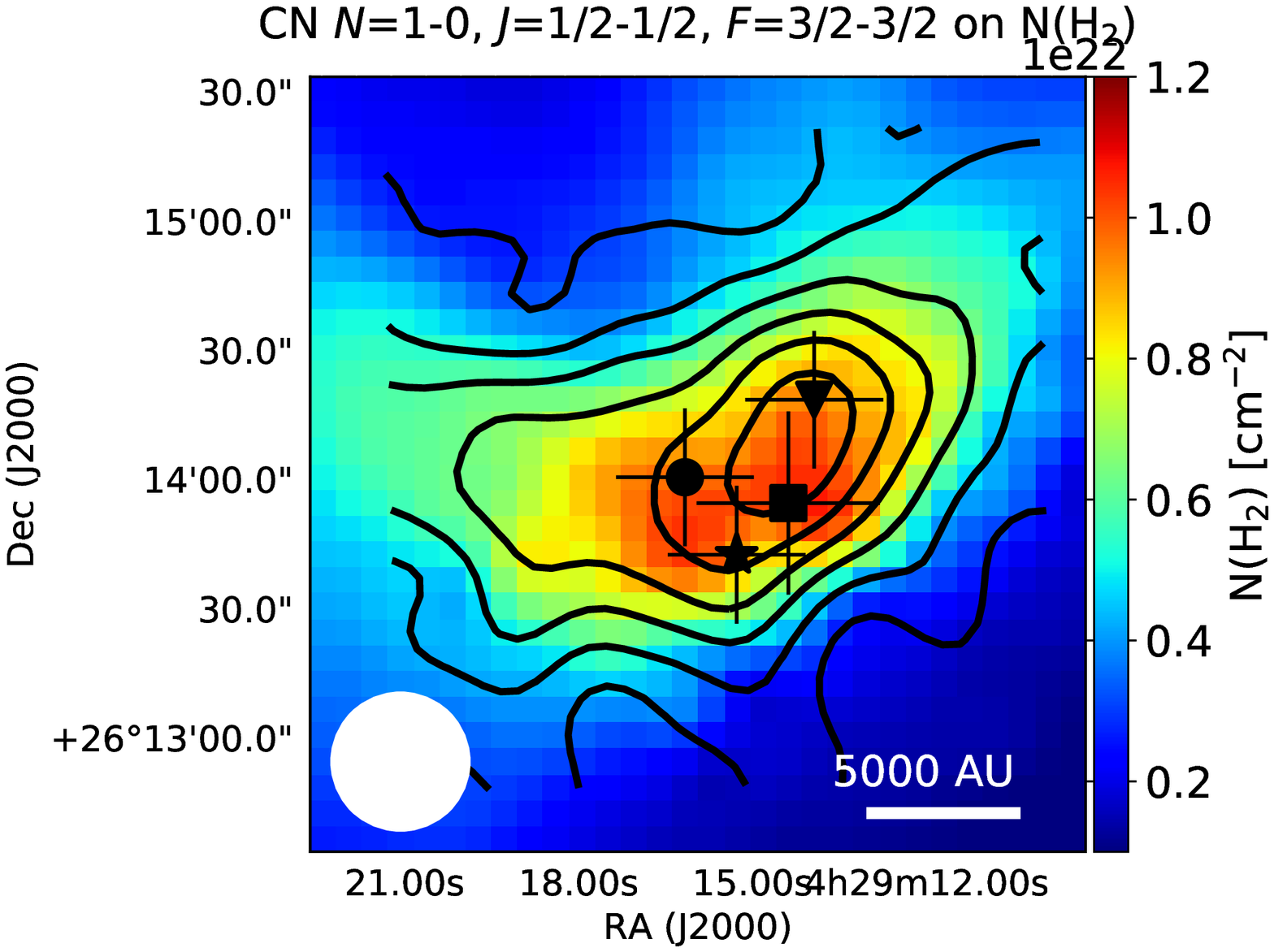} 
\includegraphics[width=6.2cm, trim=0cm 6.5cm 1.5cm 8cm,clip=true]{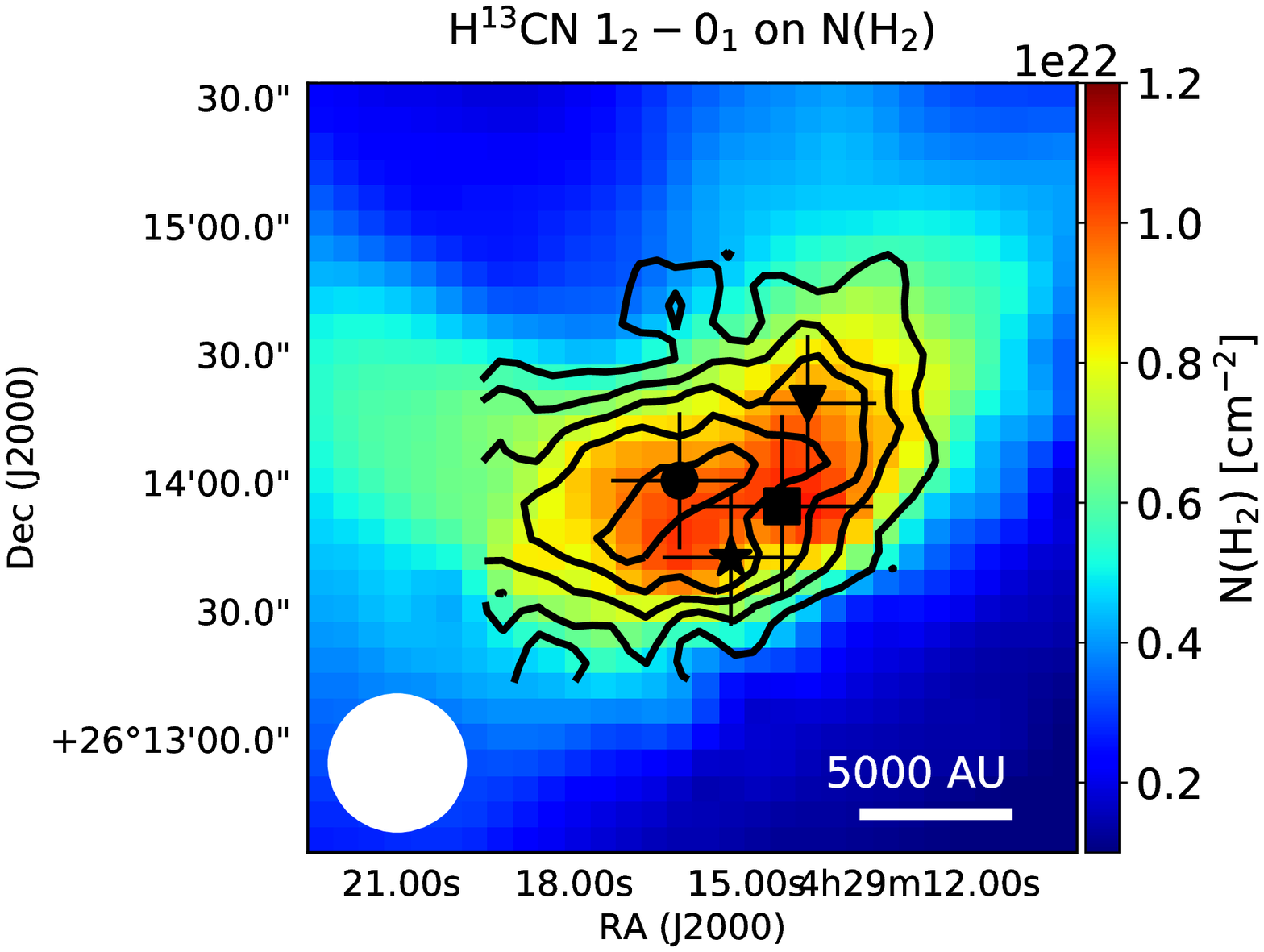}
\includegraphics[width=6.2cm, trim=0cm 6.5cm 1.5cm 8cm,clip=true]{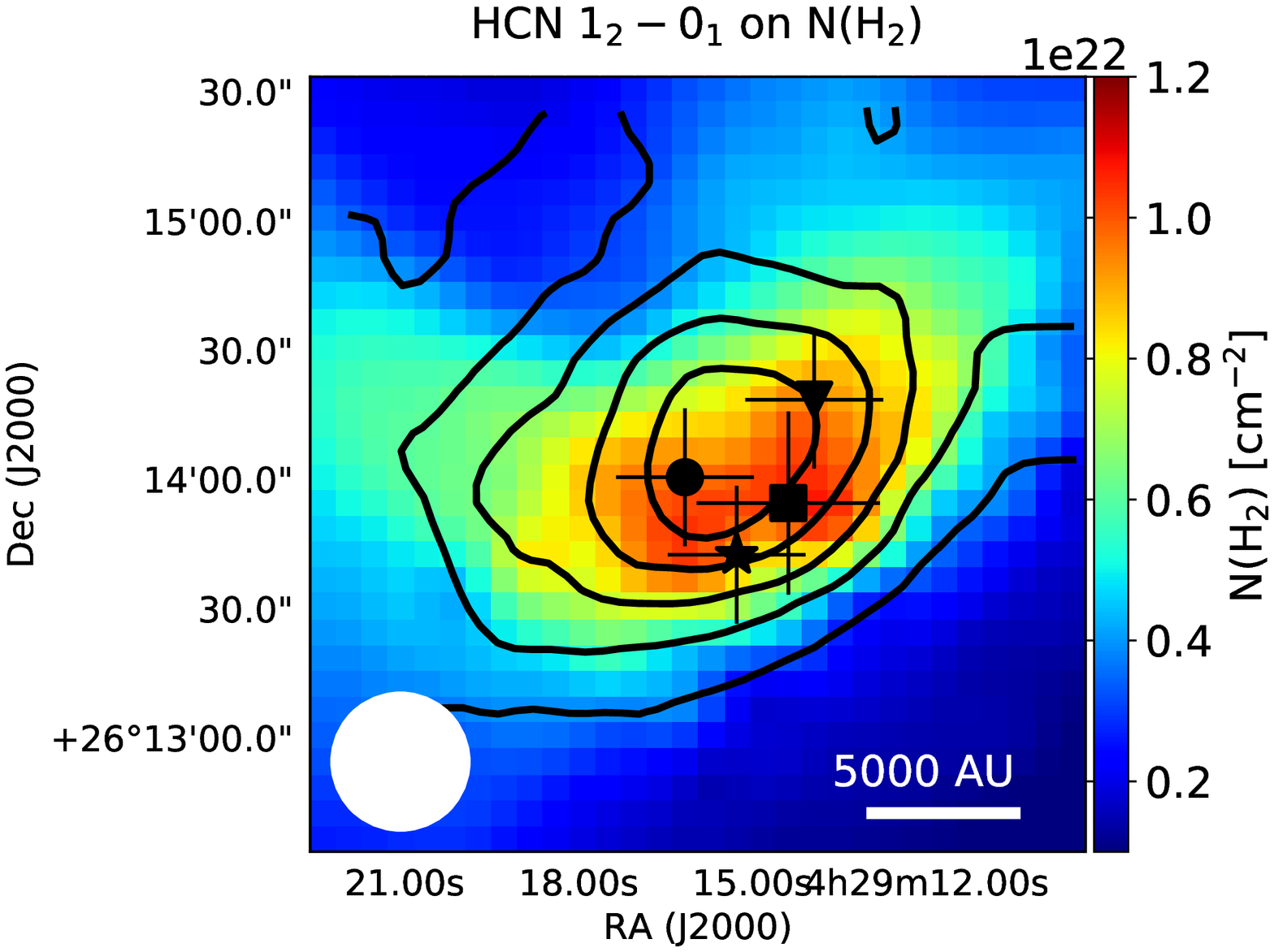}
\includegraphics[width=6.2cm, trim=0cm 6.5cm 1.5cm 8cm,clip=true]{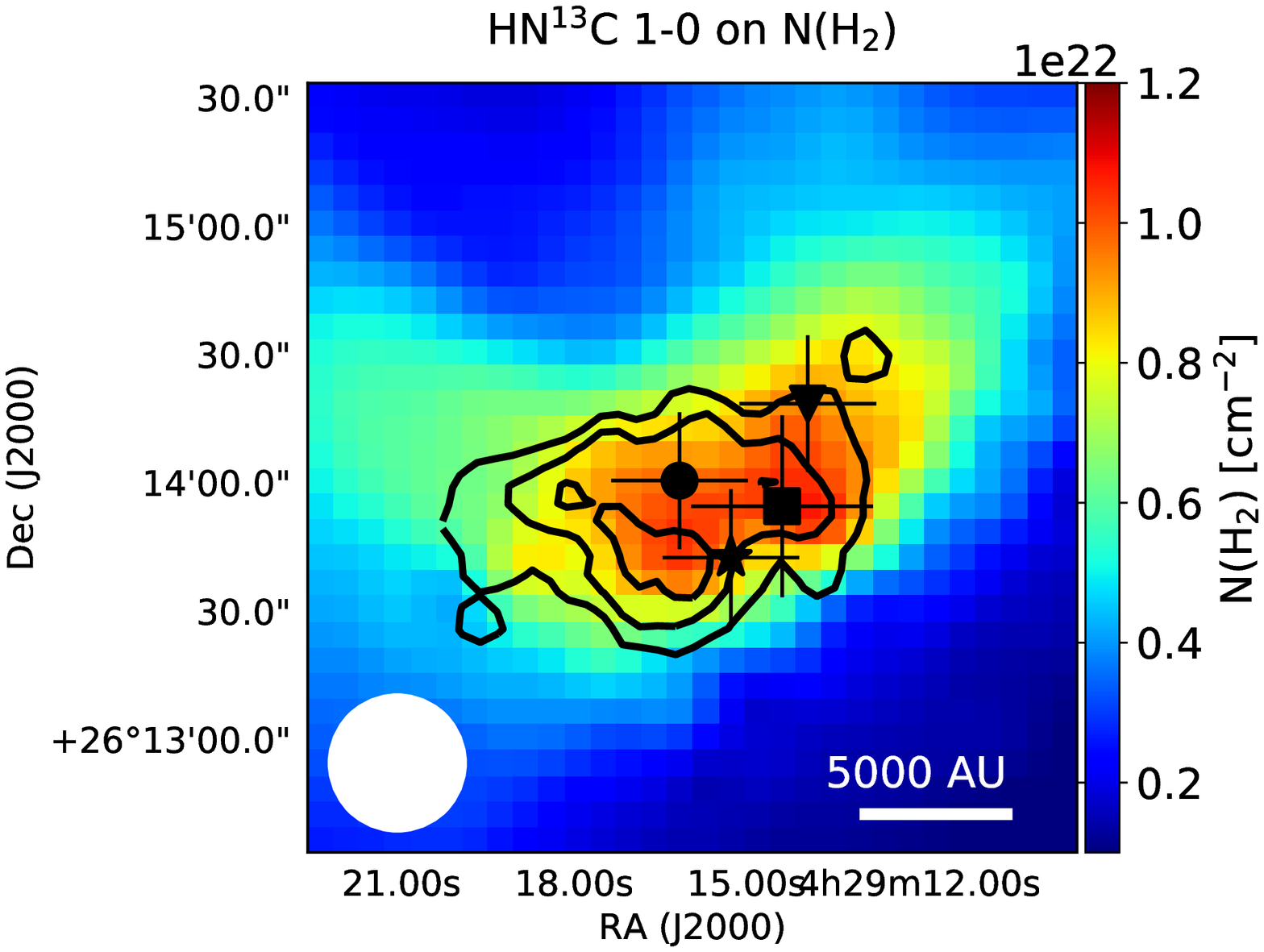} 
\includegraphics[width=6.2cm, trim=0cm 6.5cm 1.5cm 8cm,clip=true]{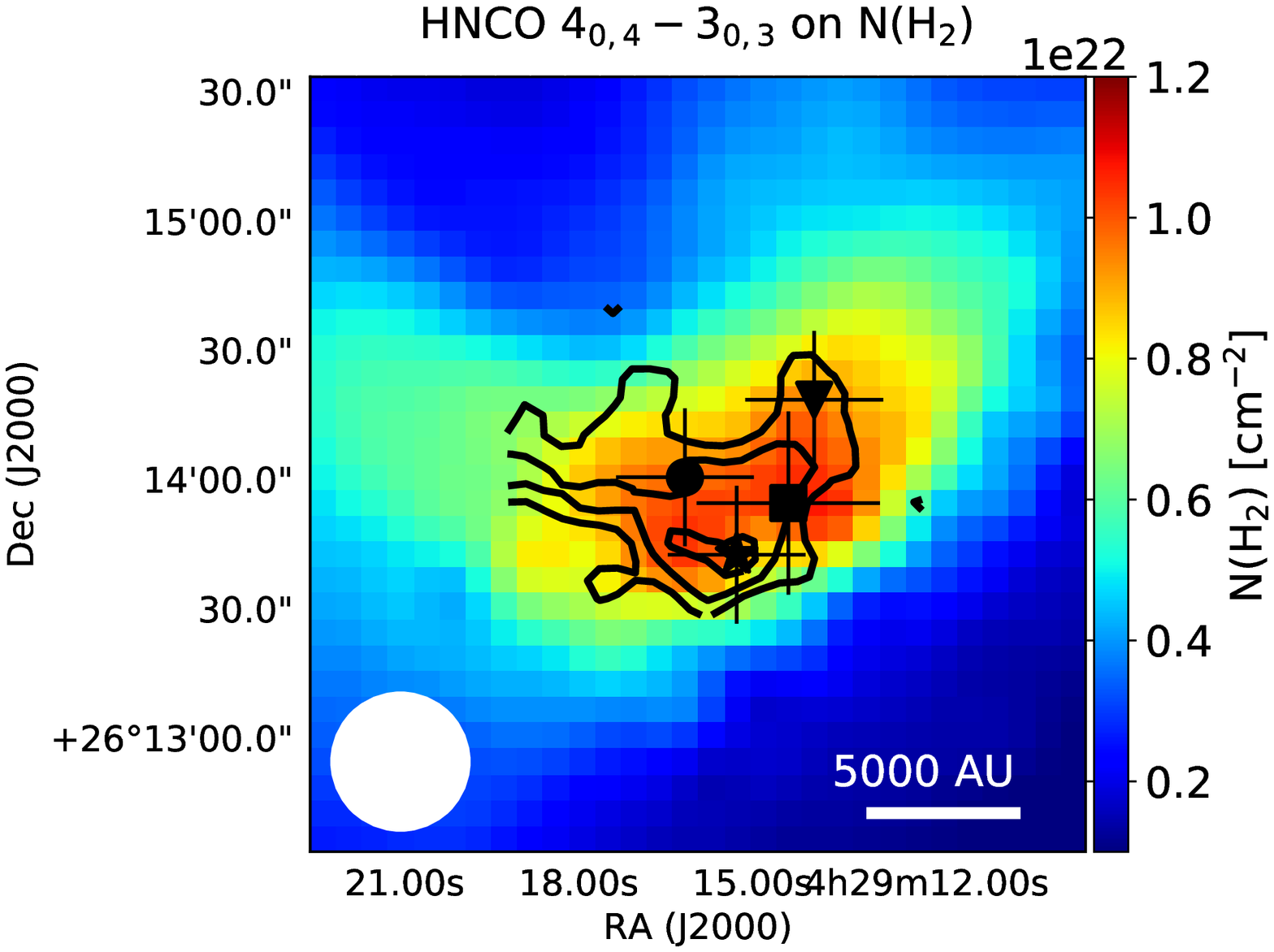} 
\includegraphics[width=6.2cm, trim=0cm 6.5cm 1.5cm 8cm,clip=true]{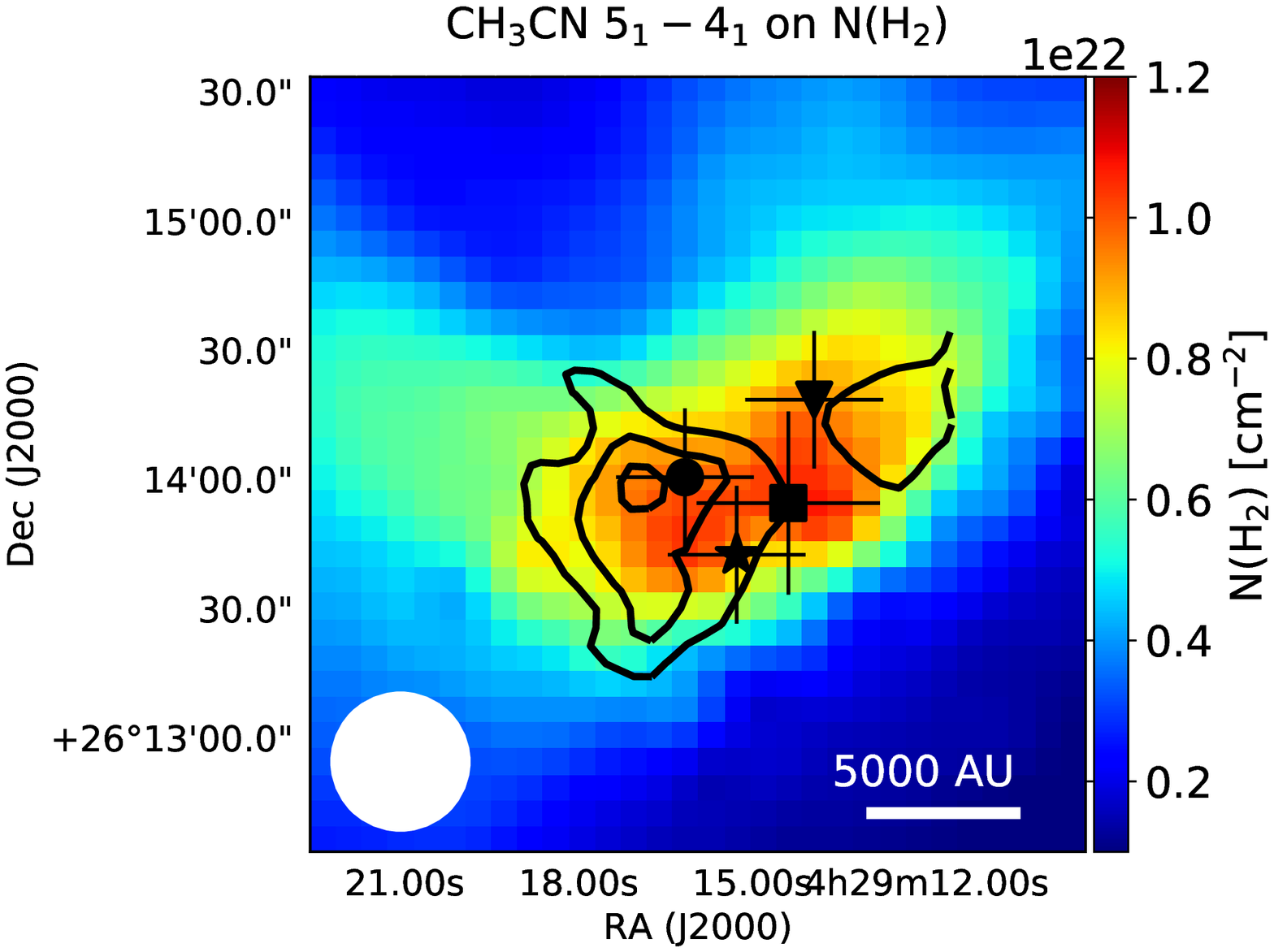}  
\includegraphics[width=6.2cm, trim=0cm 6.5cm 1.5cm 8cm,clip=true]{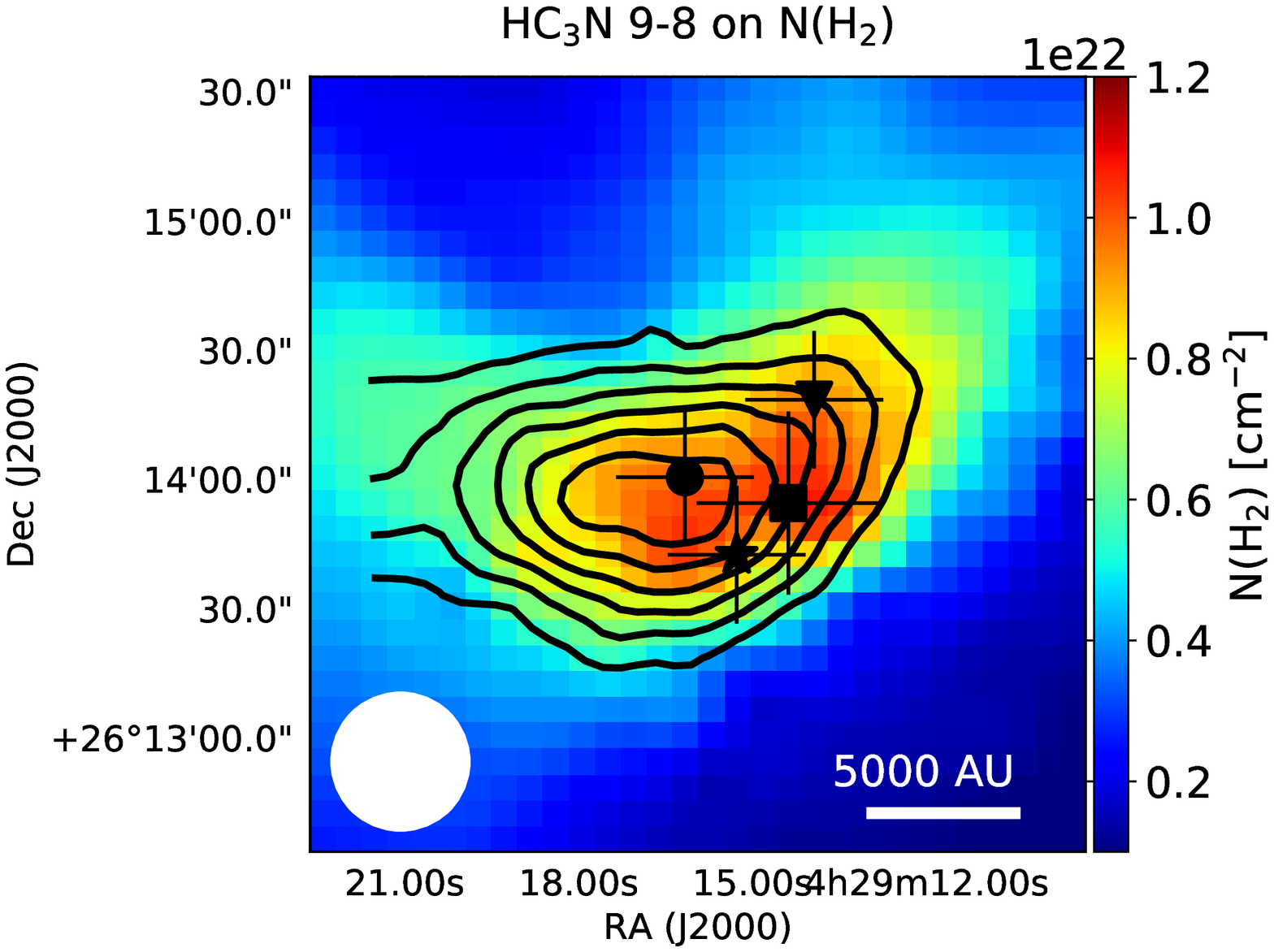}
\includegraphics[width=6.2cm, trim=0cm 6.5cm 1.5cm 8cm,clip=true]{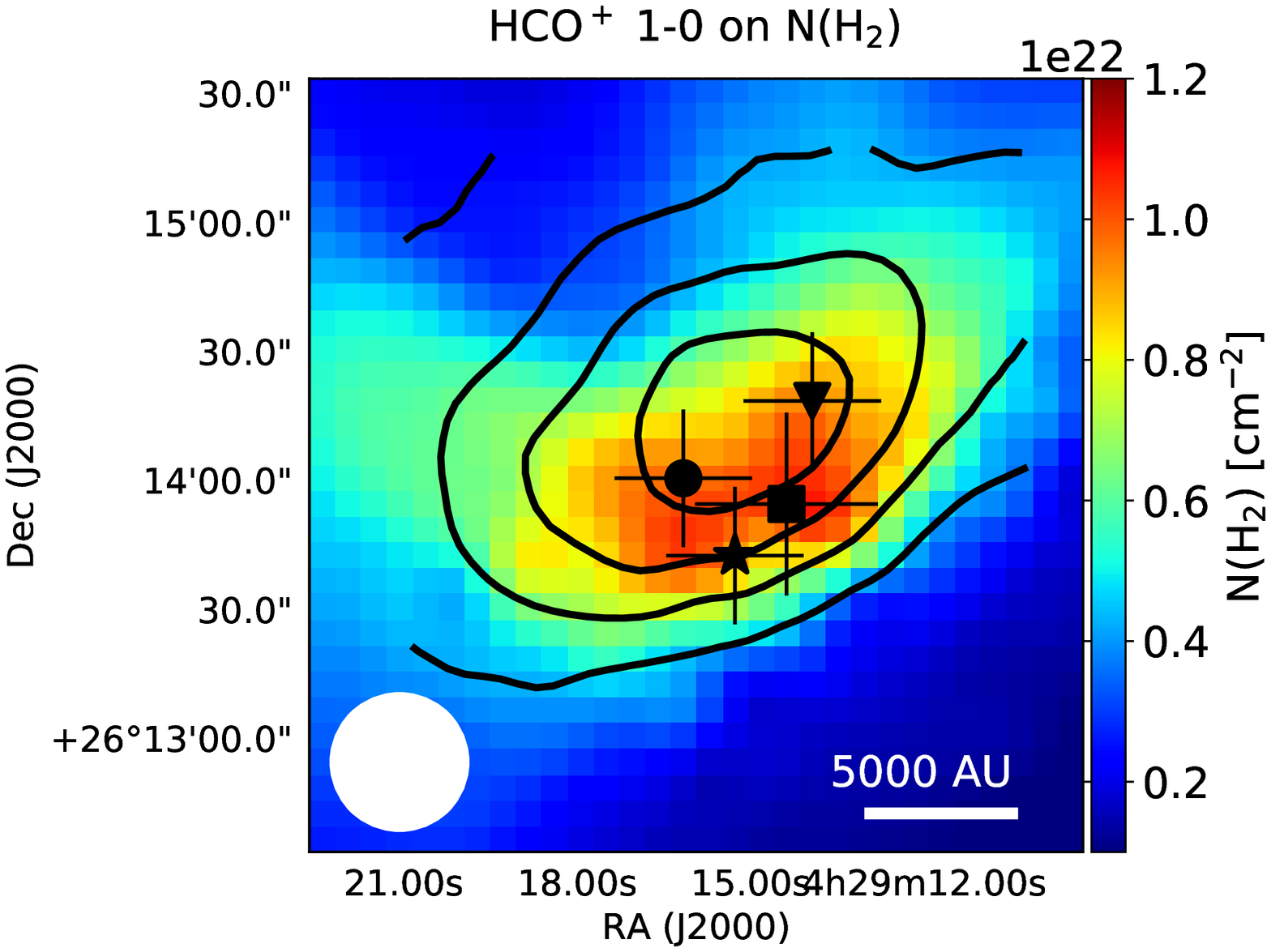}  
\includegraphics[width=6.2cm, trim=0cm 6.5cm 1.5cm 8cm,clip=true]{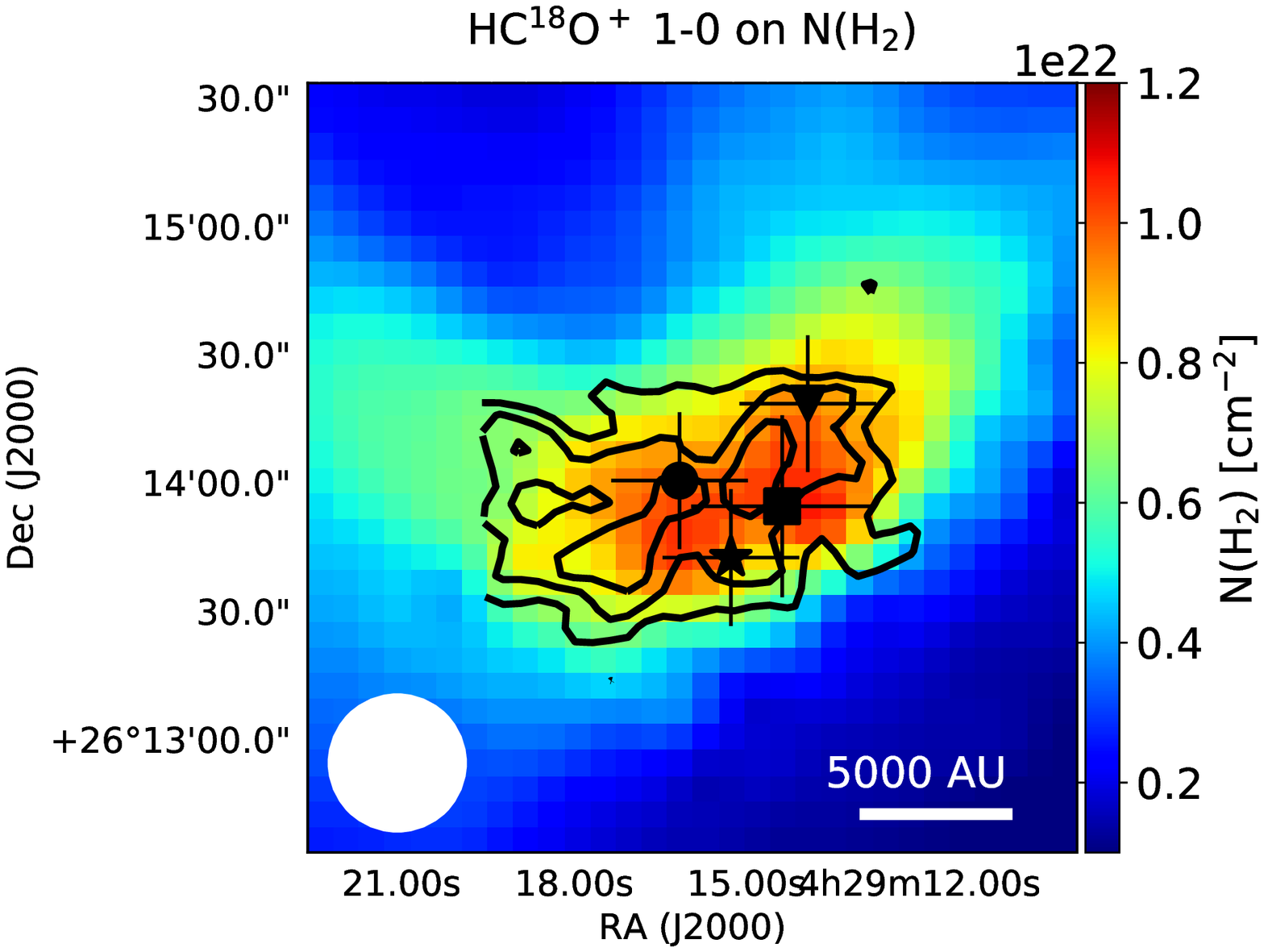}
\includegraphics[width=6.2cm, trim=0cm 6.5cm 1.5cm 8cm,clip=true]{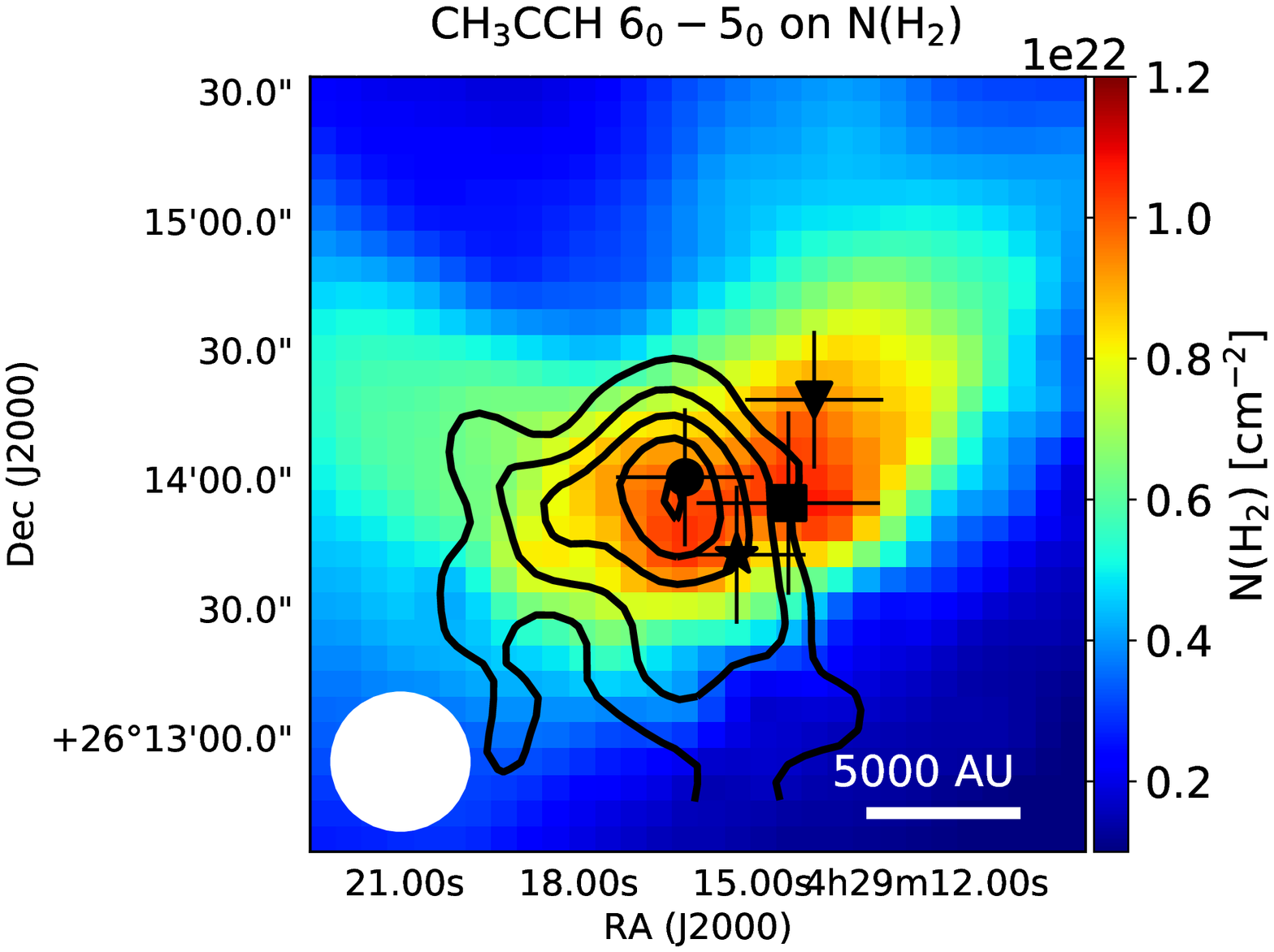}    
\caption{
Spatial distributions of molecules detected toward L1521E (black contours) overplotted on the $N$(H$_2$) map (colors) derived from \textit{Herschel}/SPIRE. The black dot, triangle, and asterisk show the \textit{c}-C$_3$H$_2$, CH$_3$OH, and HNCO peaks, respectively. The black square shows the \textit{Herschel} dust peak.
The beam size corresponds to the IRAM-30m data.
For C$^{17}$O, CN, H$^{13}$CN, HCN, and HCO$^+$ the contour levels start from 3-$\sigma$ rms in steps of 3-$\sigma$ rms with 3-$\sigma$ rms values of 0.05 K km/s, 0.05 K km/s, 0.04 K km/s, 0.17 K km/s, and 0.34 K km/s, respectively. For N$_2$H$^+$, HN$^{13}$C, HNCO, HC$_3$N, and HC$^{18}$O$^+$ the contour levels start from 6-$\sigma$ rms in steps of 4-$\sigma$ rms with 3-$\sigma$ rms levels of 0.08 K km/s, 0.08 K km/s, 0.05 K km/s, 0.12 K km/s, and 0.02 K km/s, respectively. For CH$_3$CN and CH$_3$CCH the contour levels start from 9-$\sigma$ rms in steps of 6-$\sigma$ rms with 3-$\sigma$ rms values of 0.01 K km/s and 0.03 K km/s, respectively.
}

\label{spatial_distribution_2}
\end{figure*}
\end{flushleft}

\begin{flushleft}
\begin{figure*}[ht!]
\ContinuedFloat
\captionsetup{list=off,format=cont}
\includegraphics[width=6.2cm, trim=0cm 6.5cm 1.5cm 8cm,clip=true]{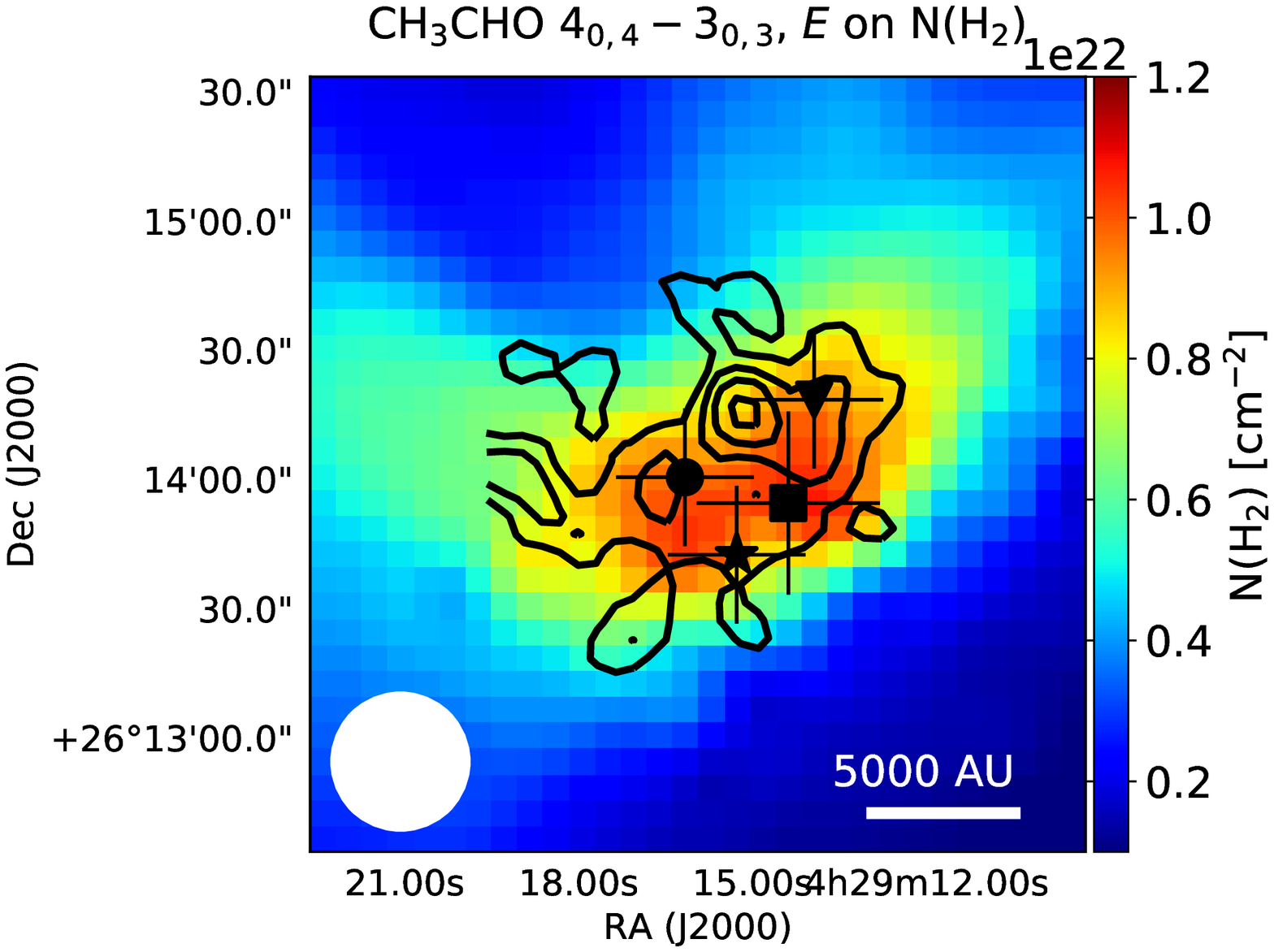}    
\includegraphics[width=6.2cm, trim=0cm 6.5cm 1.5cm 8cm,clip=true]{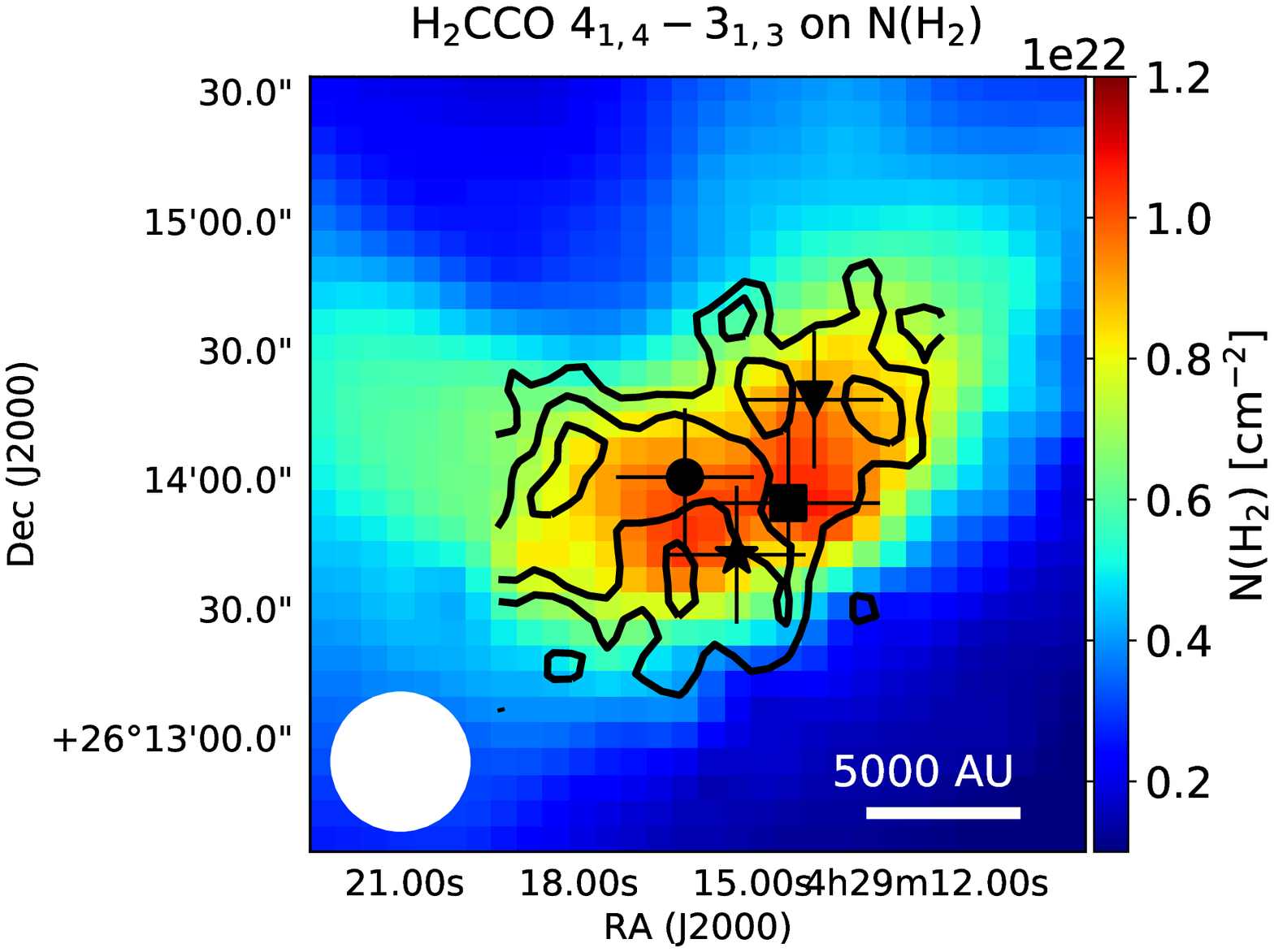}
\includegraphics[width=6.2cm, trim=0cm 6.5cm 1.5cm 8cm,clip=true]{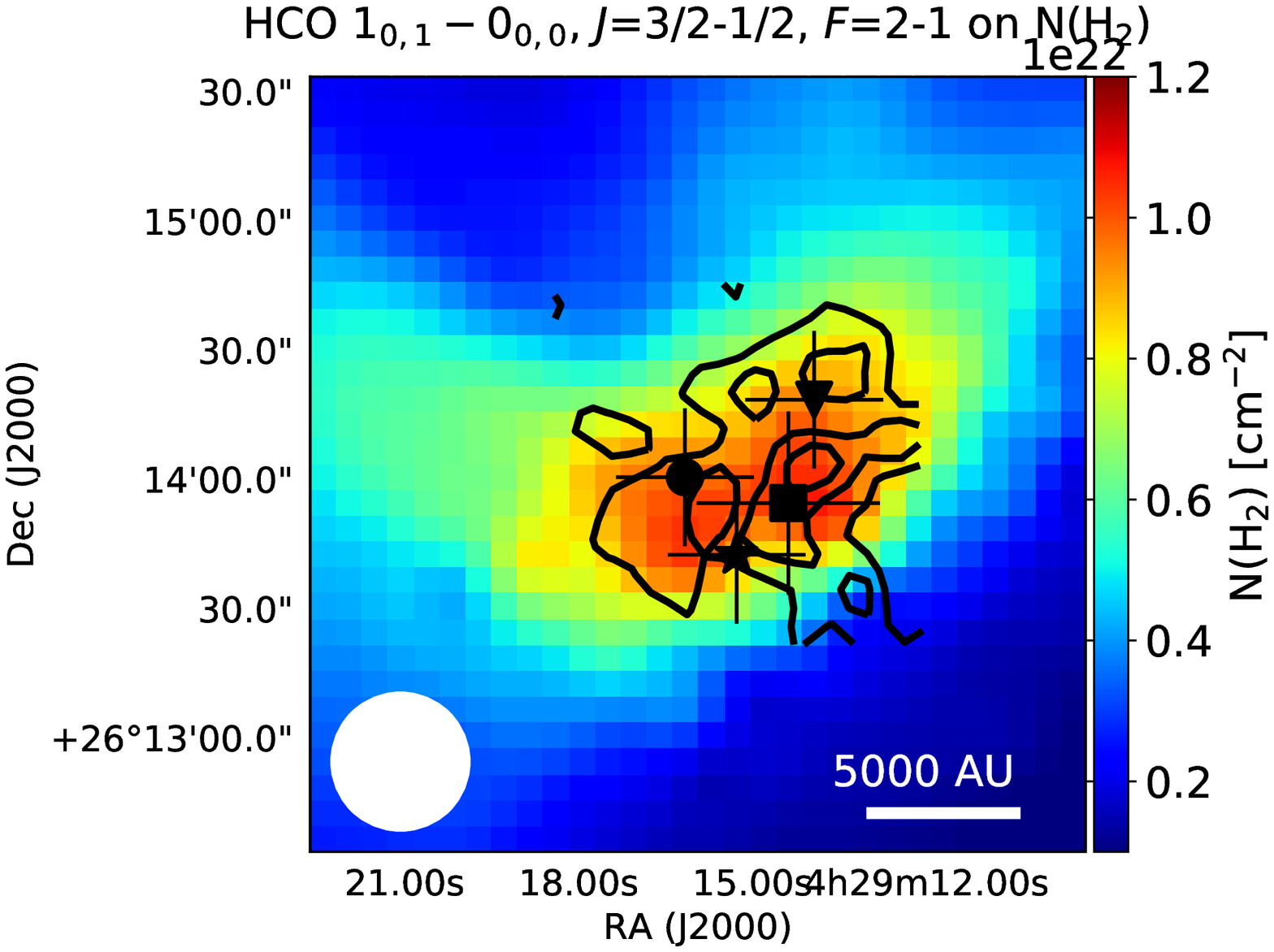}
\includegraphics[width=6.2cm, trim=0cm 6.5cm 1.5cm 8cm,clip=true]{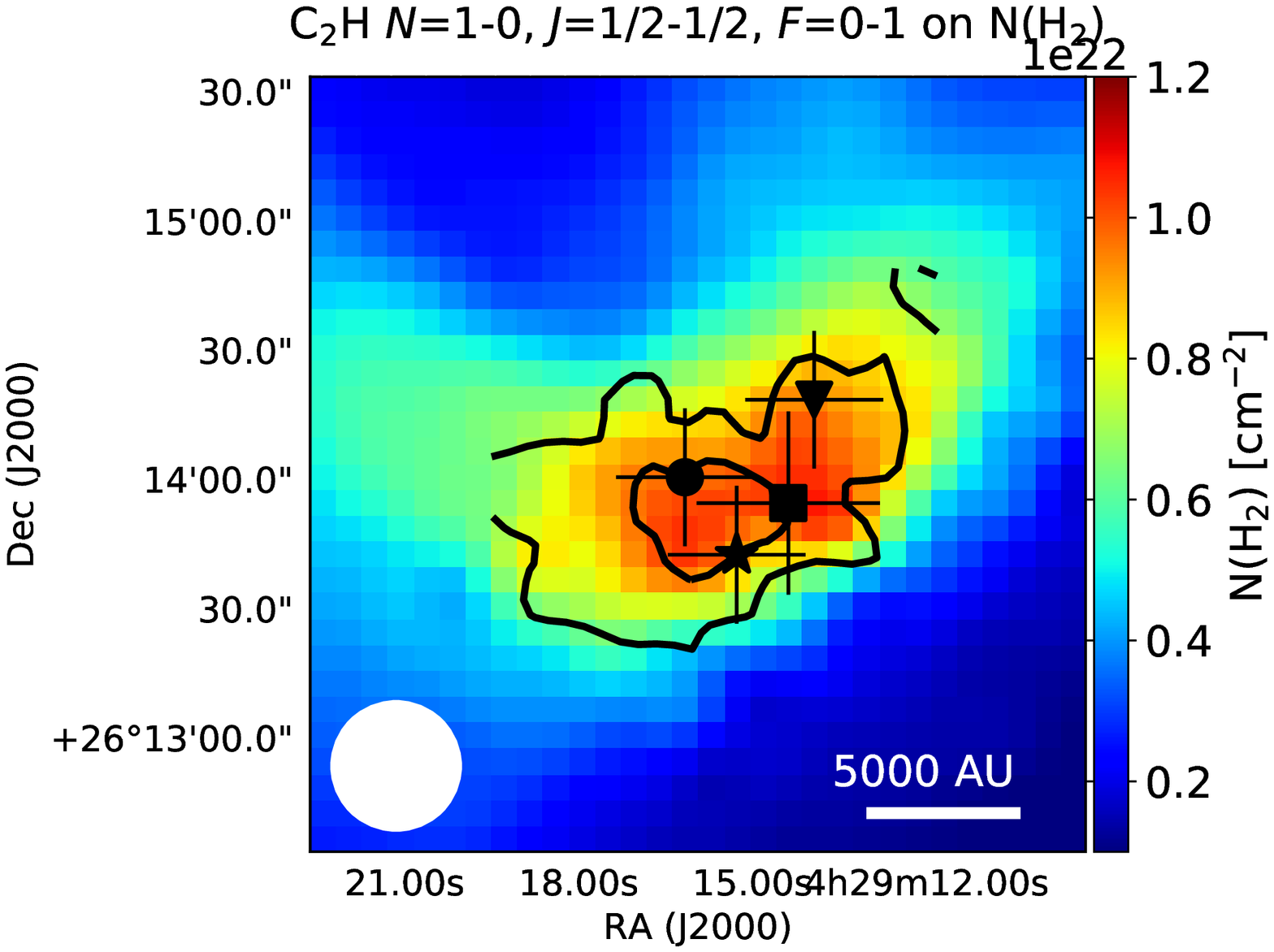} 
\includegraphics[width=6.2cm, trim=0cm 6.5cm 1.5cm 8cm,clip=true]{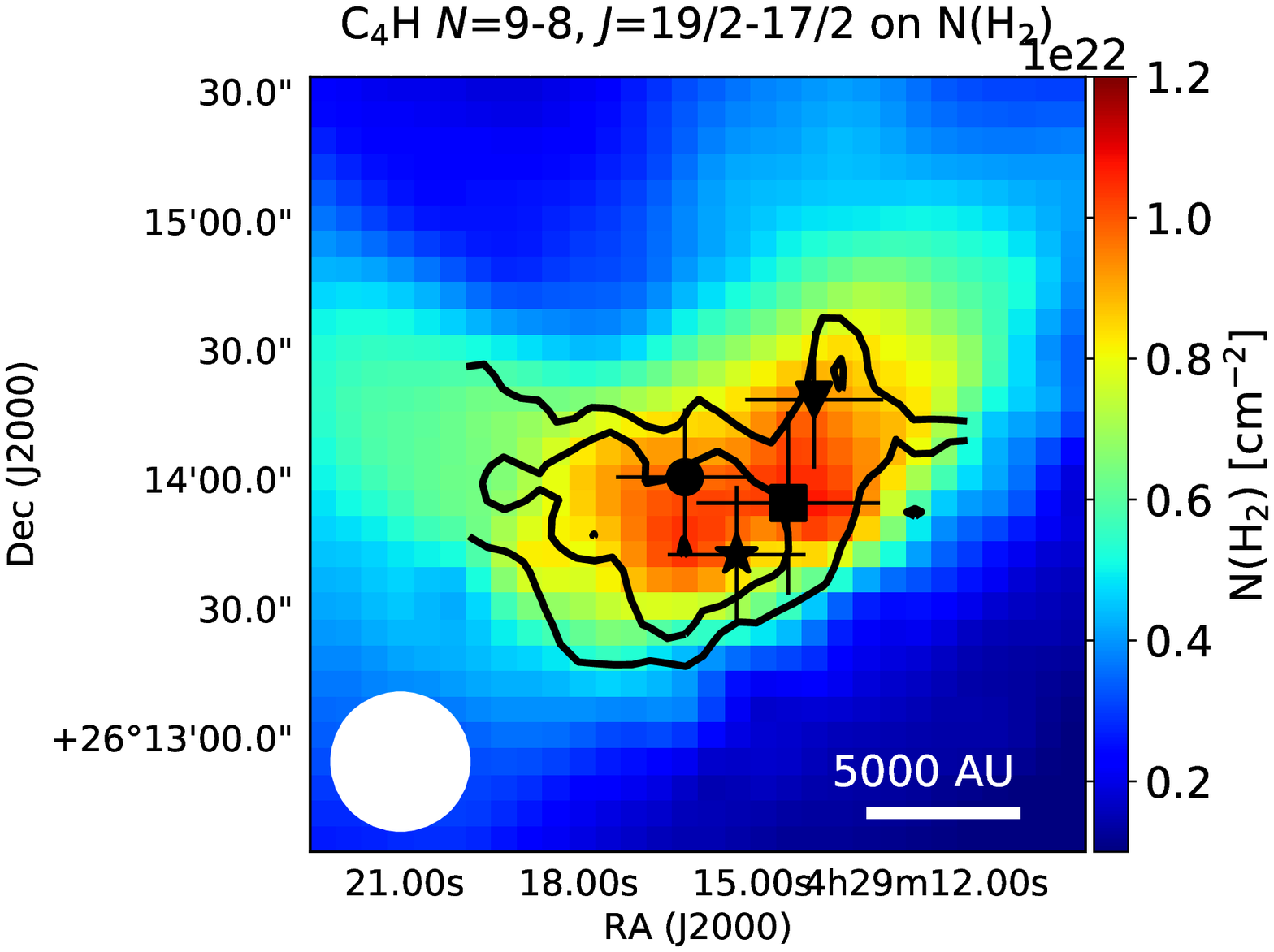}    
\includegraphics[width=6.2cm, trim=0cm 6.5cm 1.5cm 8cm,clip=true]{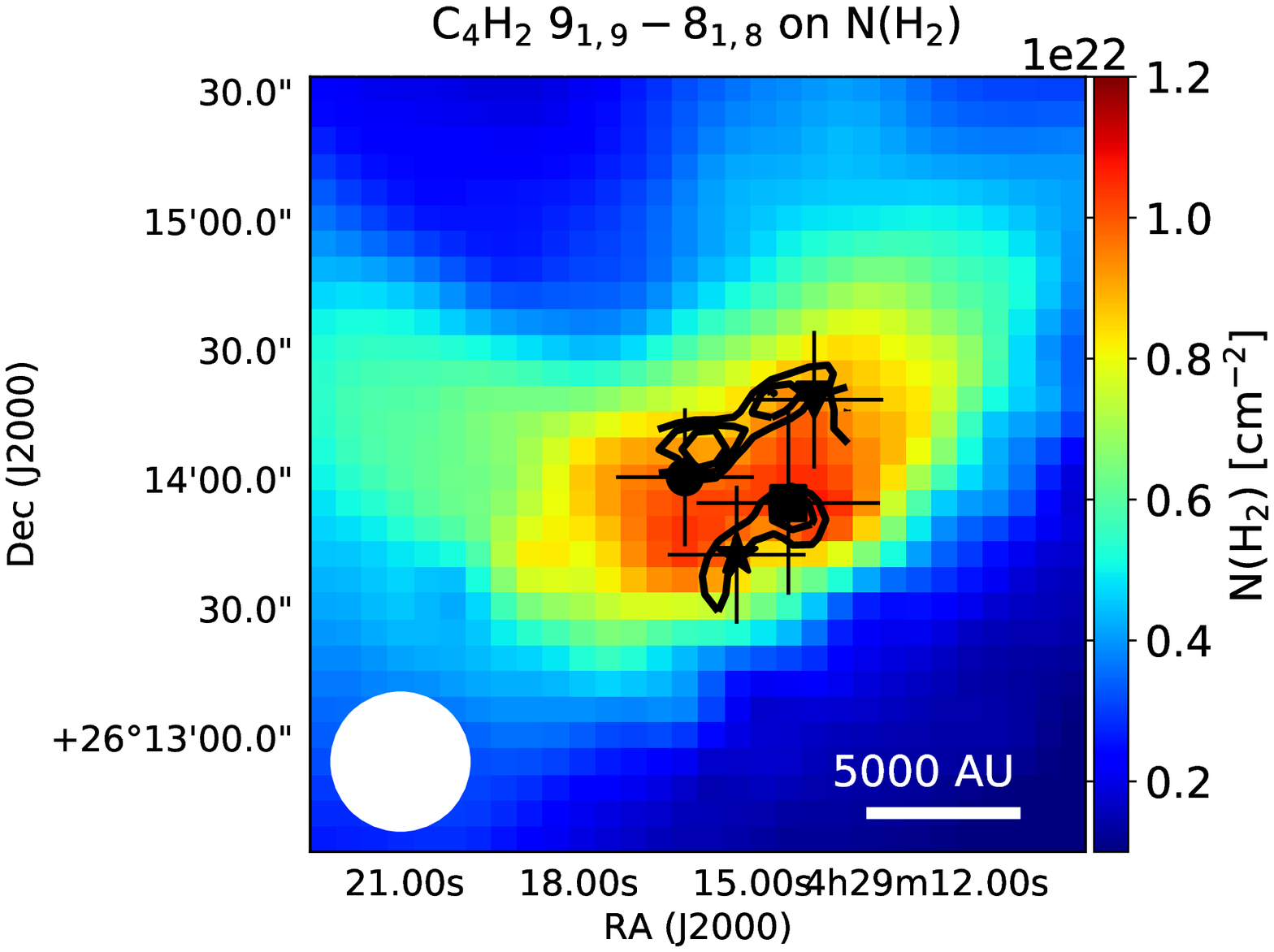}        

\caption{
Spatial distributions of molecules detected toward L1521E (black contours) overplotted on the $N$(H$_2$) map (colors) derived from \textit{Herschel}/SPIRE. The black dot, triangle, and asterisk show the \textit{c}-C$_3$H$_2$, CH$_3$OH, and HNCO peaks, respectively. The black square shows the \textit{Herschel} dust peak.
The beam size corresponds to the IRAM-30m data.
For H$_2$CCO and HCO the contour levels start from 6-$\sigma$ rms in steps of 4-$\sigma$ rms with 3-$\sigma$ rms levels of 0.02 K km/s for both species. For CH$_3$CHO, C$_2$H, C$_4$H, and C$_4$H$_2$ the contour levels start from 9-$\sigma$ rms in steps of 6-$\sigma$ rms with 3-$\sigma$ rms values of 0.01 K km/s, 0.07 K km/s, 0.03 K km/s, and 0.004 K km/s, respectively.
}

\label{spatial_distribution_3}
\end{figure*}
\end{flushleft}

\section{Results}
\label{sect_results}

\subsection{Spatial distribution of the molecules}
\label{spatial_dist}

\begin{figure}[t]
\includegraphics[width=9.0cm, trim=0cm 6.5cm 1.5cm 8cm,clip=true]{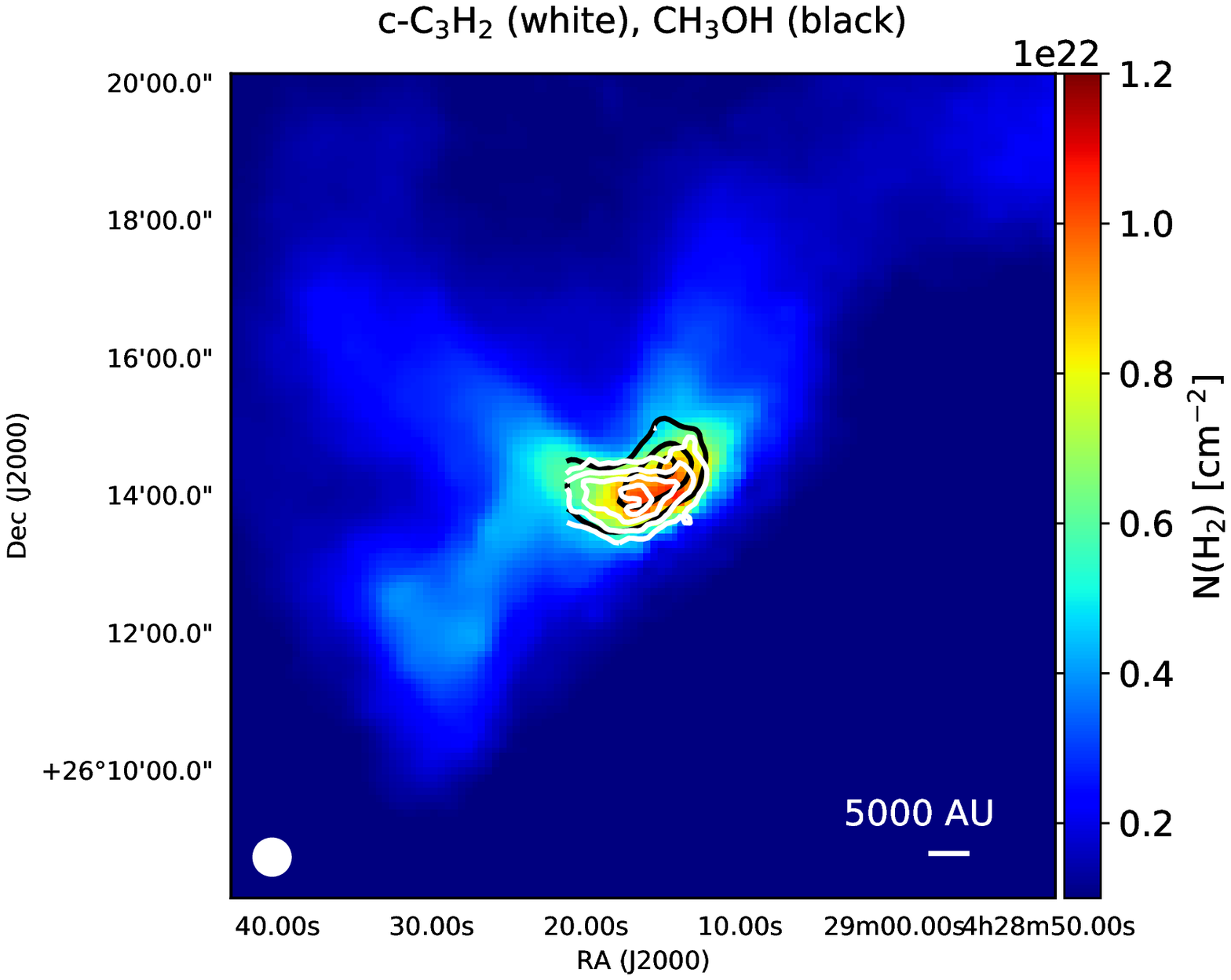}
\includegraphics[width=9.0cm, trim=0cm 6.5cm 1.5cm 8cm,clip=true]{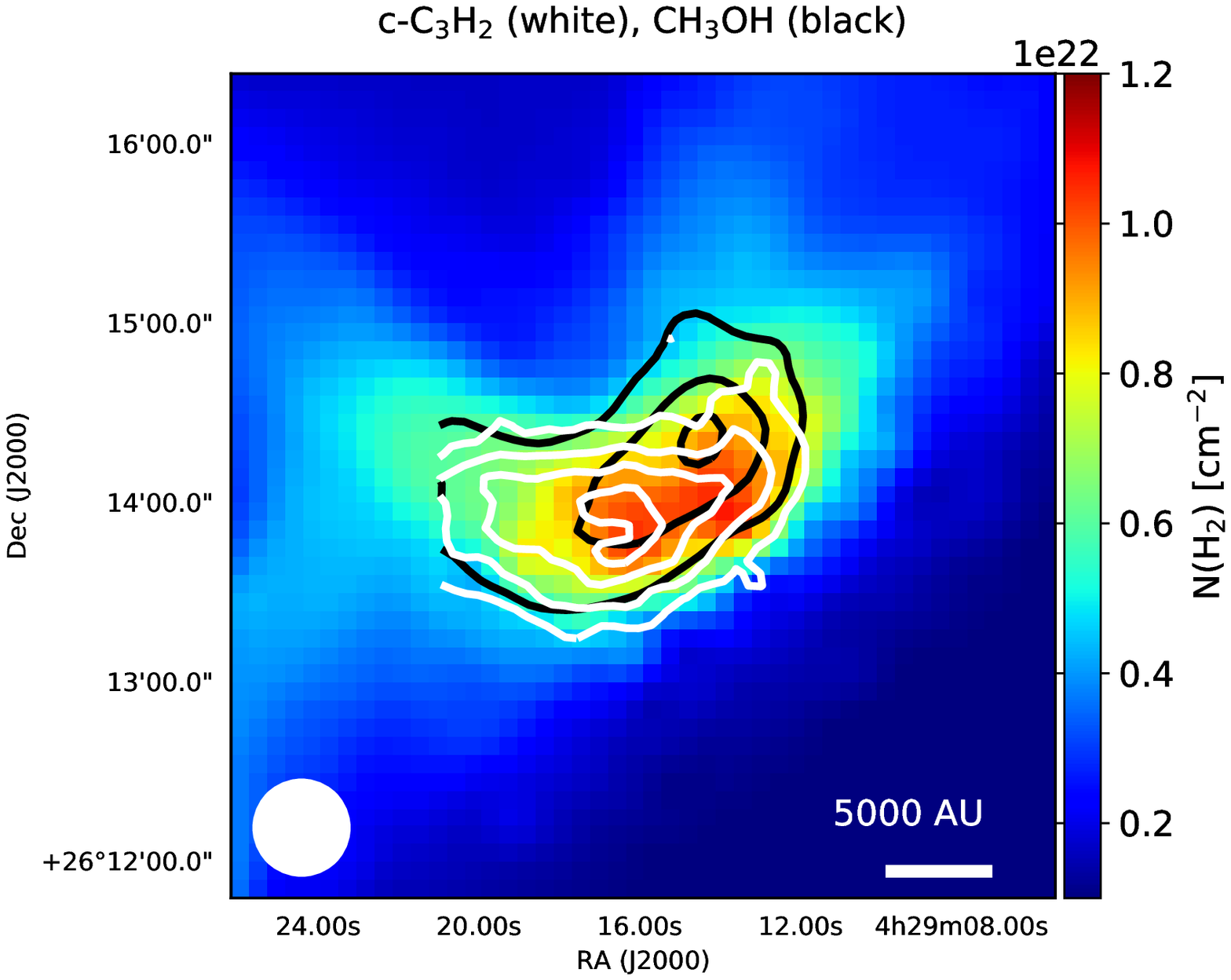}
\caption{The spatial distribution of $c$-C$_3$H$_2$ $2_{0,2}-1_{1,1}$ (white contours) and CH$_3$OH ($E_2$) (black contours) detected toward L1521E overplotted on the $N$(H$_2$) map derived from \textit{Herschel}/SPIRE (color map), showing the large-scale environment of L1521E (\textit{top}) and the spatial structure of molecular line emission close to the center of the core (\textit{bottom}). The two overplotted lines show the location of the cuts along which the line intensities shown in Fig. \ref{L1521E_cuts} were extracted. 
The contour levels start from 6-$\sigma$ rms in steps of 6-$\sigma$ rms, with 3-$\sigma$ rms levels of 0.07 K km/s and 0.06 K km/s for CH$_3$OH and $c$-C$_3$H$_2$, respectively.
}
\label{c3h2_ch3oh}
\end{figure}

\begin{figure}[h]
\centering
\includegraphics[width=7.5cm, trim=0cm 1cm 0cm 0cm,clip=true]{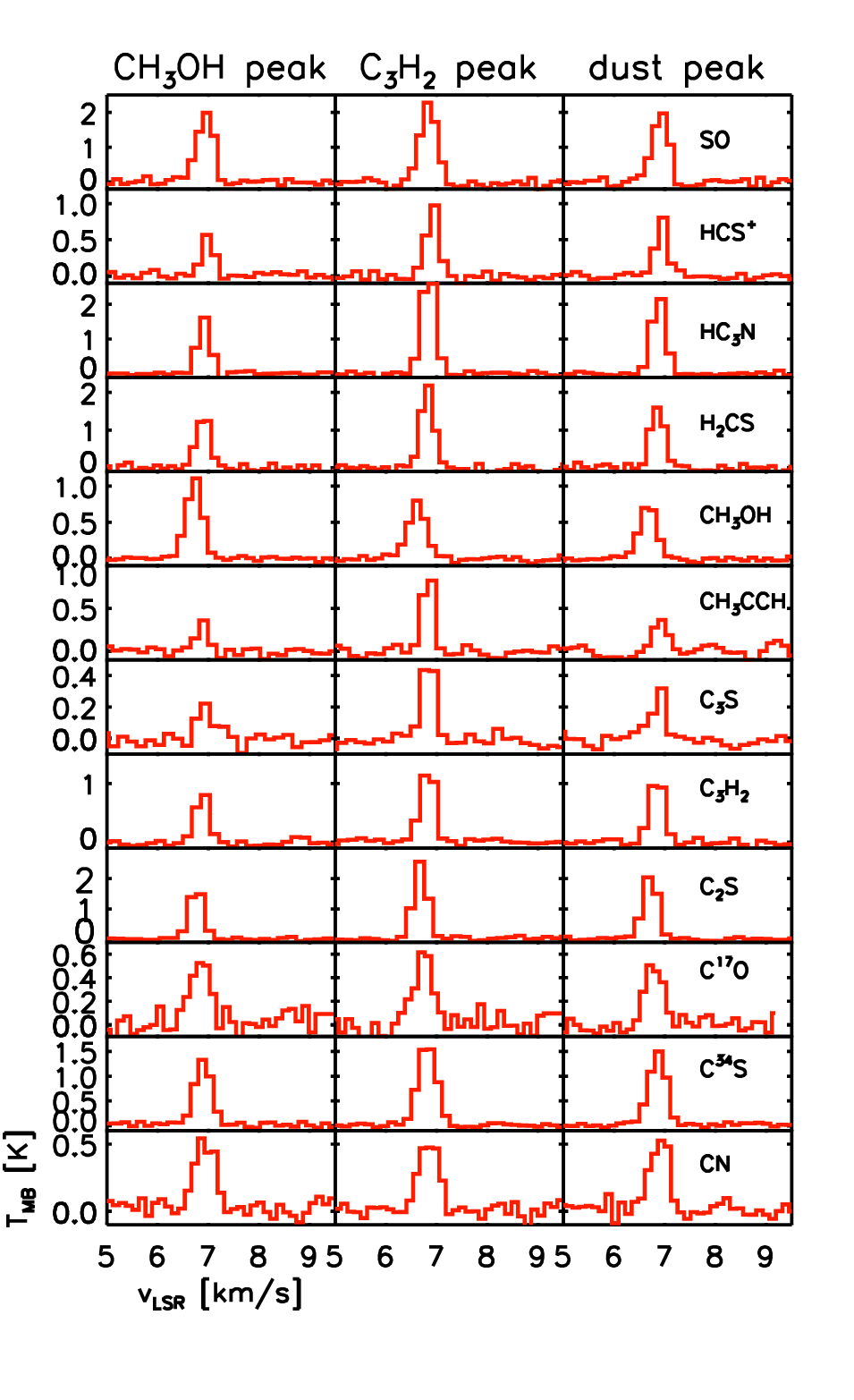}   
\caption{Line profiles of selected molecules observed toward the CH$_3$OH peak, the $c$-C$_3$H$_2$ peak, and the \textit{Herschel} dust peak. The used transitions for each molecule are marked in Table \ref{table:observations}.}
\label{line_profiles}
\end{figure}

\begin{figure}[h]
\centering   
\includegraphics[width=8cm]{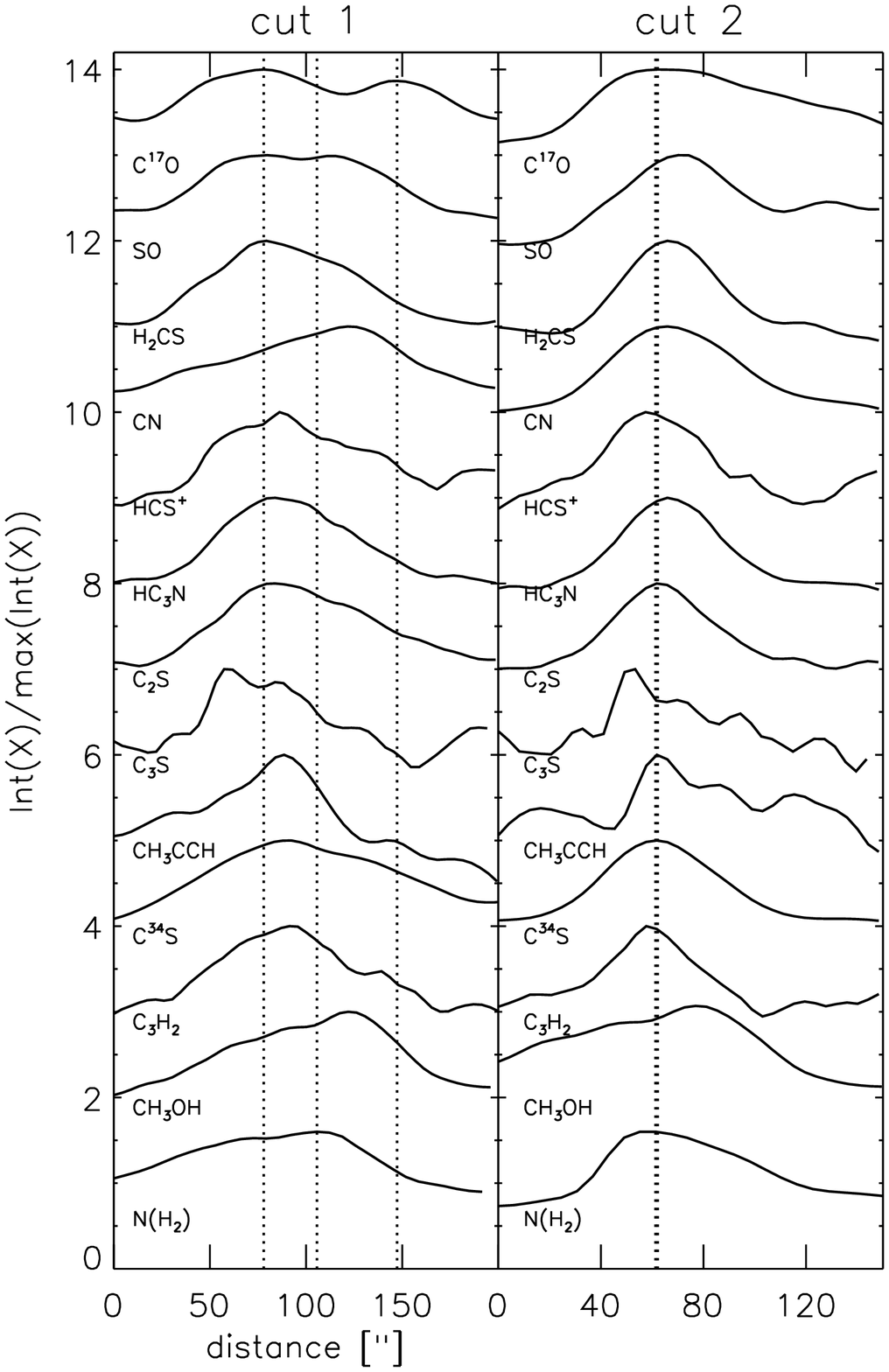}
\caption{\textit{Left}: Intensity cuts of selected molecules extracted along a line which crosses the two C$^{17}$O peaks (which are close to the $c$-C$_3$H$_2$ and CH$_3$OH peaks) for various species (cut 1 in Fig. 1). \textit{Right}: Intensity cuts extracted along a line which is perpendicular to the other cut and crosses the \textit{Herschel} dust peak (cut 2 in Fig. 1). The intensity values (y-axis) are normalized to the maximum value for each species, and a constant has been added to compare the normalized intensity distributions of the different species. The vertical dotted lines show the locations of the C$^{17}$O and $N$(H$_2$) peaks.}
\label{L1521E_cuts}
\end{figure}

Integrated intensity maps of the detected species are shown in Fig. \ref{spatial_distribution_1} and \ref{c3h2_ch3oh}. The spatial distribution of the different species is further shown as line profiles extracted toward the $c$-C$_3$H$_2$ and methanol peaks in Fig. \ref{line_profiles}. 
Furthermore, intensity cuts extracted toward a line which crosses the two molecular peaks and toward a line which is perpendicular to that, and crosses the \textit{Herschel} dust peak are shown in Fig. \ref{L1521E_cuts}. The position of the two cuts is shown in Fig. \ref{c3h2_ch3oh}.
Similar to what was observed in the more evolved L1544 core \citep{spezzano2016}, methanol and $c$-C$_3$H$_2$ peak at different positions around the center of L1521E (Fig. \ref{c3h2_ch3oh}). 
Most species peak where \textit{c}-C$_3$H$_2$ peaks (RA(J2000)=$04^{\rm{h}}29^{\rm{m}}16.0^{\rm{s}}$ Dec(J2000)=$+26^\circ14'0.9''$), such as C$_2$S, C$_3$S, HCS$^+$, HC$_3$N, H$_2$CS, CH$_3$CCH, C$^{34}$S.
The CN emission peak is between the \textit{Herschel} dust peak and the methanol peak (RA(J2000)=$04^{\rm{h}}29^{\rm{m}}14.4^{\rm{s}}$ Dec(J2000)=$+26^\circ14'17.7''$), and considering the \textit{Herschel} and IRAM-30m beam sizes, it is consistent with both peaks. N$_2$H$^+$ also peaks near the \textit{Herschel} dust peak.

\begin{figure*}[h]
\centering
\includegraphics[width=8.5cm, trim=1cm 7cm 1cm 8cm,clip=true]{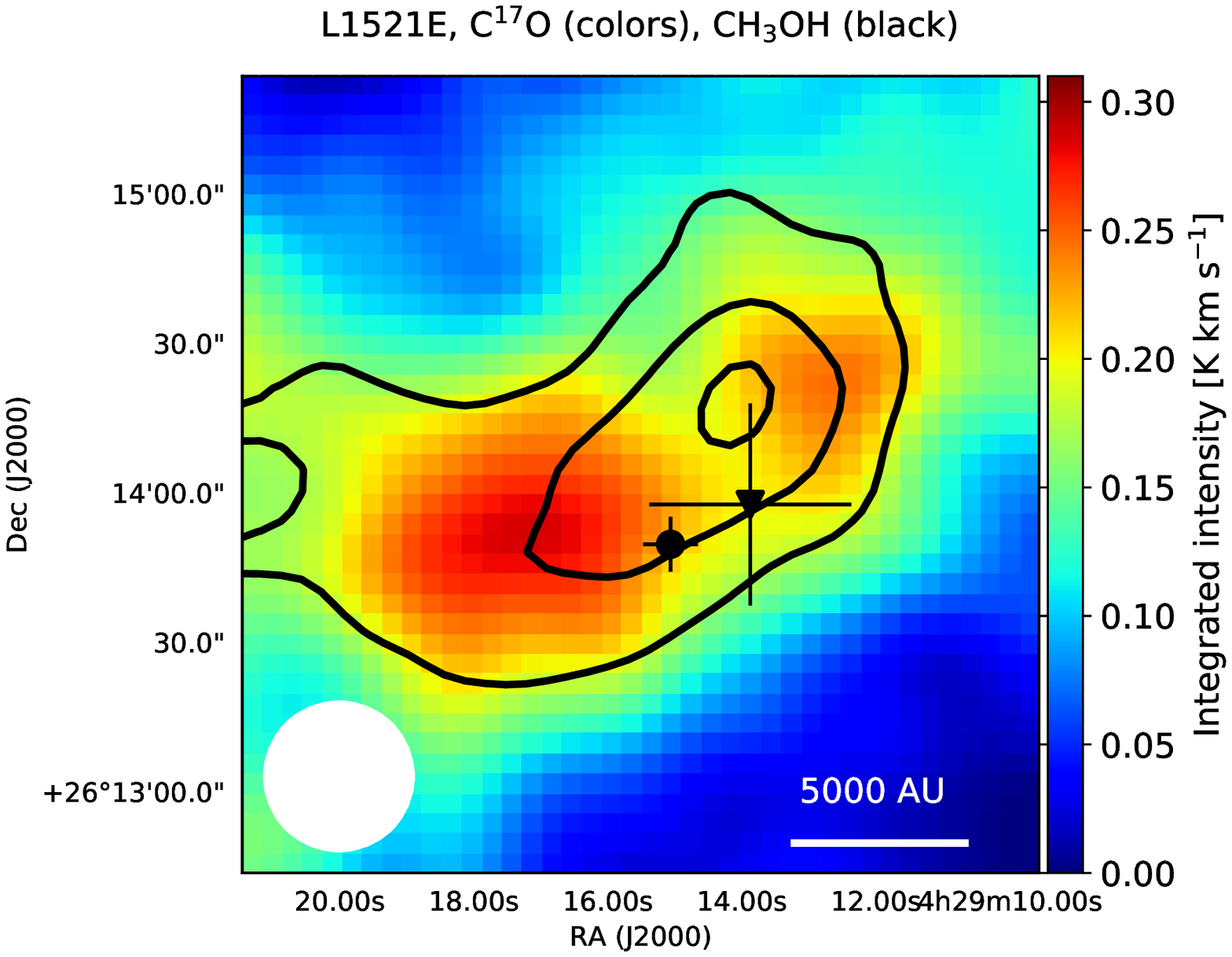}   
\includegraphics[width=8.5cm, trim=1cm 7cm 1cm 8cm,clip=true]{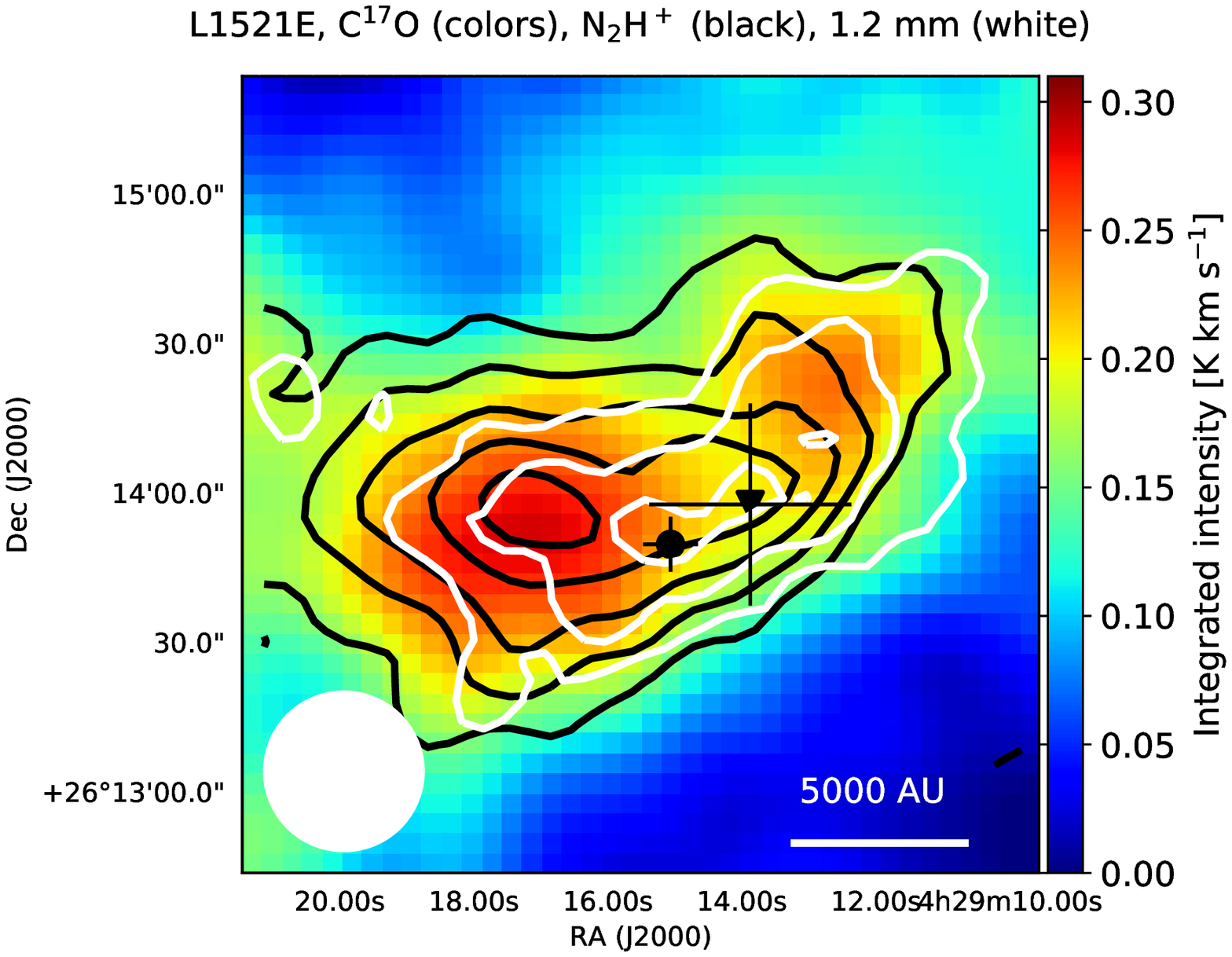}
\caption{The structure of L1521E around the center of the core. \textit{Left}: The integrated intensity of CH$_3$OH (black contours) overplotted on the integrated intensity of C$^{17}$O. The black dot shows the peak of the 1.2 mm emission. The black triangle shows the peak of the H$_2$ column density derived from the \textit{Herschel}/SPIRE data. The CH$_3$OH contour levels start from 6-$\sigma$ rms in steps of 6-$\sigma$ rms, with a 3-$\sigma$ rms level of 0.07 K km/s.
\textit{Right:} The integrated intensity of N$_2$H$^+$ (black contours) and the distribution of the 1.2 mm emission (white contours) overplotted on the integrated intensity of C$^{17}$O. The black dot shows the peak of the 1.2 mm emission. The black triangle shows the peak of the H$_2$ column density derived from the \textit{Herschel}/SPIRE data. The N$_2$H$^+$ contour levels start from 6-$\sigma$ rms in steps of 4-$\sigma$ rms, with 3-$\sigma$ rms level of 0.08 K km/s. For the 1.2 mm data the contours start from 6-$\sigma$ rms in steps of 3-$\sigma$ rms with a 3-$\sigma$ rms level of 5.38 mJy.}
\label{spatial_distribution}
\end{figure*}

The spatial distribution of gas traced by the detected species compared to the location of the dust peak traced by the 1.2 mm and the \textit{Herschel} continuum data is shown in Fig. \ref{spatial_distribution}. The dust peak based on the \textit{Herschel}/SPIRE data (RA(J2000)=$04^{\rm{h}}29^{\rm{m}}14.0^{\rm{s}}$ Dec(J2000)=$+26^\circ13'56''$) is slightly offset from the 1.2 mm peak (RA(J2000)=$04^{\rm{h}}29^{\rm{m}}15.15^{\rm{s}}$ Dec(J2000)=$+26^\circ13'50.8''$), but considering the large SPIRE beam, it is consistent with both the 1.2 mm peak and the CH$_3$OH peak. Also, SPIRE is probably probing a different dust population compared to the 1.2 mm data.
The distribution of N$_2$H$^+$ is similar to the distribution of dust, as it is also the case for L1544 \citep{caselli1999}. C$^{17}$O shows signs of depletion toward the dust peak, which was also observed for L1544 \citep{caselli1999}. CH$_3$OH peaks in a region where CO is depleted, but not exactly towards the dust peak or towards the region where the CO freeze-out is the most prominent.
The spatial distribution of species shown in Fig. \ref{spatial_distribution} and the depletion of CO is further discussed in Sect. \ref{sect:depl}.

\subsection{Principal Component Analysis}
\label{sect:pca}

\begin{figure*}[h]
\centering   
\includegraphics[height=8.5cm, trim=-2.5cm 0cm 0cm -1.5cm,clip=true]{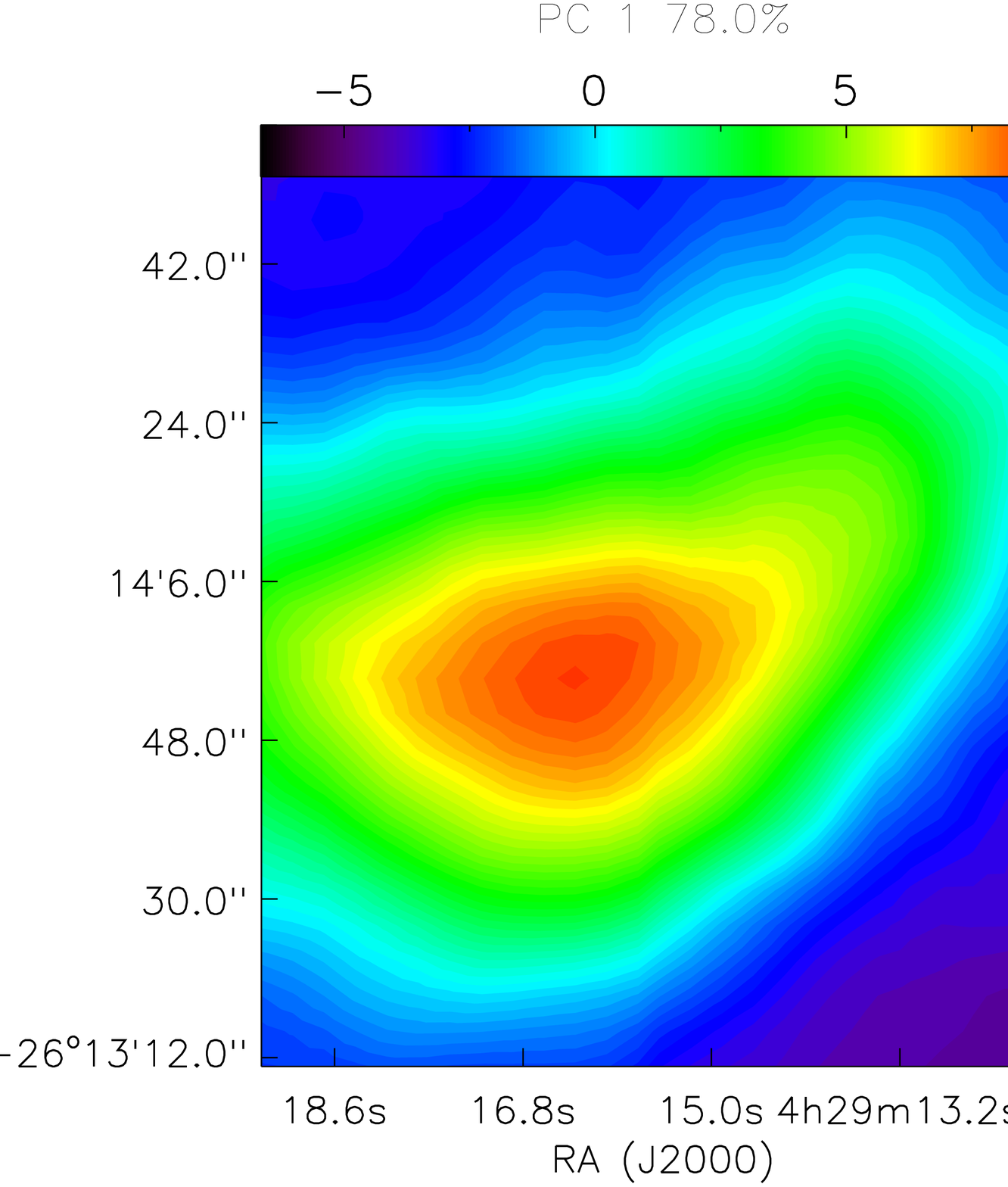}
\includegraphics[height=8.5cm, trim=-2.5cm 0cm 0cm -1.5cm,clip=true]{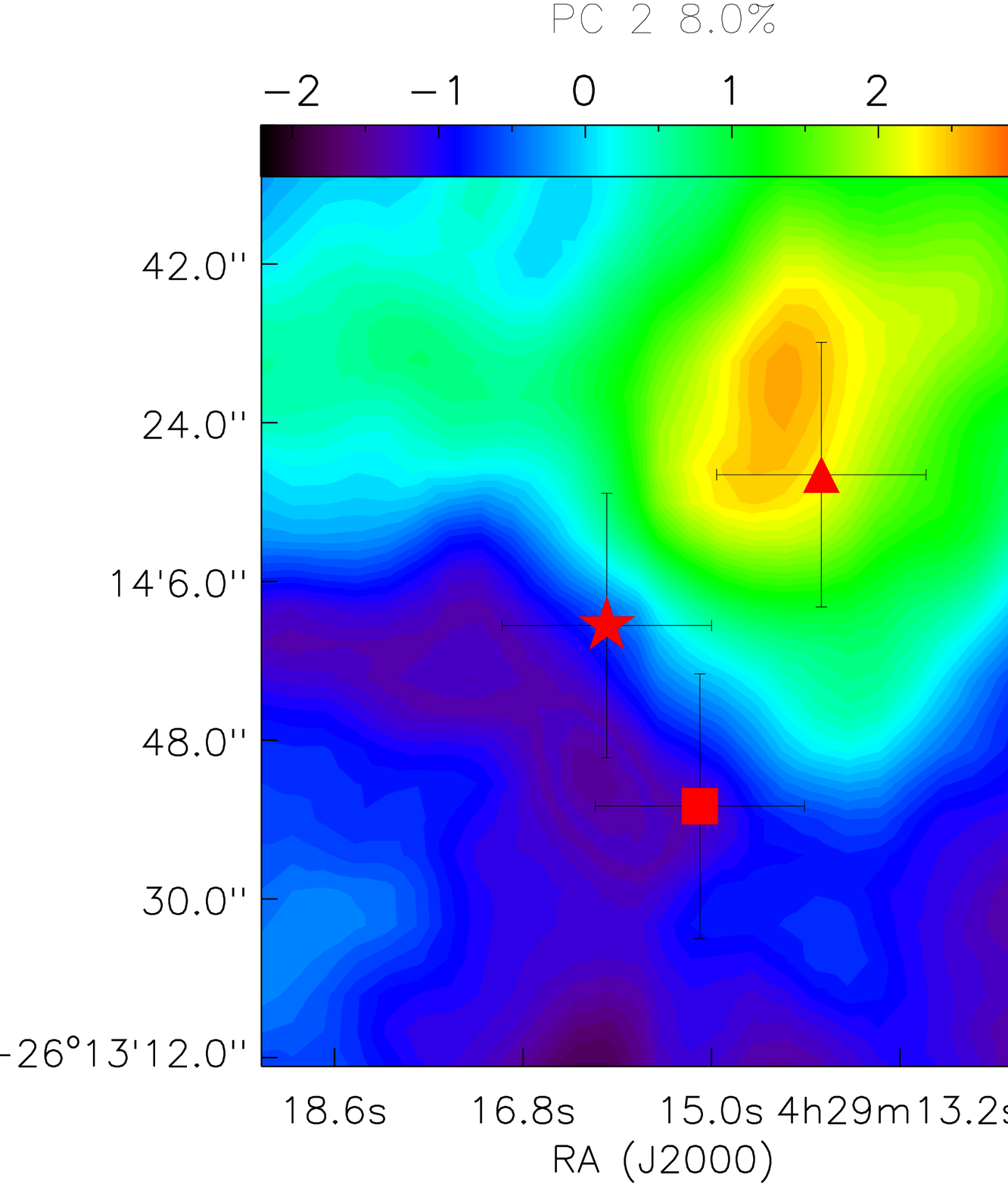}
\includegraphics[height=8.5cm, trim=-2.5cm 0cm 0cm -1.5cm,clip=true]{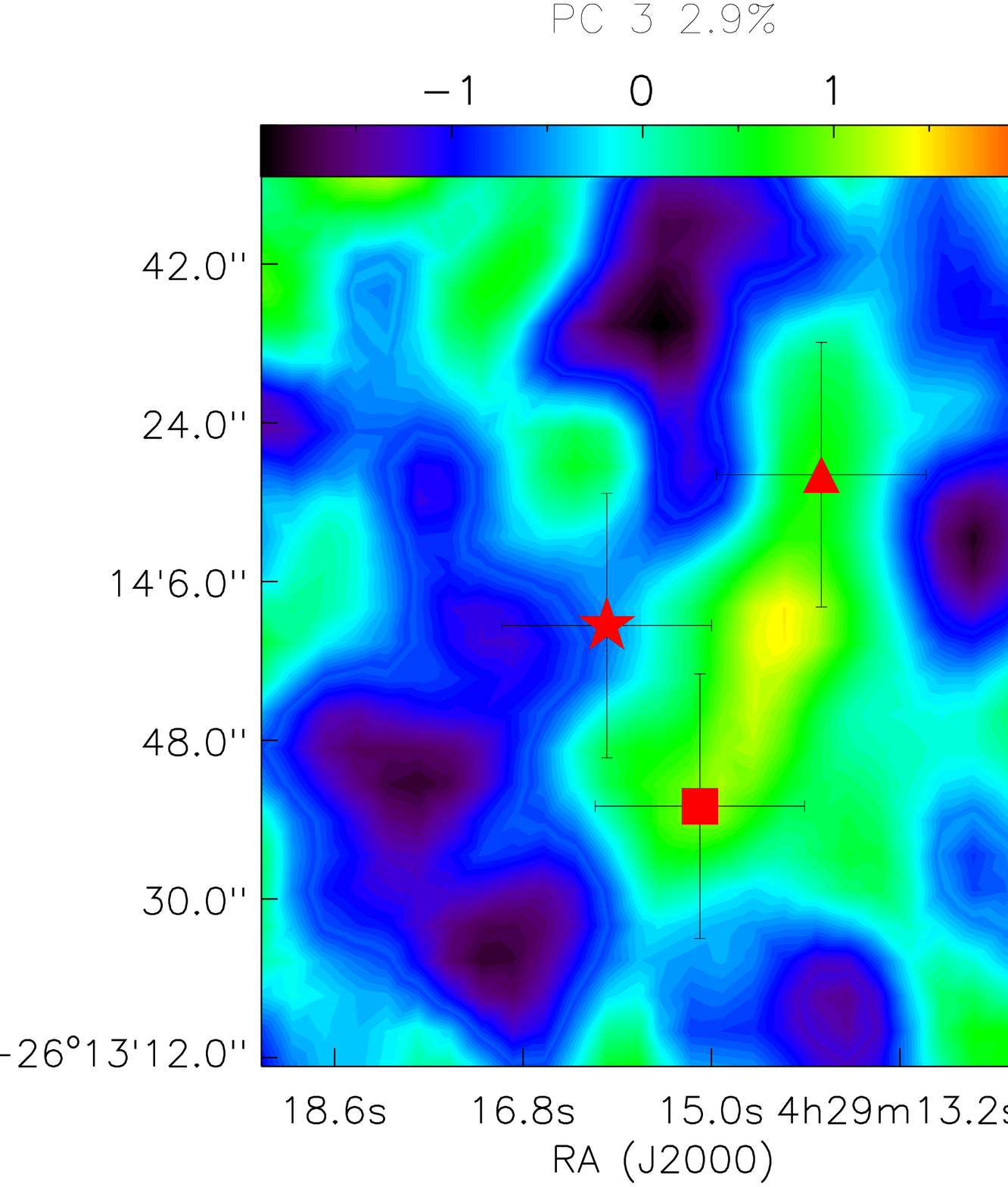}
\includegraphics[height=8.5cm, trim=-2.5cm 0cm 0cm -1.5cm,clip=true]{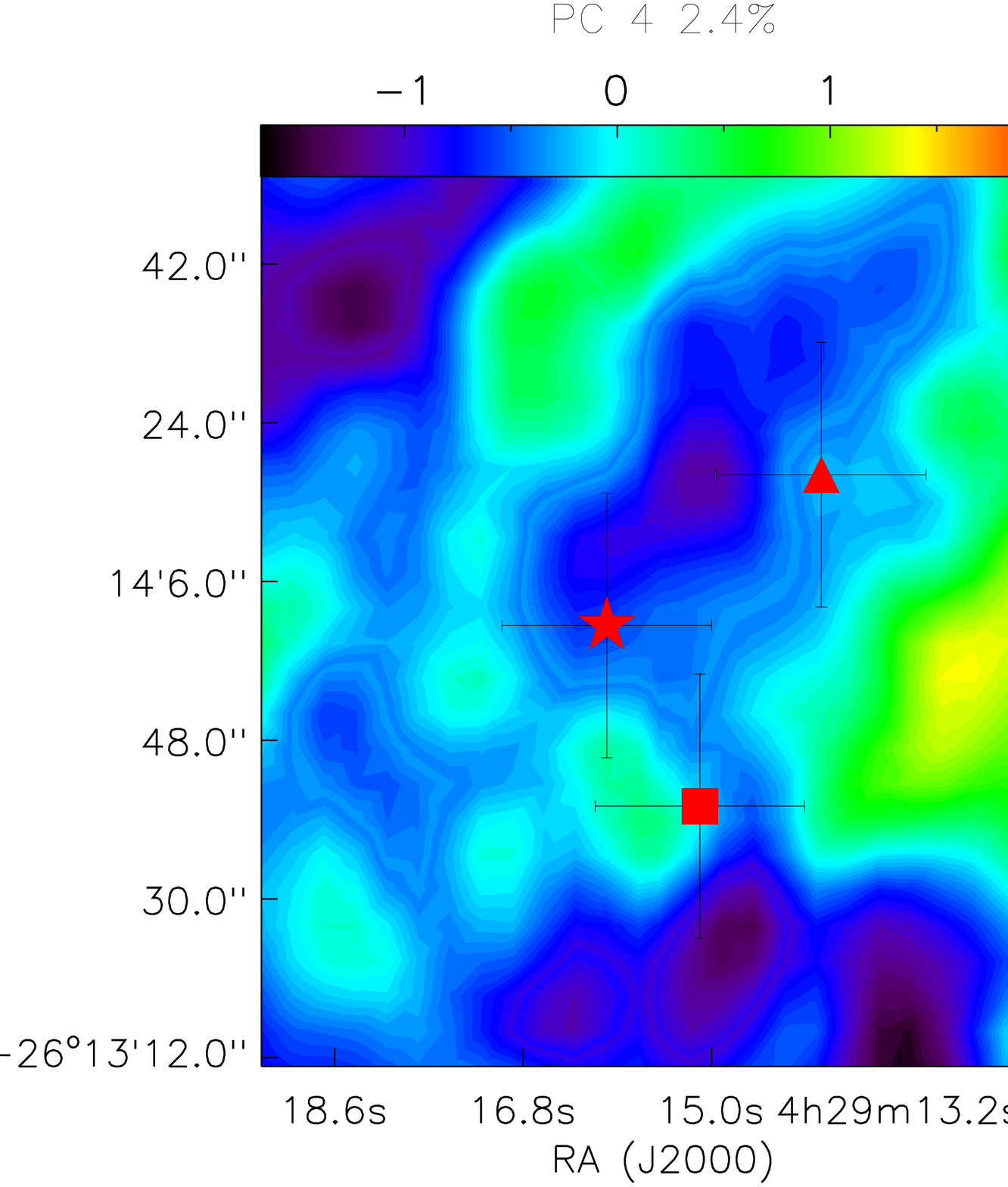}

\caption{Maps of the first four principal components obtained by performing the PCA. The percentages represent the amount of correlation that can be reproduced by the single principal component. The red triangle represents the CH$_3$OH peak, the asterisk the $c$-C$_3$H$_2$ peak, and the square the HNCO peak.}
\label{pca_maps}
\end{figure*}

\begin{figure*}[h]
\centering   
\includegraphics[width=\textwidth, trim=0cm 2.5cm 0cm 0.5cm,clip=true]{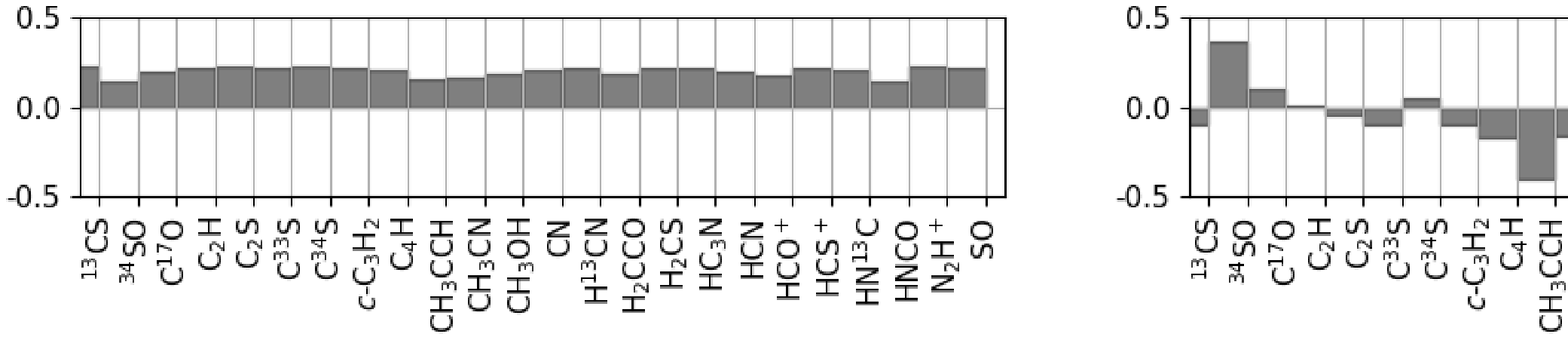}
\caption{Contribution of each transition to the first two PC, obtained by performing the PCA on the standardised data.}
\label{pca_barplot}
\end{figure*}

\begin{figure}[h]
\centering   
\includegraphics[width=9cm, trim=1.5cm 3cm 2.5cm 9cm,clip=true]{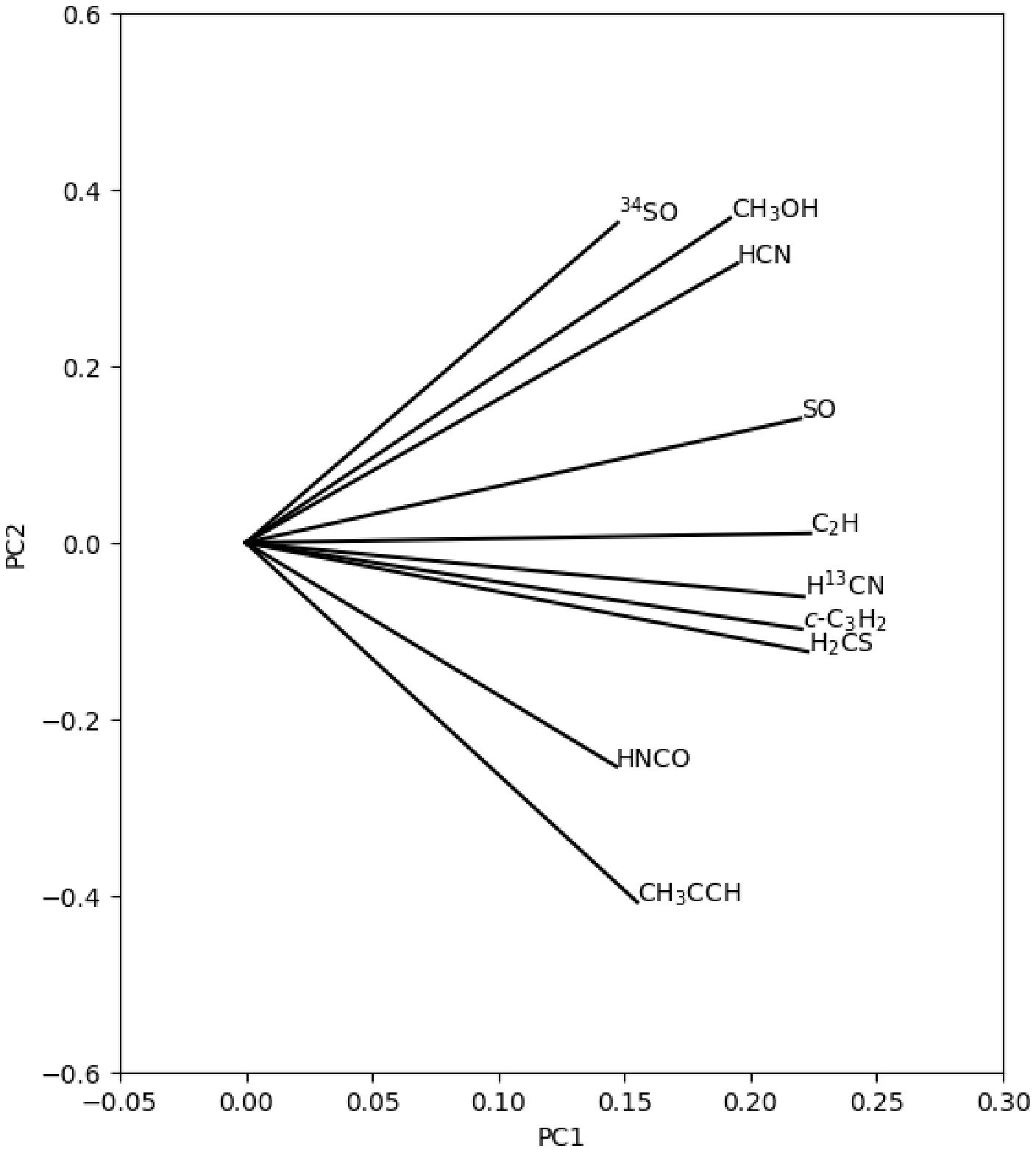}
\caption{Correlation wheels where the coordinates for each molecule are their contribution to each PC, obtained by performing the PCA on the standardised data.
}
\label{pca_corr_wheels}
\end{figure}

Another method to study the structure of L1521E and probe correlations between the observed molecular species is the Principal Component Analysis (PCA), which was done earlier for L1544 by \citet{spezzano2017}. 
A detailed description of the method is shown in \citet{ungerechts1997}. The PCA explains the total variability of correlated variables through the use of the same number of orthogonal principal components. Here the number of correlated variables is the number of maps used for the analysis: 24.  
Each map consists of 27$\times$27 pixels in a space with 24 dimensions (the number of molecular transitions used).
We used a transition for most molecular species included in this analysis, except for the species with the lowest signal-to-noise ratio: C$_3$S, CC$^{34}$S, CH$_3$CHO, HC$^{18}$O$^+$, HCO, OCS, and C$_4$H$_2$. For the species included in the analysis, we used the highest S/N transition in each case.
Similar to \citet{spezzano2017}, we performed the PCA on a standardized (mean-centered and normalized) data set, with the standardized value defined as $x_{\rm{std}}=(x-\mu) / \sigma$, where $\mu$ is the mean value and $\sigma$ is the standard deviation for each map.
While it is possible to compute 24 principal components from our data, we analyze the first four PCs in this paper, as those account for most of the correlation (91.3\%) in the data. Figure \ref{pca_maps} shows maps of the first four PCs. The first PC correlates with the continuum emission, as shown in Figure \ref{spatial_distribution_1}. Similar to the case of L1544 \citep{spezzano2017}, the second PC shows the two most prominent molecular peaks, the $c$-C$_3$H$_2$ and CH$_3$OH peaks. The third and fourth PCs account for only 2.9\% and 2.4\% of the correlation and are likely to be dominated by noise. 
Figure \ref{pca_barplot} shows the contribution of each molecule to the four PCs, and Figure \ref{pca_corr_wheels} shows correlation wheels where the coordinates for each molecule are given by their correlation coeffcients to each PC, obtained by performing the PCA on the standardised data.

The PCA results for L1521E and those presented by \citet{spezzano2017} for L1544 are similar, as seen in Fig. \ref{pca_maps} and in Fig. 2 in \citet{spezzano2017}. The first PC follows the distribution of the continuum in both cases, however, it represents 78.0\% of the correlation in the dataset for L1521E, and only 67\% for L1544. The second PC shows the methanol and $c$-C$_3$H$_2$ peaks in case of both sources, but it represents a higher correlation (12.2\%) for L1544 than for L1521E (8.0\%). 
The differences between the results for the two cores may be related to the fact that while 28 different molecular line maps were included in the PCA for L1544, only 24 species were available with a sufficient S/N level toward L1521E.

\subsection{Molecular column densities}
\label{sect:coldens}

We derived column densities for the observed species toward the \textit{Herschel} dust peak in an optically thin approximation, assuming a single excitation temperature and negligible background radiation. 
We measured the excitation temperature by fitting the hyperfine structure of the N$_2$H$^+$ (1-0) line toward the \textit{Herschel} dust peak, using Gildas/CLASS. The excitation temperature given by the fit is $\sim$4.5 K. We used this excitation temperature to derive the column density of all the species, except for C$^{17}$O. Due to its low critical density ($\sim$2$\times$10$^3$ cm$^{-3}$), C$^{17}$O is supposed to be thermalized, therefore we assumed the excitation temperature for C$^{17}$O to be 10 K, which is close to the dust temperature of 9.8 K derived by \citet{makiwa2016}.
The measured integrated intensities of the lines ($\int T_{\mathrm{MB}} \mathrm{d}V$) can be converted to the column densities of the molecules ($N_\mathrm{tot}$) in optically thin approximation using:
\begin{equation}
\label{coldens_lte}
N_{\rm{tot}}=\frac{8\pi k \nu^2}{h c^3} \frac{Q(T_{\rm{ex}})}{g_u A_{\rm{ul}}} e^{E_u/k T_{\rm{ex}}} \int{T_{\rm{mb}} d{\rm{v}}},
\end{equation}
where $g_\mathrm{u}$ is the statistical weight of the upper level, $Q(T_{\mathrm{ex}})$ is the partition function for $T_{\mathrm{ex}}$, $E_\mathrm{u}$ is the upper level energy, $\nu$ is the frequency of the used transition, and $A_{\rm{ul}}$ is the spontaneous decay rate. Equation \ref{coldens_lte} also assumes a beam filling factor of 1, which means that the source emission is assumed to be more extended than the beam size.
The column densities of the species are summarized in Table \ref{table:coldens}. The transitions used to calculate column densities are marked with asterisk in Table \ref{table:observations}. The error of the column densities is dominated by calibration error and error given by the uncertainty of the assumed excitation temperature.

\begin{table}[h]
\begin{minipage}[]{\linewidth}\centering
\setlength{\arrayrulewidth}{1pt}	
\begin{center}
\caption{Molecular column densities derived toward the L1521E and L1544 dust peaks. The column densities toward L1544 are based on data from the ASAI \citep{lefloch2018}. The assumed excitation temperature was 10 K for CO and 4.5 K for the other species for L1521E, and 5 K for L1544.
The used transitions are shown in Table \ref{table:observations} with asterisk, except for C$_3$S (92488.5 MHz), CH$_3$CHO (95947.4 MHz), and H$_2$CCO (101981.4 MHz), to be consistent between the two sources.
The SO column density is based on the $^{34}$SO column density and an isotopic ratio of 22 \citep{wilsonrood1994}.
}
\begin{tabular}{lcc}
\hline\\[-0.2cm]
Molecule& \multicolumn{2}{c}{Column density (cm$^{-2}$)}\\
        & L1521E& L1544\\
\hline\\[-0.25cm]    

SO&                   $(1.6\pm0.5)\times10^{13}$&   
                      $(1.5\pm0.5)\times10^{13}$\\

$^{13}$CS&            $(1.6\pm0.5)\times10^{12}$&
                      $(2.5\pm0.8)\times10^{11}$\\

OCS&                  $(2.3\pm0.7)\times10^{13}$&
                      $(8.9\pm2.7)\times10^{12}$\\

C$_2$S&               $(1.4\pm0.4)\times10^{13}$&   
                      $(4.8\pm1.4)\times10^{12}$\\

C$_3$S&               $(8.2\pm2.5)\times10^{13}$&                          
                      $(5.7\pm1.7)\times10^{12}$\\

C$^{33}$S&            $(1.4\pm0.4)\times10^{12}$&   
                      $(2.3\pm0.7)\times10^{11}$\\

C$^{34}$S&            $(2.9\pm0.9)\times10^{12}$&   
                      $(6.1\pm1.8)\times10^{11}$\\

CC$^{34}$S&           $(8.3\pm2.5)\times10^{11}$&                      
                                                -\\                       

HCS$^+$&              $(1.1\pm0.3)\times10^{12}$&   
                      $(4.4\pm1.3)\times10^{11}$\\
 
H$_2$CS&              $(1.8\pm0.5)\times10^{13}$&                        
                      $(6.0\pm1.8)\times10^{12}$\\

C$^{17}$O&            $(1.2\pm0.4)\times10^{14}$&                          
                      -\\
HC$_3$N&              $(8.4\pm2.5)\times10^{12}$&   
                      $(1.1\pm0.3)\times10^{13}$\\

$c$-C$_3$H$_2$&       $(1.3\pm0.4)\times10^{12}$&   
                      $(4.3\pm1.3)\times10^{12}$\\

C$_4$H&               $(1.6\pm0.5)\times10^{14}$&   
                      $(2.0\pm0.6)\times10^{14}$\\

HN$^{13}$C&           $(6.0\pm1.8)\times10^{11}$&   
                      $(2.0\pm0.6)\times10^{12}$\\

CH$_3$OH ($E_2$)&     $(1.9\pm0.6)\times10^{13}$&   
                      $(2.2\pm0.7)\times10^{13}$\\

CH$_3$OH ($A^+$)&     $(5.7\pm1.7)\times10^{12}$&   
                      $(7.3\pm2.2)\times10^{12}$\\

CH$_3$CCH&            $(2.1\pm0.6)\times10^{13}$&                          
                      $(4.9\pm1.5)\times10^{13}$\\    

CH$_3$CHO&            $(8.3\pm2.5)\times10^{10}$&
                      $(6.2\pm1.9)\times10^{10}$\\   

$l$-C$_4$H$_2$&       $(4.1\pm1.2)\times10^{11}$&
                                                -\\

H$_2$CCO&             $(1.4\pm0.4)\times10^{12}$&
                                                -\\                                                 

CH$_3$CN&             $(4.8\pm1.4)\times10^{11}$&
                      $(6.1\pm1.8)\times10^{11}$\\

HC$^{18}$O$^+$&       $(6.7\pm2.0)\times10^{10}$&
                      $(8.3\pm2.5)\times10^{10}$\\

H$^{13}$CN&           $(1.6\pm0.5)\times10^{12}$&
                      $(2.6\pm0.8)\times10^{12}$\\

HCO&                  $(7.0\pm2.1)\times10^{11}$&
                      $(5.0\pm1.5)\times10^{11}$\\

HCN&                  $(2.4\pm0.7)\times10^{12}$&
                      $(1.3\pm0.4)\times10^{12}$\\

HCO$^+$&              $(1.3\pm0.4)\times10^{12}$&
                      $(3.2\pm1.0)\times10^{11}$\\         

N$_2$H$^+$&           $(2.5\pm0.8)\times10^{13}$&
                      $(2.0\pm0.6)\times10^{14}$\\
                       
HNCO&                 $(9.3\pm2.8)\times10^{12}$&
                      $(1.6\pm0.5)\times10^{13}$\\
                       
C$_2$H&               $(7.6\pm2.3)\times10^{13}$&
                      $(9.5\pm2.9)\times10^{13}$\\                                              

CN&                   $(1.2\pm0.4)\times10^{13}$&
                      -\\

\hline

\end{tabular}
\label{table:coldens}
\end{center}
\end{minipage}
\end{table}	

\subsection{CO depletion}
\label{sect:depl}

As already seen in Fig. \ref{spatial_distribution}, the C$^{17}$O line intensities decrease around the \textit{Herschel} dust peak, which suggests that there is some CO depletion.
To calculate the CO depletion factors we used the observed C$^{17}$O map and the $N$(H$_2$) map derived from the \textit{Herschel}/SPIRE data. We use the $J$=1$-$0, $F$=5/2$-$5/2 transition of C$^{17}$O at 112.36 GHz to derive the C$^{17}$O column density in optically 
thin LTE approximation assuming an excitation temperature of 10~K \citep{tafallasantiago2004}. For this analysis we smoothed the 
C$^{17}$O map from its original spatial resolution ($\sim$30$''$) to the resolution of the $N$(H$_2$) map from the \textit{Herschel}/SPIRE data 
($\sim40''$).
We used a $^{16}$O/$^{17}$O abundance ratio of 2044 (\citealp{penzias1981}, \citealp{wilsonrood1994}) and a canonical CO abundance of 8.5$\times$10$^{-5}$ \citep{frerking1982}.

To derive $N$(H$_2$) from the \textit{Herschel}/SPIRE fluxes we used a dust emissivity index of $\beta$=1.5. The value of $\beta$=1.5 gives a good comparison with L1544 as it was used to derive $N$(H$_2$) for that source \citep{spezzano2016}. Other studies, 
such as \citet{kirk2007} assumed a value of $\beta$=2.0 for L1544. A $\beta$=2.6$\pm$0.9, which was derived from \textit{Herschel}/SPIRE data by 
\citet{makiwa2016} is also consistent with this value.

\begin{figure*}[h]
\centering   
\includegraphics[width=9cm, trim=-1.5cm 0cm 0cm -1.0cm,clip=true]{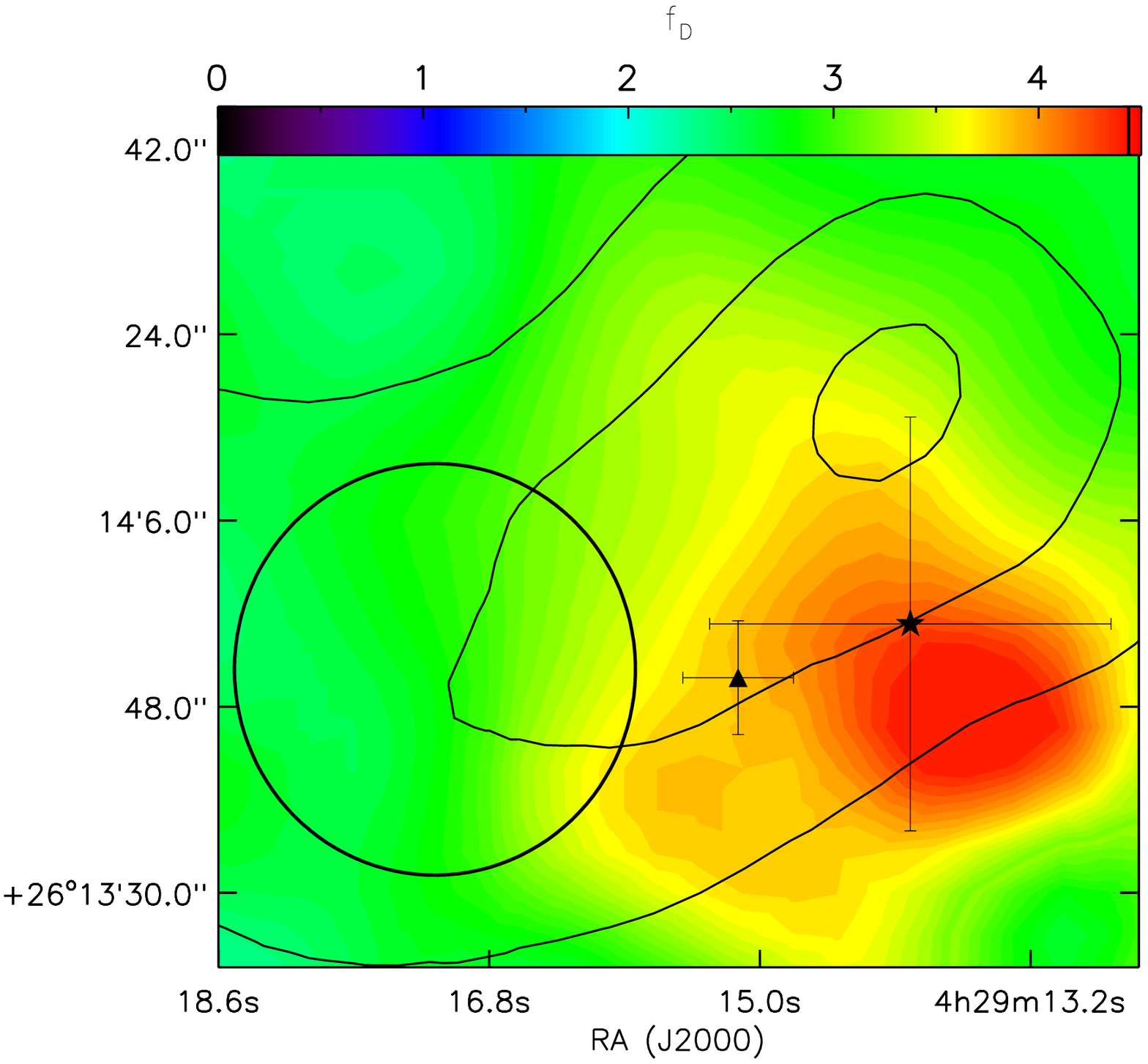}
\includegraphics[width=9cm, trim=-1.5cm 0cm 0cm -1.0cm,clip=true]{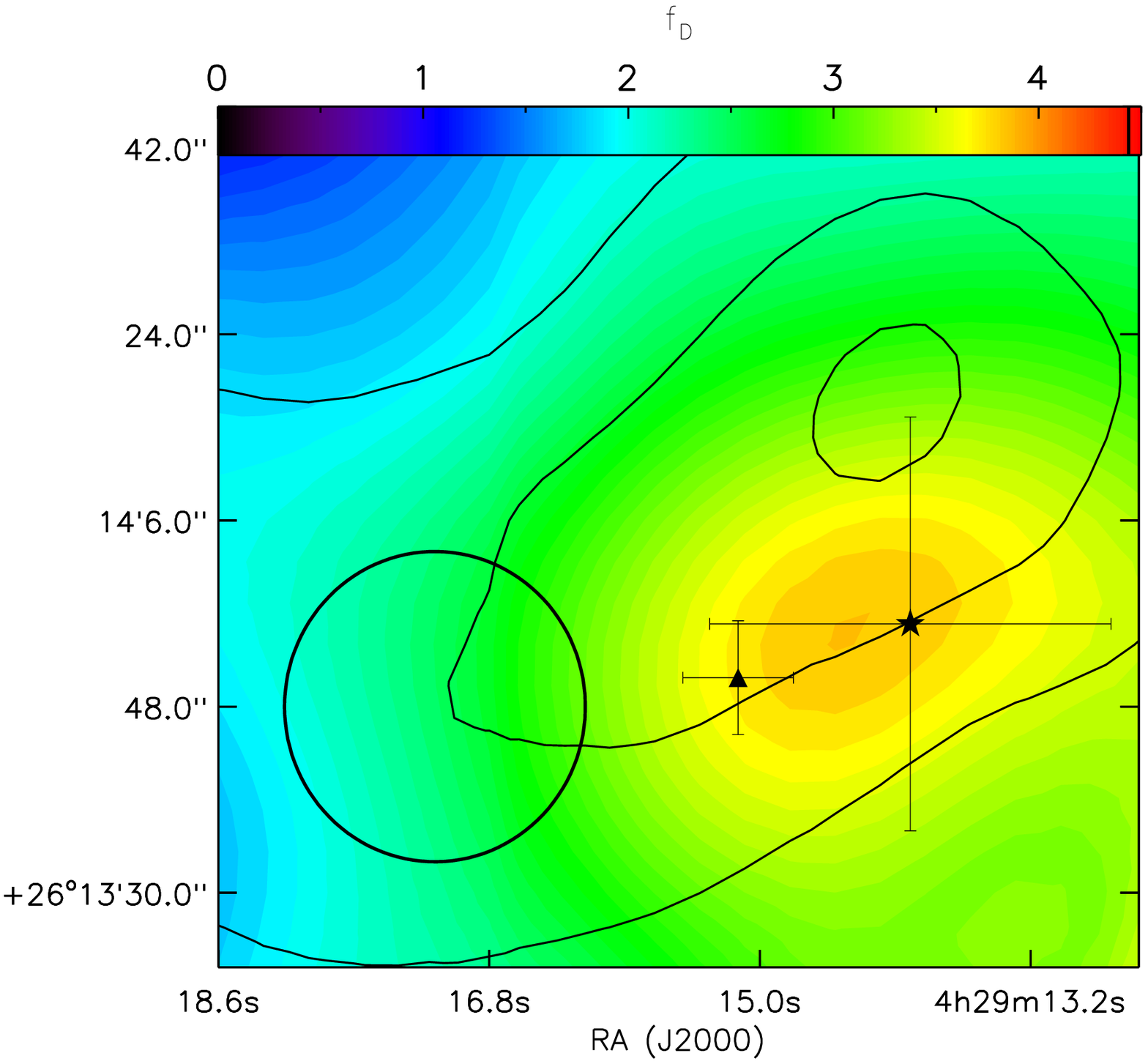}
\caption{
Map of the CO depletion factors toward L1521E calculated using H$_2$ column densities based on the \textit{Herschel}/SPIRE data with $\beta$=1.5 (left panel) and based on the 1.2 mm map (right panel), the derived C$^{17}$O abundances and a reference value of X(C$^{17}$O)$=$5.4$\times$10$^{-8}$. 
The contours show the spatial distribution of methanol and start from 6-$\sigma$ rms noise level in steps of 6-$\sigma$ rms. The triangle shows the peak of the 1.2 mm emission and the asterisk symbol shows the \textit{Herschel} dust peak.
The depletion factor toward the \textit{Herschel} dust peak is about 4.3 based on the \textit{Herschel} $N$(H$_2$) values and about 3.9 based on the $N$(H$_2$) values calculated from the 1.2 mm data. The black circles show the beam sizes which were used to calculate the depletion factors.
}
\label{depl_map}
\end{figure*}

The calculation for $\beta$=1.5 results in a CO depletion factor of 4.3$\pm$1.6 toward the \textit{Herschel} dust peak. 
The derived depletion factors are shown in Fig. \ref{depl_map}, with the methanol integrated intensity contours overplotted. Even though the depletion factors don't peak at the same position as the methanol intensities, the depletion factors are increasing with increasing methanol abundance, which was expected as CH$_3$OH is produced on the surface of dust grains by CO hydrogenation \citep{watanabekouchi2002} and then released to the gas phase via 
reactive desorption \citep{vasyunin2017}.

We also used the 1.2 mm data to derive $N$(H$_2$) from a different method describe in \citet{crapsi2005}. We smoothed the 1.2 mm map to the resolution of the IRAM-30m data ($\sim30''$). Similar to \citet{crapsi2005}, for the dust opacity per unit mass we used a value of $\kappa_{\rm{1.2mm}}=0.005$ cm$^2$ g$^{-1}$. This value was chosen following \citet{crapsi2005}, who derived depletion factors for a number of pre-stellar / starless cores. Originally, \citet{andre1996} chose this value based on the recommendation of \citet{henning1995} and \citet{preibisch1993} for clouds at intermediate densities ($\sim$10$^5$ cm$^{-3}$) and gas-to-dust ratios of 100.
The CO depletion factors resulting from this method are similar to the values derived using $N$(H$_2$) values from the \textit{Herschel} data, with a value of $\sim$3.9$\pm$1.4 toward the \textit{Herschel} dust peak. 
The large depletion factors based on the 1.2 mm data in the southern part of the map in Fig. \ref{depl_map} are probably an artifact due to the filtering of the extended cloud structure by MAMBO-2. The signal-to-noise ratio in these parts of the map is below 3.

\citet{crapsi2005} calculated depletion factors for several pre-stellar cores which are in the range between 3.4$\pm$2.1 and 15.5$\pm$3.7, including the value 14$\pm$2.2 for L1544, which is about three times the value derived for L1521E (4.3$\pm$1.6) when assuming $\beta$=1.5 for the H$_2$ column density. Based on the comparison of C$^{18}$O observations of L1521E to chemical models, \citet{tafallasantiago2004} concluded that no CO depletion is present in L1521E, given that the data are consistent with a constant abundance profile. 
Similarly, \citet{fordshirley2011} found little or no C$^{18}$O depletion toward L1521E by the comparison of observations and chemical models.
\citet{tafallasantiago2004} found some differences between the distribution of the C$^{18}$O and continuum emission outside the core central region (which is also seen in Figure 1 of \citealp{fordshirley2011}), but attributed  them to the low sensitivity of the data at those positions, and not to CO depletion.
The depletion factor of 4.3$\pm$1.6 found using the data presented here and the spatial distribution of C$^{17}$O (1-0) suggest that some CO depletion is present in L1521E near the dust peak.
   
\subsection{Comparison to models}

\begin{figure}[h]
\centering   
\includegraphics[width=9cm, trim=-1cm -1cm 0cm 0cm,clip=true]{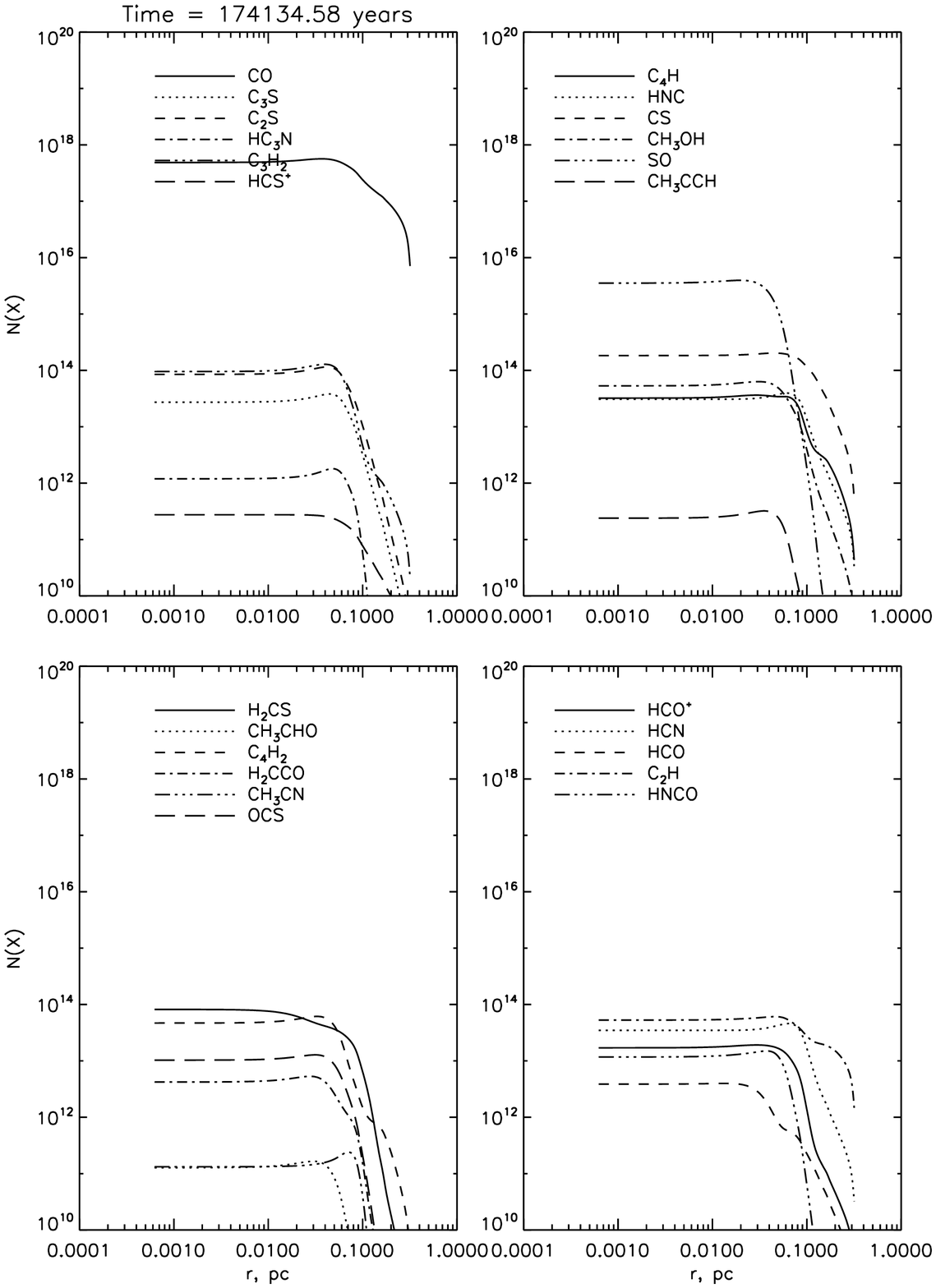}
\caption{Column densities of the different species as a function of the radius predicted in the model.}
\label{model_coldens}
\end{figure}

To understand the measured column densities, we preformed chemical modeling using the 1D code described in \citet{vasyunin2017}. The model represents the core with a spherically symmetrical structure. We used the density structure reported by \citet{tafallasantiago2004} and a constant temperature profile at $T$=10 K, also used by \citet{tafallasantiago2004}. 
In the model we calculate time-dependent fractional abundances and column densities of species. Then the calculated column densities are convolved with a 26$''$ gaussian beam, to be comparable to those measured with the IRAM-30m. In the model, only the main isotopologues of the species are available, therefore, we used isotopic ratios based on \citet{wilsonrood1994} to compare the modelled and observed column densities. We ran several models with different inital abundances and compared the resulting column densities to those obtained from the observations, to find a model which fits most observed column within an order of magnitude. The best fit model uses ''low metals'' (EA1) initial abundances from \citet{wakelamherbst2008}, except for the fractional abundance of S$^+$, which needs to be increased to 10$^{-6}$ instead of 8$\times$10$^{-8}$ to fit the observations.
The model was ran in two stages. First, we ran the evolution of a diffuse cloud for 10$^6$ 
years. Then, we took the final abundances of species from the first stage as the initial chemical composition for the next stage, and ran the evolution of L1521E itself.
The best fit column densities from the models are shown in Table \ref{table:modelresults} and in Fig. \ref{model_coldens} as a function of radius.
This model is able to reproduce most measured column densities, especially those of the sulfur-bearing molecules except for SO. This supports that significant sulfur depletion is taking place during the evolution of dense cores. The best fit model suggests a chemical age of $\sim$1.7$\times$10$^5$ years, which is close to what was found by \citet{tafallasantiago2004}. 
For this age estimate $t$=0 in the start of the second stage of the model run mentioned above, when the adopted 1D physical structure of the core has been formed.
Since our physical model is not evolving with time, the obtained age is rather a timescale of the relaxation for chemical evolution of the model core up to the observed chemical composition rather than an estimate of the physical age of the real core.
The CO depletion factor at the moment of best agreement towards the center of L1521E in the model is $\sim$3, which is close to the value of 4.3$\pm$1.6 derived from the observations.

\begin{table}[h]
\begin{minipage}[]{\linewidth}\centering
\setlength{\arrayrulewidth}{1pt}	
\begin{center}
\caption{
Best fit column densities derived for L1521E with the chemical code described in \citet{vasyunin2017} and their agreement with the measured values given in Sect. \ref{coldens_l1521e_l1544}.
}
\begin{tabular}{lcc}
\hline\\[-0.2cm]
Molecule& Column density (cm$^{-2}$)& agree(+)/disagree(-)\\
\hline\\[-0.25cm]    

SO&             3.5$\times$10$^{15}$&     -\\

CS&             1.8$\times$10$^{14}$&     +\\

OCS&            1.0$\times$10$^{13}$&     +\\

C$_2$S&         8.5$\times$10$^{13}$&     +\\

C$_3$S&         2.7$\times$10$^{13}$&     +\\

HCS$^+$&        2.7$\times$10$^{11}$&     +\\

H$_2$CS&        8.2$\times$10$^{13}$&     +\\

CO&             4.9$\times$10$^{17}$&     +\\

HC$_3$N&        1.2$\times$10$^{12}$&     +\\

$c$-C$_3$H$_2$& 9.5$\times$10$^{13}$&     -\\

C$_4$H&         3.2$\times$10$^{13}$&     +\\

HNC&            3.1$\times$10$^{13}$&     +\\

CH$_3$OH&       5.3$\times$10$^{13}$&     +\\

CH$_3$CCH&      2.4$\times$10$^{11}$&     -\\

CH$_3$CHO&      1.3$\times$10$^{11}$&     +\\

$l$-C$_4$H$_2$& 4.7$\times$10$^{13}$&     -\\

H$_2$CCO&       4.2$\times$10$^{12}$&     +\\

CH$_3$CN&       1.3$\times$10$^{11}$&     +\\

HCO$^+$&        1.7$\times$10$^{13}$&     +\\

HCN&            3.5$\times$10$^{13}$&     +\\

HCO&            3.9$\times$10$^{12}$&     +\\

C$_2$H&         5.3$\times$10$^{13}$&     +\\

HNCO&           1.2$\times$10$^{13}$&     +\\

N$_2$H$^+$&     4.8$\times$10$^{12}$&     +\\    

CN&             6.2$\times$10$^{13}$&     +\\                  
      
\hline

\end{tabular}
\label{table:modelresults}
\end{center}
\end{minipage}
\end{table}	

\section{Discussion}
\label{sect_discussion}

We have analyzed IRAM-30m observations of various molecules toward the starless core L1521E. We found that, similarly to L1544, CH$_3$OH and $c$-C$_3$H$_2$ peak at two different positions, but the distance between the two peaks is smaller and similar to the size of the CO depletion zone (in both cases). Most observed species peak at the $c$-C$_3$H$_2$ peak. The C$^{17}$O column densities calculated toward L1521E suggest that CO is depleted toward the \textit{Herschel} dust peak. The depletion factor based on the C$^{17}$O data and $N$(H$_2$) calculated from \textit{Herschel}/SPIRE data was found to be 4.3$\pm$1.6 and the depletion factor based on the 1.2 mm emission was found to be 3.9$\pm$1.4.

\begin{figure*}[h]
\centering   
\includegraphics[height=16cm, trim=0cm 0cm 0cm 0cm, angle=-90,clip=true]{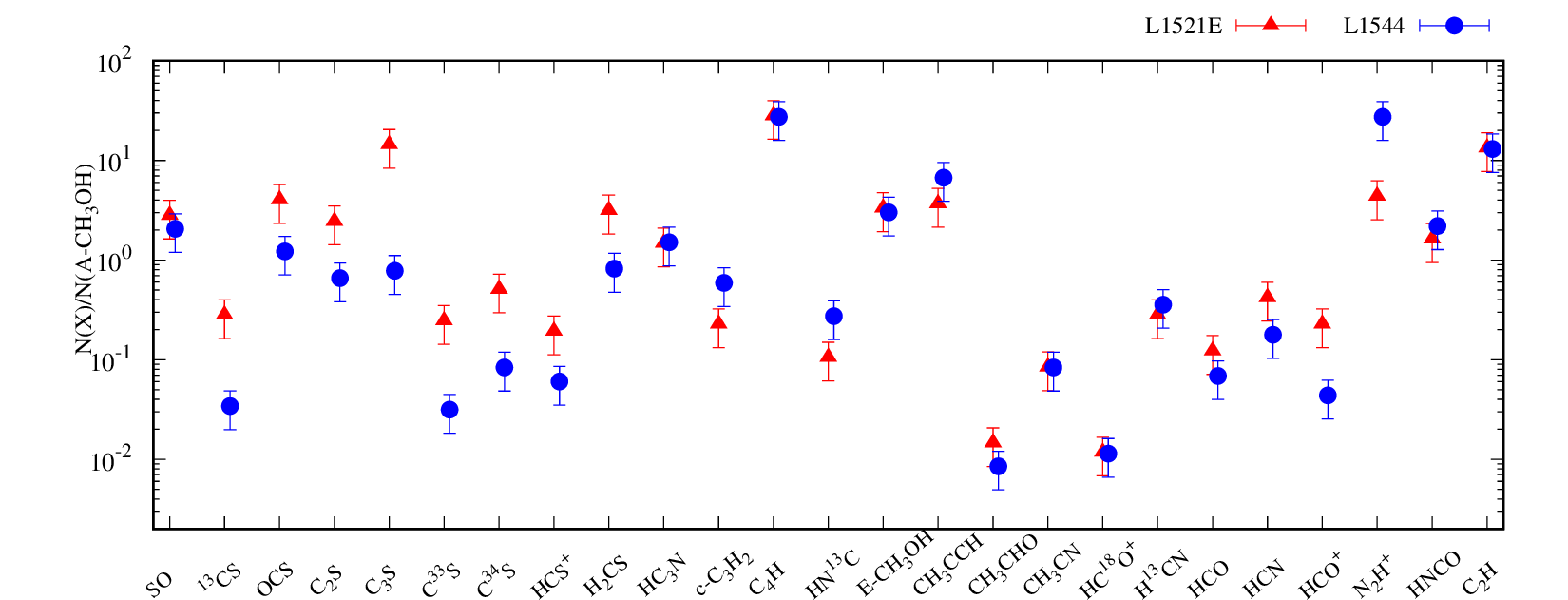}
\caption{Comparison of the abundances of species observed toward the dust peak of both L1521E and L1544. As explained in Sect. 4.1, the abundances were derived using the column densities for an excitation temperature of 4.5~K for L1521E and 5~K for L1544.}
\label{coldens_l1521e_l1544}
\end{figure*}

We compared the column densities / abundances of the species measured toward the dust peak of L1521E with those measured toward the dust peak of L1544. To estimate column densities toward L1544 we used the spectrum measured toward L1544 in the 81-106 GHz range as part of the ASAI \citep{lefloch2018}.  
The column densities of the species toward L1544 were derived with the same method as for L1521E described in Sect. \ref{sect:coldens}, using the same transitions of the species toward both cores. We assumed an excitation temperature of 5 K which was derived by \citet{crapsi2005} by fitting the N$_2$H$^+$ (1-0) hyperfine structure.
Since the H$_2$ column densities reported toward L1544 define a significant range, from 2.8$\times$10$^{22}$ cm$^{-2}$ \citep{spezzano2016} derived from \textit{Herschel}/SPIRE data to (9.4$\pm$1.6)$\times$10$^{22}$ cm$^{-2}$ derived from the 1.2 mm continuum \citep{crapsi2005}, we compared column density ratios toward the two cores. 
It is important to note that the column densities were derived from data measured with similar beam sizes. 
The column density ratios with the reference column density of A-CH$_3$OH toward the two cores are shown in Fig. \ref{coldens_l1521e_l1544}. 
One of the main differences between the column density ratios calculated toward the two sources is seen for the sulfur-bearing species with carbon, such as C$_2$S, HCS$^+$, C$^{34}$S, C$^{33}$S, and HCS$^+$, which have higher abundances toward L1521E than toward L1544. 
These C \& S-bearing molecules were found to follow the $c$-C$_3$H$_2$ peak in L1544 \citep{spezzano2017}, i.e. toward L1544 they appear to trace chemically young gas, probably affected by the interstellar radiation field. The relatively higher abundance of these molecules toward L1521E may indicate that sulfur depletion continues to take place during the dynamical evolution of dense cores, as also recently found by \citet{laascaselli2019}, therefore, organic S-bearing molecules can be considered good tracers of chemically young dense cores.
Among the S-bearing species studied here, SO was also found to trace cores in an early evolutionary state by other studies, such as \citet{hacar2013}, who found extended SO emission toward the L1495/B213 complex in Taurus. They interpreted SO to trace gas in regions of early evolutionary state as it was found to be bright in regions with bright C$^{18}$O emission and intense dust millimeter continuum together with little N$_2$H$^+$ emission. \citet{tafalla2006} studied the abundances of molecules toward the L1498 and L1517B starless cores in Taurus. Based on the comparison of observations to radiative transfer models they found SO and C$_2$S to be good tracers of molecular depletion.

The CH$_3$OH ($E$) relative abundances are similar in the two cores. 
The lower N$_2$H$^+$ abundance toward L1521E than toward L1544 confirms that L1521E is indeed in an earlier stage of core evolution, due to the relatively long formation route of N$_2$H$^+$ starting from atomic nitrogen (e.g. \citealp{hilyblant2010}).

In addition to the abundances, the spatial distribution of species helps us to understand better the difference between the two cores. 
\citet{spezzano2017} found four different molecular peaks toward L1544: dust peak, $c$-C$_3$H$_2$ peak, methanol peak, and HNCO peak. These peaks are also present toward L1521E, but they are closer together, compared to the more dynamically evolved L1544 pre-stellar core. N$_2$H$^+$ and CN peak close to the dust peak toward L1521E, and similarly N$_2$H$^+$ and $^{13}$CN also peak close to the dust peak toward L1544 \citep{spezzano2017}. Not many species were found to peak toward the methanol peak in L1521E. SO, which peaks toward the methanol peak in L1544 shows two peaks in L1521E, one toward the $c$-C$_3$H$_2$, and one close to the CH$_3$OH peak. \citet{spezzano2017} found that O- and S-bearing molecules follow the methanol distribution toward L1544, and therefore trace more chemically evolved gas or gas more protected from the interstellar radiation field (see also \citealp{swade1989}).
 
Most species, mostly carbon chain molecules, peak toward the $c$-C$_3$H$_2$ peak in both cores. One difference is CH$_3$CCH, which peaks toward the $c$-C$_3$H$_2$ peak in L1521E, and toward the HNCO peak in L1544.

\begin{figure}[h]
\centering   
\includegraphics[width=9cm, trim=0cm 0cm 0cm 0cm,clip=true]{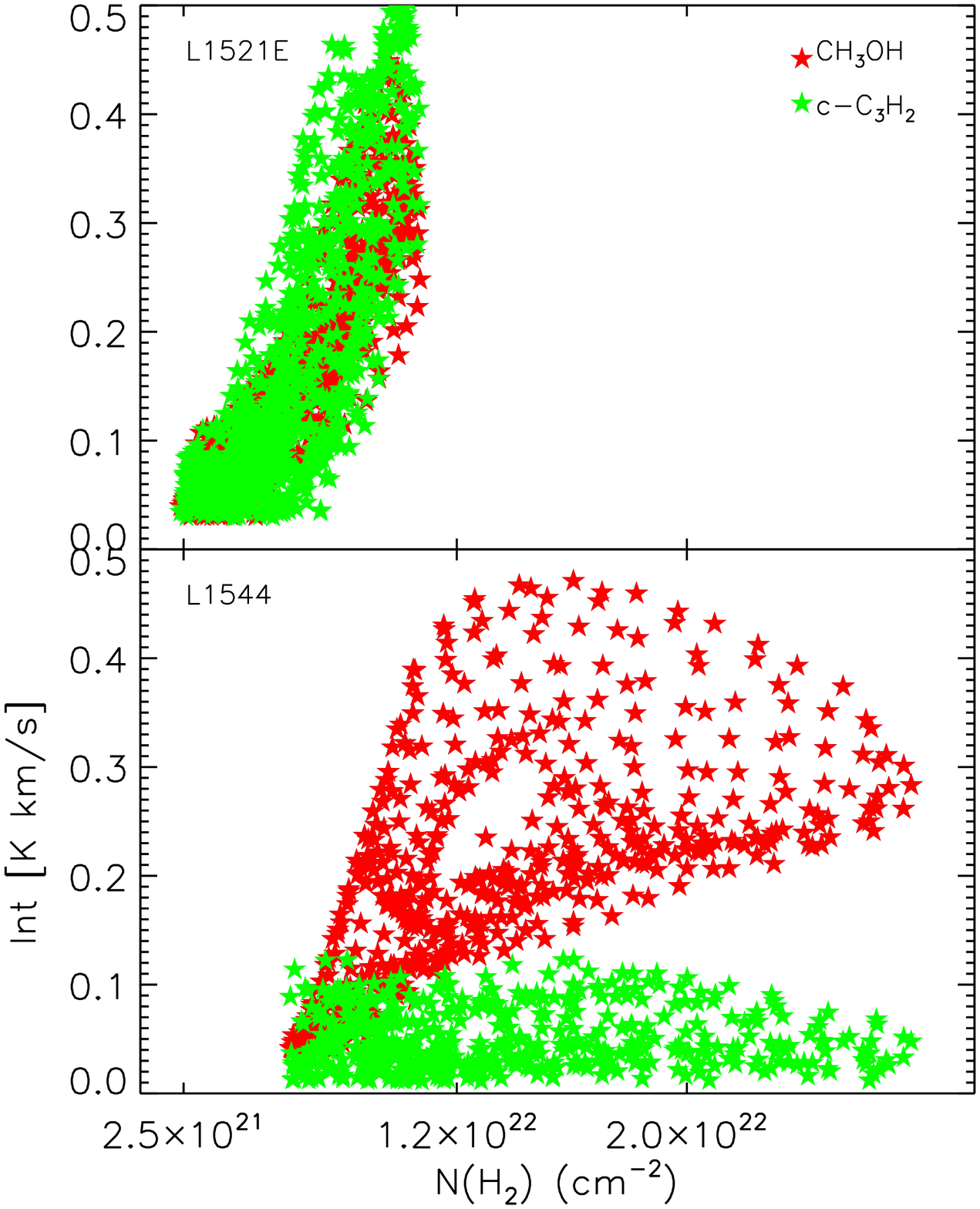}
\caption{Integrated intensities of \textit{c}-C$_3$H$_2$ (green symbols) and CH$_3$OH (red symbols) as a function of the H$_2$ column densities toward L1521E and L1544. The data plotted for L1544 are from \citep{spezzano2016}.}
\label{nh2_ch3oh_c3h2}
\end{figure}

As seen in Sect. \ref{spatial_dist} $c$-C$_3$H$_2$ and CH$_3$OH peak at different positions, similar to what was observed toward L1544 \citep{spezzano2016}. However, the interpretation of the different spatial distributions is different for L1521E than for L1544. As it was shown by \citet{spezzano2016} for L1544, as the $c$-C$_3$H$_2$ intensities do not correlate with the H$_2$ column densities, but rather show a flat distribution, unlike CH$_3$OH (Fig. \ref{nh2_ch3oh_c3h2}), $c$-C$_3$H$_2$ probably originates in a layer-like structure, tracing the external core envelope. However in the case of L1521E, even though they peak at different positions toward the core, $c$-C$_3$H$_2$ and CH$_3$OH both show a correlation with the H$_2$ column densities, which can be understood if the central density of the L1521E core is not high enough for catastrophic freeze-out to take place.

Overall, the chemical segregation in L1521E is not as developed as in L1544, which is probably related to the earlier evolutionary state of L1521E, and the fact that L1521E has not developed a high density central region, where freeze-out of C- and O-bearing bearing molecules dominates.

\section{Summary}
\label{sect_summary}

We presented observations toward the starless core L1521E carried out with the IRAM-30m telescope. The main results can be summarized as follows:

\begin{itemize}

\item The lower limit on the CO depletion factor toward the dust peak of L1521E was found to be 4.3$\pm$1.6, which is about one third of the value measured toward the more evolved L1544 core.  

\item $c$-C$_3$H$_2$ and CH$_3$OH peak at different positions in both L1521E and L1544 (based on \citealp{spezzano2016}), but the origin of the distributions toward the two cores is different, as toward L1521E both molecules show an increasing trend with increasing $N$(H$_2$) while toward L1544 only the CH$_3$OH intensities increase with increasing $N$(H$_2$), while \textit{c}-C$_3$H$_2$ shows a flat intensity profile. This difference is due to the lower central density of L1521E compared to L1544 and the fact that the region of catastrophic CO freeze-out is smaller in L1521E compared to L1544.

\item Comparing the relative abundances toward L1521E and L1544 we found that most sulfur bearing species such as C$_2$S, HCS$^+$, C$^{34}$S, C$^{33}$S, and HCS$^+$ are more abundant toward L1521E than toward L1544. This is confirmed by a chemical model which reproduces most column densities measured toward L1521E within an order of magnitude. 
This suggests that significant sulfur depletion is taking place during the dynamical evolution of dense cores.

\end{itemize}

\begin{acknowledgements}
"This work is based on observations carried out under project numbers 005-16 and 100-17 with the IRAM-30m telescope. IRAM is supported by INSU/CNRS (France), MPG (Germany) and IGN (Spain)."
The work by AV is supported by the Russian Science Foundation via the project 18-12-00351.
MT acknowledges funding from project AYA2016-79006-P.
\end{acknowledgements}

\begin{appendix}
\onecolumn

\begin{landscape}

\section{The observed lines and their parameters}

\setlength\LTcapwidth{22cm}

\begin{small}

\renewcommand{\footnoterule}{}
\renewcommand{\thefootnote}{\alph{footnote}}

\begin{longtable}{lrrrrrrrrccc}

\caption[]{
Molecules and transitions detected toward L1521E. 
The line parameters are measured toward the \textit{Herschel} dust peak using Gaussian fitting.
Some of the detections toward L1544 are based on the spectrum measured in the $\sim$80-106 GHz frequency range toward the dust peak of L1544 as part of ASAI (\citealp{lefloch2018}, \citealp{vastel2018}).
The transitions marked with asterisk were used to calculate column densities (Section \ref{sect:coldens}) and for the principal component analysis (Section \ref{sect:pca}).
}

\label{table:observations} \\

\hline
Molecule& Transition& $\nu$& $A_{ij}$& $E_{\rm{up}}$& $\int T_{\rm A^*}{\mathrm d}V$ & $V_{\rm LSR}$& $\Delta V$& $T_{\rm peak}$& 3-$\sigma$ rms& Critical& Detected toward\\
& & (MHz)& (s$^{-1}$)& (K)& (K km s$^{-1}$)& (km s$^{-1}$)& (km s$^{-1}$)& (K)& (K)& density (cm$^{-3}$)& L1544\footnotemark[1]\\
\hline\\[-0.25cm]

\endhead

\hline \multicolumn{12}{|r|}{{Continued on next page}} \\ \hline
\endfoot

\hline \hline
\endlastfoot

\hline

C$_4$H& $N$=8-7, $J$=15/2-13/2, $F$=7-6 \& 8-7& 76156.0& 1.82$\times$10$^{-6}$& 16.5& 
0.19$\pm$0.01& 6.76$\pm$0.01& 0.37$\pm$0.02& 0.47& 0.13& $-$& $-$\\

CH$_3$CHO$^*$& 	$4_{0,4}-3_{0,3}$, $E$& 76866.4& 1.43$\times$10$^{-5}$&  9.3& 
0.05$\pm$0.01& 6.79$\pm$0.04& 0.40$\pm$0.09& 0.12& 0.03& $-$& $-$\\

CH$_3$CHO& $4_{0,4}-3_{0,3}$, $A$& 76879.0& 1.43$\times$10$^{-5}$&  9.2& 
0.06$\pm$0.01& 7.07$\pm$0.03& 0.39$\pm$0.07& 0.15& 0.03& $-$& $-$\\

C$_2$S&          $N$=6-5, $J$=6-5& 77731.7& 2.03$\times$10$^{-5}$& 21.8& 
0.23$\pm$0.01& 6.77$\pm$0.01& 0.36$\pm$0.02& 0.62& 0.18& $-$& $-$\\

CC$^{34}$S$^*$&  $N$=6-5, $J$=7-6& 79827.5& 2.31$\times$10$^{-5}$& 15.1& 
0.05$\pm$0.01& 7.03$\pm$0.03& 0.32$\pm$0.05& 0.16& 0.04& $-$& $-$\\

$l$-C$_4$H$_2^*$& $9_{1,9}-8_{1,8}$& 80046.7& 4.69$\times$10$^{-5}$& 32.7& 
0.03$\pm$0.01& 6.81$\pm$0.05& 0.35$\pm$0.08& 0.08& 0.01& $-$& $-$\\

H$_2$CCO$^*$&  $J_{K_a;K_c}=4_{1,4}-3_{1,3}$& 80076.7& 5.04$\times$10$^{-6}$& 22.7& 
0.07$\pm$0.01&  7.04$\pm$0.02& 0.34$\pm$0.05& 0.20& 0.05& $-$& $-$\\ 

H$_2$CCO&      $J_{K_a;K_c}=4_{0,4}-3_{0,3}$& 80832.1& 5.52$\times$10$^{-6}$& 9.7& 
0.05$\pm$0.01&  6.93$\pm$0.07& 0.53$\pm$0.13& 0.09& 0.03& $-$& $-$\\

C$_3$S$^*$&              $J$=14-13& 80928.2& 3.86$\times$10$^{-5}$& 29.1& 
0.10$\pm$0.02& 6.89$\pm$0.02& 0.26$\pm$0.06& 0.35& 0.07&  $-$& $-$\\


C$_2$S$^*$& $N$=6-5, $J$=7-6& 81505.2& 2.46$\times$10$^{-5}$& 15.4& 
0.81$\pm$0.02& 6.71$\pm$0.01& 0.35$\pm$0.01& 2.18& 0.35& $-$& Yes\\

HC$_3$N$^*$&                  $J$=9-8& 81881.5& 4.20$\times$10$^{-5}$& 19.7& 
0.80$\pm$0.01& 6.87$\pm$0.01& 0.33$\pm$0.01& 2.30& 0.34& 4.2$\times$10$^{5}$& Yes\\

\textit{c}-C$_3$H$_2^*$& $J_{K_a;K_c}=2_{0,2}-1_{1,1}$& 82093.5& 2.07$\times$10$^{-5}$& 6.4& 
0.41$\pm$0.01& 6.84$\pm$0.01& 0.33$\pm$0.01& 1.16& 0.17& 1.1$\times$10$^{6}$& Yes\\

OCS$^*$&                       $J$=7-6& 85139.1& 1.71$\times$10$^{-6}$& 16.3& 
0.06$\pm$0.01& 6.82$\pm$0.04& 0.43$\pm$0.09& 0.13& 0.04& 2.3$\times$10$^{4}$& Yes\\ 

HC$^{18}$O$^{+*}$&             $J$=1-0& 85162.2& 3.64$\times$10$^{-5}$& 4.1& 
0.07$\pm$0.01& 6.80$\pm$0.02& 0.33$\pm$0.05& 0.21& 0.05& 1.4$\times$10$^{5}$& Yes\\

$c$-C$_3$H$_2$& $J_{K_a;K_c}=2_{1,2}-1_{0,1}$& 85338.9& 2.55$\times$10$^{-5}$& 6.5& 
0.81$\pm$0.01& 6.88$\pm$0.01& 0.39$\pm$0.01& 1.95& 0.51& 1.2$\times$10$^{6}$& Yes\\

HCS$^{+*}$&                   $J$=2-1& 85347.9& 1.33$\times$10$^{-5}$& 6.1& 
0.24$\pm$0.01& 6.94$\pm$0.01& 0.27$\pm$0.02& 0.83& 0.11& 3.5$\times$10$^{4}$& Yes\\

CH$_3$CCH&            $J_K=5_1$-$4_1$& 85455.6& 1.78$\times$10$^{-6}$& 19.5& 
0.16$\pm$0.01& 6.60$\pm$0.01& 0.34$\pm$0.02& 0.45& 0.12& $-$& Yes\\

CH$_3$CCH&            $J_K=5_0$-$4_0$& 85457.3& 2.03$\times$10$^{-6}$& 12.3& 
0.18$\pm$0.01& 6.84$\pm$0.01& 0.34$\pm$0.02& 0.48& 0.12& $-$& Yes\\

C$_4$H$^*$&    $N$=9-8, $J$=19/2-17/2, $F$=10-9 \& 9-8& 85634.0& 2.81$\times$10$^{-6}$& 20.6& 
0.15$\pm$0.01& 6.87$\pm$0.01& 0.31$\pm$0.03& 0.46& 0.10& $-$& Yes\\

C$_4$H&        $N$=9-8, $J$=17/2-15/2, $F$=9-8 \& 8-7& 85672.6& 2.79$\times$10$^{-6}$& 20.6& 
0.14$\pm$0.01& 6.93$\pm$0.02& 0.27$\pm$0.02& 0.48& 0.10& $-$& Yes\\

SO&            $N_J=2_2$-$1_1$& 86094.0& 5.25$\times$10$^{-6}$& 19.3&  
0.10$\pm$0.01& 7.06$\pm$0.02& 0.36$\pm$0.03& 0.24& 0.05& 8.2$\times$10$^{4}$& Yes\\

C$_2$S&        $N_J=7_6$-$6_5$& 86181.4& 2.78$\times$10$^{-5}$& 23.4&  
0.24$\pm$0.01& 6.83$\pm$0.01& 0.31$\pm$0.01& 0.71& 0.14& $-$& Yes\\
    
H$^{13}$CN&     $J_F=1_1$-$0_1$& 86338.8& 2.22$\times$10$^{-5}$& 4.1&  
0.11$\pm$0.01& 7.06$\pm$0.03& 0.42$\pm$0.06& 0.24& 0.10& $-$& Yes\\

H$^{13}$CN$^*$& $J_F=1_2$-$0_1$& 86340.2& 2.22$\times$10$^{-5}$& 4.1&  
0.15$\pm$0.02& 6.95$\pm$0.03& 0.41$\pm$0.06& 0.34& 0.10& $-$& Yes\\

H$^{13}$CN&     $J_F=1_0$-$0_1$& 86342.3& 2.22$\times$10$^{-5}$& 4.1& 
0.04$\pm$0.01& 6.97$\pm$0.03& 0.29$\pm$0.07& 0.14& 0.10& $-$& Yes\\

HCO$^*$&  $N_{K_a,K_c}=1_{0,1}-0_{0,0}$, $J$=3/2–1/2, $F$=2-1& 86670.8& 4.69$\times$10$^{-6}$& 4.2&  
0.07$\pm$0.01& 7.17$\pm$0.03& 0.39$\pm$0.06& 0.17& 0.05& $-$& Yes\\

HN$^{13}$C$^*$&               $J$=1-0&  87090.9& 1.87$\times$10$^{-5}$& 4.2& 
0.31$\pm$0.04& 7.13$\pm$0.03& 0.58$\pm$0.08& 0.49& 0.17& $-$& Yes\\

C$_2$H&	    $N$=1-0, $J$=3/2-1/2, $F$=1-1&  87284.2&  3.75$\times$10$^{-7}$&  4.2& 	
0.33$\pm$0.05& 6.98$\pm$0.02& 0.31$\pm$0.05& 0.98& 0.20& 7.5$\times$10$^4$& Yes\\

C$_2$H& $N$=1-0, $J$=3/2-1/2, $F$=2-1&	87316.9&  2.21$\times$10$^{-6}$&  4.2&	
0.79$\pm$0.05& 6.96$\pm$0.01& 0.40$\pm$0.03& 1.88& 0.46& 1.8$\times$10$^5$& Yes\\

C$_2$H&	    $N$=1-0, $J$=3/2-1/2, $F$=1-0& 	87328.6&  1.83$\times$10$^{-6}$&  4.2&
0.58$\pm$0.04& 7.00$\pm$0.01& 0.38$\pm$0.03& 1.46& 0.34& 2.3$\times$10$^5$& Yes\\

C$_2$H&	    $N$=1-0, $J$=1/2-1/2, $F$=1-1&	87402.0&  1.83$\times$10$^{-6}$&  4.2&	
0.44$\pm$0.05& 6.94$\pm$0.02& 0.33$\pm$0.04& 1.24& 0.38& 1.7$\times$10$^5$& Yes\\

C$_2$H$^*$&	$N$=1-0, $J$=1/2-1/2, $F$=0-1&	87407.2&  2.22$\times$10$^{-6}$&  4.2&	
0.38$\pm$0.04& 6.87$\pm$0.01& 0.30$\pm$0.04& 1.18& 0.21& 1.4$\times$10$^5$& Yes\\

C$_2$H&	    $N$=1-0, $J$=1/2-1/2, $F$=1-0&	87446.5&  3.77$\times$10$^{-7}$&  4.2&	
0.21$\pm$0.04& 7.01$\pm$0.02& 0.26$\pm$0.09& 0.76& 0.18& 7.1$\times$10$^4$& Yes\\

HNCO$^*$&   $J_{Ka,Kc}=4_{0,4}-3_{0,3}$&  87925.3& 8.46$\times$10$^{-6}$&	10.6&
0.21$\pm$0.04& 6.81$\pm$0.05& 0.54$\pm$0.12& 0.37& 0.11& 1.2$\times$10$^6$& Yes\\

HCN&      $J_F=1_1$-$0_1$&  88630.4& 2.43$\times$10$^{-5}$& 4.3& 
0.62$\pm$0.01& 6.86$\pm$0.01& 0.55$\pm$0.01& 1.06& 0.41& 4.3$\times$10$^{6}$& Yes\\

HCN$^*$&  $J_F=1_2$-$0_1$&  88631.9& 2.43$\times$10$^{-5}$& 4.3& 
0.85$\pm$0.08& 7.14$\pm$0.03& 0.49$\pm$0.06& 1.61& 0.41& 4.3$\times$10$^{6}$& Yes\\

HCN&      $J_F=1_0$-$0_1$&  88633.9& 2.43$\times$10$^{-5}$& 4.3& 
0.65$\pm$0.01& 6.73$\pm$0.01& 0.52$\pm$0.01& 1.19& 0.41& 4.3$\times$10$^{6}$& Yes\\

HCO$^{+*}$&                 $J$=1-0&  89188.5& 4.19$\times$10$^{-5}$& 4.3& 
1.46$\pm$0.01& 6.86$\pm$0.01& 0.55$\pm$0.01& 2.50& 0.79& 1.6$\times$10$^{5}$& Yes\\

CH$_3$CN$^*$&         $J_K=5_1$-$4_1$& 91985.3& 6.08$\times$10$^{-5}$& 20.4& 
0.03$\pm$0.01& 6.76$\pm$0.06& 0.38$\pm$0.10& 0.08& 0.03& 5.3$\times$10$^{5}$& Yes\\

CH$_3$CN&             $J_K=5_0$-$4_0$& 91987.1& 6.33$\times$10$^{-5}$& 13.2& 
0.04$\pm$0.01& 6.98$\pm$0.05& 0.48$\pm$0.08& 0.09& 0.03& 5.4$\times$10$^{5}$& Yes\\

C$_3$S&               $J$=16-15& 92488.5& 6.13$\times$10$^{-5}$& 37.7& 
0.06$\pm$0.01& 6.91$\pm$0.03& 0.40$\pm$0.07& 0.13& 0.03& $-$& Yes\\

$^{13}$CS$^*$&        $J_F$=$2_0-1_0$& 92494.3& 1.41$\times$10$^{-5}$&  6.7& 
0.30$\pm$0.01& 6.95$\pm$0.01& 0.38$\pm$0.01& 0.73& 0.16& $-$& Yes\\

N$_2$H$^+$& $J$=1-0, $F_1$=1-1, $F$=0-1& 93171.6&	1.45$\times$10$^{-6}$& 4.5& 
0.04$\pm$0.01& 6.80$\pm$0.06& 0.40\footnotemark[2]& 0.09& 0.23& 9.1$\times$10$^{3}$& Yes\\	

N$_2$H$^+$& $J$=1-0, $F_1$=1-1, $F$=2-2& 93171.9&	6.89$\times$10$^{-6}$& 4.5&	
0.10$\pm$0.01& 6.83$\pm$0.02& 0.40\footnotemark[2]& 0.24& 0.23& 1.7$\times$10$^{5}$& Yes\\

N$_2$H$^+$& $J$=1-0, $F_1$=1-1, $F$=1-0& 93172.1&   3.99$\times$10$^{-6}$& 4.5&
0.07$\pm$0.01& 7.09$\pm$0.03& 0.40\footnotemark[2]& 0.16& 0.23& 1.0$\times$10$^{5}$& Yes\\

N$_2$H$^+$& $J$=1-0, $F_1$=2-1, $F$=2-1& 93173.5&   6.53$\times$10$^{-6}$& 4.5&	
0.10$\pm$0.01& 6.95$\pm$0.01& 0.40$\pm$0.03&        0.24& 0.23& 1.6$\times$10$^{5}$& Yes\\

N$_2$H$^+$& $J$=1-0, $F_1$=2-1, $F$=3-2& 93173.8& 	9.43$\times$10$^{-6}$& 4.5&	
0.16$\pm$0.01& 6.98$\pm$0.01& 0.41$\pm$0.03&        0.36& 0.23& 5.9$\times$10$^{4}$& Yes\\

N$_2$H$^+$& $J$=1-0, $F_1$=2-1, $F$=1-1& 93174.0&	3.99$\times$10$^{-6}$& 4.5&	
0.06$\pm$0.01& 7.02$\pm$0.03& 0.38$\pm$0.06&        0.15& 0.23& 8.9$\times$10$^{5}$& Yes\\

N$_2$H$^{+*}$& $J$=1-0, $F_1$=0-1, $F$=1-2&  93176.3&  3.99$\times$10$^{-6}$& 4.5&
0.07$\pm$0.01& 6.91$\pm$0.02& 0.35$\pm$0.04&        0.19& 0.23& 2.2$\times$10$^{5}$& Yes\\

C$_4$H&   $N$=10-9, $J$=21/2-19/2, $F$=10-9 \& 11-10& 95150.4& 3.60$\times$10$^{-6}$& 25.1& 
0.10$\pm$0.01& 6.93$\pm$0.01& 0.30$\pm$0.02& 0.32& 0.07& $-$& Yes\\

C$_4$H&   $N$=10-9, $J$=19/2-17/2, $F$=9-8 \& 10-9& 95189.0& 3.58$\times$10$^{-6}$& 25.1& 
0.10$\pm$0.01& 7.03$\pm$0.02& 0.37$\pm$0.05& 0.24& 0.07& $-$& Yes\\

CH$_3$CHO&   $5_{0,5}-4_{0,4}$, $(E)$& 95947.4& 2.84$\times$10$^{-5}$& 13.9& 
0.03$\pm$0.01& 6.65$\pm$0.05& 0.40$\pm$0.05& 0.08& 0.02& $-$& Yes\\

C$^{34}$S$^*$&                $J$=2-1& 96412.9& 1.61$\times$10$^{-5}$&
6.3& 0.64$\pm$0.01& 6.86$\pm$0.01& 0.40$\pm$0.01& 1.51& 0.30& $-$& Yes\\

CH$_3$OH$^*$&   $2_{-1,0}-1_{-1,0}$ ($E_2$)& 96739.3& 2.56$\times$10$^{-6}$& 12.5& 
0.32$\pm$0.01& 6.67$\pm$0.01& 0.40$\pm$0.01& 0.76& 0.20& 2.9$\times$10$^{4}$& Yes\\

CH$_3$OH$^*$&   $2_{0,2}-1_{0,1}$ ($A^+$)& 96741.4& 3.41$\times$10$^{-6}$& 7.0& 
0.43$\pm$0.01& 6.93$\pm$0.01& 0.39$\pm$0.01& 1.05& 0.20& 3.1$\times$10$^{4}$& Yes\\

C$^{33}$S$^*$&  $J$=2-1, $F$=5/2-3/2, 7/2-5/2, \& 1/2-1/2& 97171.8& 1.63$\times$10$^{-5}$& 6.3& 
0.12$\pm$0.01& 6.83$\pm$0.01& 0.29$\pm$0.03& 0.39& 0.08& $-$& Yes\\

$^{34}$SO$^*$&           $N_J=2_3-1_2$& 97715.4& 1.09$\times$10$^{-5}$& 9.1& 
0.09$\pm$0.02& 6.81$\pm$0.02& 0.26$\pm$0.07& 0.32& 0.03& $-$& Yes\\

CS&            	         $J$=2-1& 97981.0& 1.70$\times$10$^{-5}$& 7.1& 
1.53$\pm$0.01& 7.00$\pm$0.01& 0.57$\pm$0.01& 2.51& 0.92& 3.4$\times$10$^{5}$& Yes\\

SO$^*$&                  $J$=3-2, $N$=2-1& 99299.9& 1.15$\times$10$^{-5}$& 9.2& 
0.82$\pm$0.02& 6.92$\pm$0.01& 0.38$\pm$0.01& 2.05& 0.39& 2.8$\times$10$^{5}$& Yes\\

C$_2$S&                  $N$=8-7, $J$=7-6& 99866.5& 4.47$\times$10$^{-5}$& 28.1& 
0.14$\pm$0.01& 6.84$\pm$0.01& 0.31$\pm$0.02& 0.41& 0.09& $-$& Yes\\

HC$_3$N&                $J$=11-10& 100076.4& 7.74$\times$10$^{-5}$& 28.8& 
0.43$\pm$0.01& 6.87$\pm$0.01& 0.36$\pm$0.01& 1.12& 0.31& 8.8$\times$10$^{5}$& Yes\\

H$_2$CS&        $J_{Ka,Kc}=3_{1,3}-2_{1,2}$& 101477.8& 1.26$\times$10$^{-5}$& 22.9& 
0.63$\pm$0.01& 6.79$\pm$0.01& 0.33$\pm$0.01& 1.81& 0.41& 1.6$\times$10$^{5}$& Yes\\

H$_2$CCO&       $J_{Ka,Kc}=5_{1,4}-4_{1,3}$& 101981.4& 1.09$\times$10$^{-5}$& 27.7& 
0.09$\pm$0.01& 6.77$\pm$0.02& 0.37$\pm$0.07& 0.24& 0.04& $-$& Yes\\

CH$_3$CCH&              $J_K=6_1-5_1$& 102546.0& 3.17$\times$10$^{-6}$& 24.4& 
0.13$\pm$0.02& 6.91$\pm$0.03& 0.32$\pm$0.06& 0.38& 0.10& $-$& Yes\\ 

CH$_3$CCH$^*$&          $J_K=6_0-5_0$& 102548.0& 3.26$\times$10$^{-6}$& 17.2& 
0.12$\pm$0.02& 6.79$\pm$0.02& 0.29$\pm$0.04& 0.40& 0.10& $-$& Yes\\

H$_2$CS$^*$&    $J_{Ka,Kc}=3_{0,3}-2_{0,2}$& 103040.5& 1.48$\times$10$^{-5}$& 9.8& 
0.58$\pm$0.02& 6.85$\pm$0.01& 0.33$\pm$0.02& 1.65& 0.22& 1.4$\times$10$^{5}$& Yes\\

H$_2$CS&        $J_{Ka,Kc}=3_{1,2}-2_{1,1}$& 104617.0& 1.38$\times$10$^{-5}$& 23.2& 
0.59$\pm$0.01& 6.75$\pm$0.01& 0.34$\pm$0.01& 1.65& 0.38& 2.3$\times$10$^{5}$& Yes\\

C$_4$H&   $N$=11-10, $J$=23/2-21/2, $F$=11-10 \& 12-11& 104666.6& 4.83$\times$10$^{-6}$& 30.1& 
0.07$\pm$0.01& 6.95$\pm$0.03& 0.42$\pm$0.08& 0.16& 0.04& $-$& Yes\\


C$^{17}$O& $J$=1-0 $F$=3/2-5/2&	112358.8& 6.70$\times$10$^{-8}$& 5.4& 
0.12$\pm$0.02& 6.89$\pm$0.02& 0.31$\pm$0.05& 0.36& 0.15& 2.0$\times$10$^{3}$& $-$\\

C$^{17}$O& $J$=1-0 $F$=7/2-5/2&	112359.0& 6.70$\times$10$^{-8}$& 5.4& 
0.31$\pm$0.02& 6.86$\pm$0.01& 0.41$\pm$0.03& 0.71& 0.15& 2.0$\times$10$^{3}$& $-$\\  

C$^{17}$O$^*$& $J$=1-0 $F$=5/2-5/2&	112360.0& 6.70$\times$10$^{-8}$& 5.4& 
0.24$\pm$0.01& 6.78$\pm$0.01& 0.42$\pm$0.03& 0.54& 0.15& 2.0$\times$10$^{3}$& $-$\\

CN& $N$=1-0, $J$=1/2-1/2, $F$=1/2-3/2& 113144.2& 1.05$\times$10$^{-5}$& 5.4& 
0.27$\pm$0.02& 6.90$\pm$0.02& 0.46$\pm$0.03& 0.55& 0.14& 3.4$\times$10$^{5}$& $-$\\

CN& $N$=1-0, $J$=1/2-1/2, $F$=3/2-1/2& 113170.5& 5.14$\times$10$^{-6}$& 5.4& 
0.21$\pm$0.02& 6.85$\pm$0.02& 0.40$\pm$0.04& 0.48& 0.14& 6.3$\times$10$^{4}$& $-$\\

CN$^*$& $N$=1-0, $J$=1/2-1/2, $F$=3/2-3/2& 113191.3& 6.68$\times$10$^{-6}$& 5.4& 
0.26$\pm$0.02& 6.89$\pm$0.01& 0.44$\pm$0.03& 0.55& 0.14& 1.3$\times$10$^{5}$& $-$\\

\hline

\end{longtable}

\footnotetext[1]{The transitions which are marked with '$-$' were not covered in the available dataset for L1544.}
\footnotetext[2]{Fixed parameter in the fit.}

\end{small}

\end{landscape}

\end{appendix}

\end{document}